\documentclass[traditabstract,longauth]{aa}
\usepackage{txfonts}
\usepackage{graphicx}
\usepackage{natbib}
\usepackage{color}
\usepackage{lscape}
\usepackage{fixltx2e}
\usepackage{afterpage}
\usepackage{verbatim}
\usepackage[colorlinks=true,citecolor=blue]{hyperref}

\def\setsymbol#1#2{\expandafter\def\csname #1\endcsname{#2}}
\def\getsymbol#1{\csname #1\endcsname}

\def\Planck{\textit{Planck}}

\def\all2013resultspapers{\nocite{planck2013-p01, planck2013-p02, planck2013-p02a, planck2013-p02d, planck2013-p02b, planck2013-p03, planck2013-p03c, planck2013-p03f, planck2013-p03d, planck2013-p03e, planck2013-p06, planck2013-p03a, planck2013-pip88, planck2013-p08, planck2013-p11, planck2013-p12, planck2013-p13, planck2013-p14, planck2013-p15, planck2013-p05b, planck2013-p17, planck2013-p09, planck2013-p09a, planck2013-p20, planck2013-p19, planck2013-pipaberration, planck2013-p05, planck2013-p05a, planck2013-pip56, planck2013-p06b, planck2013-p01a}}

\newbox\tablebox    \newdimen\tablewidth
\def\leaderfil{\leaders\hbox to 5pt{\hss.\hss}\hfil}
\def\endPlancktablewide{\tablewidth=\textwidth 
    $$\hss\copy\tablebox\hss$$
    \vskip-\lastskip\vskip -2pt}
\def\tablenote#1 #2\par{\begingroup \parindent=0.8em
    \abovedisplayshortskip=0pt\belowdisplayshortskip=0pt
    \noindent
    $$\hss\vbox{\hsize\tablewidth \hangindent=\parindent \hangafter=1 \noindent
    \hbox to \parindent{$^#1$\hss}\strut#2\strut\par}\hss$$
    \endgroup}
\def\doubleline{\vskip 3pt\hrule \vskip 1.5pt \hrule \vskip 5pt}

\def\L2{\ifmmode L_2\else $L_2$\fi}

\def\DeltaT{\ifmmode \Delta T\else $\Delta T$\fi}
\def\deltat{\ifmmode \Delta t\else $\Delta t$\fi}
\def\fknee{\ifmmode f_{\rm knee}\else $f_{\rm knee}$\fi}
\def\Fmax{\ifmmode F_{\rm max}\else $F_{\rm max}$\fi}
\def\solar{\ifmmode{\rm M}_{\mathord\odot}\else${\rm M}_{\mathord\odot}$\fi}
\def\Msolar{\ifmmode{\rm M}_{\mathord\odot}\else${\rm M}_{\mathord\odot}$\fi}
\def\Lsolar{\ifmmode{\rm L}_{\mathord\odot}\else${\rm L}_{\mathord\odot}$\fi}
\def\inv{\ifmmode^{-1}\else$^{-1}$\fi}
\def\mo{\ifmmode^{-1}\else$^{-1}$\fi}
\def\sup#1{\ifmmode ^{\rm #1}\else $^{\rm #1}$\fi}
\def\expo#1{\ifmmode \times 10^{#1}\else $\times 10^{#1}$\fi}
\def\,{\thinspace}
\def\lsim{\mathrel{\raise .4ex\hbox{\rlap{$<$}\lower 1.2ex\hbox{$\sim$}}}}
\def\gsim{\mathrel{\raise .4ex\hbox{\rlap{$>$}\lower 1.2ex\hbox{$\sim$}}}}

\def\simprop{\mathrel{\raise .4ex\hbox{\rlap{$\propto$}\lower 1.2ex\hbox{$\sim$}}}}
\def\deg{\ifmmode^\circ\else$^\circ$\fi}
\def\pdeg{\ifmmode $\setbox0=\hbox{$^{\circ}$}\rlap{\hskip.11\wd0 .}$^{\circ}
          \else \setbox0=\hbox{$^{\circ}$}\rlap{\hskip.11\wd0 .}$^{\circ}$\fi}
\def\arcs{\ifmmode {^{\scriptstyle\prime\prime}}
          \else $^{\scriptstyle\prime\prime}$\fi}
\def\arcm{\ifmmode {^{\scriptstyle\prime}}
          \else $^{\scriptstyle\prime}$\fi}
\newdimen\sa  \newdimen\sb
\def\parcs{\sa=.07em \sb=.03em
     \ifmmode \hbox{\rlap{.}}^{\scriptstyle\prime\kern -\sb\prime}\hbox{\kern -\sa}
     \else \rlap{.}$^{\scriptstyle\prime\kern -\sb\prime}$\kern -\sa\fi}
\def\parcm{\sa=.08em \sb=.03em
     \ifmmode \hbox{\rlap{.}\kern\sa}^{\scriptstyle\prime}\hbox{\kern-\sb}
     \else \rlap{.}\kern\sa$^{\scriptstyle\prime}$\kern-\sb\fi}
\def\ra[#1 #2 #3.#4]{#1\sup{h}#2\sup{m}#3\sup{s}\llap.#4}
\def\dec[#1 #2 #3.#4]{#1\deg#2\arcm#3\arcs\llap.#4}
\def\deco[#1 #2 #3]{#1\deg#2\arcm#3\arcs}
\def\rra[#1 #2]{#1\sup{h}#2\sup{m}}

\def\dots{\relax\ifmmode \ldots\else $\ldots$\fi}
\def\WHzsr{\ifmmode $W\,Hz\mo\,sr\mo$\else W\,Hz\mo\,sr\mo\fi}
\def\mHz{\ifmmode $\,mHz$\else \,mHz\fi}
\def\GHz{\ifmmode $\,GHz$\else \,GHz\fi}
\def\mKs{\ifmmode $\,mK\,s$^{1/2}\else \,mK\,s$^{1/2}$\fi}
\def\muKs{\ifmmode \,\mu$K\,s$^{1/2}\else \,$\mu$K\,s$^{1/2}$\fi}
\def\muKRJs{\ifmmode \,\mu$K$_{\rm RJ}$\,s$^{1/2}\else \,$\mu$K$_{\rm RJ}$\,s$^{1/2}$\fi}
\def\muKHz{\ifmmode \,\mu$K\,Hz$^{-1/2}\else \,$\mu$K\,Hz$^{-1/2}$\fi}
\def\MJysr{\ifmmode \,$MJy\,sr\mo$\else \,MJy\,sr\mo\fi}
\def\MJysrmK{\ifmmode \,$MJy\,sr\mo$\,mK$_{\rm CMB}\mo\else \,MJy\,sr\mo\,mK$_{\rm CMB}\mo$\fi}
\def\microns{\ifmmode \,\mu$m$\else \,$\mu$m\fi}

\def\muK{\ifmmode \,\mu$K$\else \,$\mu$\hbox{K}\fi}
\def\microK{\ifmmode \,\mu$K$\else \,$\mu$\hbox{K}\fi}
\def\muW{\ifmmode \,\mu$W$\else \,$\mu$\hbox{W}\fi}
\def\kms{\ifmmode $\,km\,s$^{-1}\else \,km\,s$^{-1}$\fi}
\def\kmsMpc{\ifmmode $\,\kms\,Mpc\mo$\else \,\kms\,Mpc\mo\fi}
\setsymbol{LFI:center:frequency:70GHz:units}{70.3\,GHz}
\setsymbol{LFI:center:frequency:44GHz:units}{44.1\,GHz}
\setsymbol{LFI:center:frequency:30GHz:units}{28.5\,GHz}

\setsymbol{LFI:center:frequency:70GHz}{70.3}
\setsymbol{LFI:center:frequency:44GHz}{44.1}
\setsymbol{LFI:center:frequency:30GHz}{28.5}

\setsymbol{LFI:center:frequency:LFI18:Rad:M:units}{71.7\GHz}
\setsymbol{LFI:center:frequency:LFI19:Rad:M:units}{67.5\GHz}
\setsymbol{LFI:center:frequency:LFI20:Rad:M:units}{69.2\GHz}
\setsymbol{LFI:center:frequency:LFI21:Rad:M:units}{70.4\GHz}
\setsymbol{LFI:center:frequency:LFI22:Rad:M:units}{71.5\GHz}
\setsymbol{LFI:center:frequency:LFI23:Rad:M:units}{70.8\GHz}
\setsymbol{LFI:center:frequency:LFI24:Rad:M:units}{44.4\GHz}
\setsymbol{LFI:center:frequency:LFI25:Rad:M:units}{44.0\GHz}
\setsymbol{LFI:center:frequency:LFI26:Rad:M:units}{43.9\GHz}
\setsymbol{LFI:center:frequency:LFI27:Rad:M:units}{28.3\GHz}
\setsymbol{LFI:center:frequency:LFI28:Rad:M:units}{28.8\GHz}
\setsymbol{LFI:center:frequency:LFI18:Rad:S:units}{70.1\GHz}
\setsymbol{LFI:center:frequency:LFI19:Rad:S:units}{69.6\GHz}
\setsymbol{LFI:center:frequency:LFI20:Rad:S:units}{69.5\GHz}
\setsymbol{LFI:center:frequency:LFI21:Rad:S:units}{69.5\GHz}
\setsymbol{LFI:center:frequency:LFI22:Rad:S:units}{72.8\GHz}
\setsymbol{LFI:center:frequency:LFI23:Rad:S:units}{71.3\GHz}
\setsymbol{LFI:center:frequency:LFI24:Rad:S:units}{44.1\GHz}
\setsymbol{LFI:center:frequency:LFI25:Rad:S:units}{44.1\GHz}
\setsymbol{LFI:center:frequency:LFI26:Rad:S:units}{44.1\GHz}
\setsymbol{LFI:center:frequency:LFI27:Rad:S:units}{28.5\GHz}
\setsymbol{LFI:center:frequency:LFI28:Rad:S:units}{28.2\GHz}

\setsymbol{LFI:center:frequency:LFI18:Rad:M}{71.7}
\setsymbol{LFI:center:frequency:LFI19:Rad:M}{67.5}
\setsymbol{LFI:center:frequency:LFI20:Rad:M}{69.2}
\setsymbol{LFI:center:frequency:LFI21:Rad:M}{70.4}
\setsymbol{LFI:center:frequency:LFI22:Rad:M}{71.5}
\setsymbol{LFI:center:frequency:LFI23:Rad:M}{70.8}
\setsymbol{LFI:center:frequency:LFI24:Rad:M}{44.4}
\setsymbol{LFI:center:frequency:LFI25:Rad:M}{44.0}
\setsymbol{LFI:center:frequency:LFI26:Rad:M}{43.9}
\setsymbol{LFI:center:frequency:LFI27:Rad:M}{28.3}
\setsymbol{LFI:center:frequency:LFI28:Rad:M}{28.8}
\setsymbol{LFI:center:frequency:LFI18:Rad:S}{70.1}
\setsymbol{LFI:center:frequency:LFI19:Rad:S}{69.6}
\setsymbol{LFI:center:frequency:LFI20:Rad:S}{69.5}
\setsymbol{LFI:center:frequency:LFI21:Rad:S}{69.5}
\setsymbol{LFI:center:frequency:LFI22:Rad:S}{72.8}
\setsymbol{LFI:center:frequency:LFI23:Rad:S}{71.3}
\setsymbol{LFI:center:frequency:LFI24:Rad:S}{44.1}
\setsymbol{LFI:center:frequency:LFI25:Rad:S}{44.1}
\setsymbol{LFI:center:frequency:LFI26:Rad:S}{44.1}
\setsymbol{LFI:center:frequency:LFI27:Rad:S}{28.5}
\setsymbol{LFI:center:frequency:LFI28:Rad:S}{28.2}

\setsymbol{LFI:white:noise:sensitivity:70GHz:units}{134.7\muKs}
\setsymbol{LFI:white:noise:sensitivity:44GHz:units}{164.7\muKs}
\setsymbol{LFI:white:noise:sensitivity:30GHz:units}{143.4\muKs}

\setsymbol{LFI:white:noise:sensitivity:70GHz}{134.7}
\setsymbol{LFI:white:noise:sensitivity:44GHz}{164.7}
\setsymbol{LFI:white:noise:sensitivity:30GHz}{143.4}

\setsymbol{LFI:white:noise:sensitivity:LFI18:Rad:M:units}{512.0\muKs}
\setsymbol{LFI:white:noise:sensitivity:LFI19:Rad:M:units}{581.4\muKs}
\setsymbol{LFI:white:noise:sensitivity:LFI20:Rad:M:units}{590.8\muKs}
\setsymbol{LFI:white:noise:sensitivity:LFI21:Rad:M:units}{455.2\muKs}
\setsymbol{LFI:white:noise:sensitivity:LFI22:Rad:M:units}{492.0\muKs}
\setsymbol{LFI:white:noise:sensitivity:LFI23:Rad:M:units}{507.7\muKs}
\setsymbol{LFI:white:noise:sensitivity:LFI24:Rad:M:units}{462.2\muKs}
\setsymbol{LFI:white:noise:sensitivity:LFI25:Rad:M:units}{413.6\muKs}
\setsymbol{LFI:white:noise:sensitivity:LFI26:Rad:M:units}{478.6\muKs}
\setsymbol{LFI:white:noise:sensitivity:LFI27:Rad:M:units}{277.7\muKs}
\setsymbol{LFI:white:noise:sensitivity:LFI28:Rad:M:units}{312.3\muKs}
\setsymbol{LFI:white:noise:sensitivity:LFI18:Rad:S:units}{465.7\muKs}
\setsymbol{LFI:white:noise:sensitivity:LFI19:Rad:S:units}{555.6\muKs}
\setsymbol{LFI:white:noise:sensitivity:LFI20:Rad:S:units}{623.2\muKs}
\setsymbol{LFI:white:noise:sensitivity:LFI21:Rad:S:units}{564.1\muKs}
\setsymbol{LFI:white:noise:sensitivity:LFI22:Rad:S:units}{534.4\muKs}
\setsymbol{LFI:white:noise:sensitivity:LFI23:Rad:S:units}{542.4\muKs}
\setsymbol{LFI:white:noise:sensitivity:LFI24:Rad:S:units}{399.2\muKs}
\setsymbol{LFI:white:noise:sensitivity:LFI25:Rad:S:units}{392.6\muKs}
\setsymbol{LFI:white:noise:sensitivity:LFI26:Rad:S:units}{418.6\muKs}
\setsymbol{LFI:white:noise:sensitivity:LFI27:Rad:S:units}{302.9\muKs}
\setsymbol{LFI:white:noise:sensitivity:LFI28:Rad:S:units}{285.3\muKs}

\setsymbol{LFI:white:noise:sensitivity:LFI18:Rad:M}{512.0}
\setsymbol{LFI:white:noise:sensitivity:LFI19:Rad:M}{581.4}
\setsymbol{LFI:white:noise:sensitivity:LFI20:Rad:M}{590.8}
\setsymbol{LFI:white:noise:sensitivity:LFI21:Rad:M}{455.2}
\setsymbol{LFI:white:noise:sensitivity:LFI22:Rad:M}{492.0}
\setsymbol{LFI:white:noise:sensitivity:LFI23:Rad:M}{507.7}
\setsymbol{LFI:white:noise:sensitivity:LFI24:Rad:M}{462.2}
\setsymbol{LFI:white:noise:sensitivity:LFI25:Rad:M}{413.6}
\setsymbol{LFI:white:noise:sensitivity:LFI26:Rad:M}{478.6}
\setsymbol{LFI:white:noise:sensitivity:LFI27:Rad:M}{277.7}
\setsymbol{LFI:white:noise:sensitivity:LFI28:Rad:M}{312.3}
\setsymbol{LFI:white:noise:sensitivity:LFI18:Rad:S}{465.7}
\setsymbol{LFI:white:noise:sensitivity:LFI19:Rad:S}{555.6}
\setsymbol{LFI:white:noise:sensitivity:LFI20:Rad:S}{623.2}
\setsymbol{LFI:white:noise:sensitivity:LFI21:Rad:S}{564.1}
\setsymbol{LFI:white:noise:sensitivity:LFI22:Rad:S}{534.4}
\setsymbol{LFI:white:noise:sensitivity:LFI23:Rad:S}{542.4}
\setsymbol{LFI:white:noise:sensitivity:LFI24:Rad:S}{399.2}
\setsymbol{LFI:white:noise:sensitivity:LFI25:Rad:S}{392.6}
\setsymbol{LFI:white:noise:sensitivity:LFI26:Rad:S}{418.6}
\setsymbol{LFI:white:noise:sensitivity:LFI27:Rad:S}{302.9}
\setsymbol{LFI:white:noise:sensitivity:LFI28:Rad:S}{285.3}

\setsymbol{LFI:knee:frequency:70GHz:units}{29.5\mHz}
\setsymbol{LFI:knee:frequency:44GHz:units}{56.2\mHz}
\setsymbol{LFI:knee:frequency:30GHz:units}{113.7\mHz}

\setsymbol{LFI:knee:frequency:70GHz}{29.5}
\setsymbol{LFI:knee:frequency:44GHz}{56.2}
\setsymbol{LFI:knee:frequency:30GHz}{113.7}

\setsymbol{LFI:knee:frequency:LFI18:Rad:M:units}{16.3\mHz}
\setsymbol{LFI:knee:frequency:LFI19:Rad:M:units}{15.1\mHz}
\setsymbol{LFI:knee:frequency:LFI20:Rad:M:units}{18.7\mHz}
\setsymbol{LFI:knee:frequency:LFI21:Rad:M:units}{37.2\mHz}
\setsymbol{LFI:knee:frequency:LFI22:Rad:M:units}{12.7\mHz}
\setsymbol{LFI:knee:frequency:LFI23:Rad:M:units}{34.6\mHz}
\setsymbol{LFI:knee:frequency:LFI24:Rad:M:units}{46.2\mHz}
\setsymbol{LFI:knee:frequency:LFI25:Rad:M:units}{24.9\mHz}
\setsymbol{LFI:knee:frequency:LFI26:Rad:M:units}{67.6\mHz}
\setsymbol{LFI:knee:frequency:LFI27:Rad:M:units}{187.4\mHz}
\setsymbol{LFI:knee:frequency:LFI28:Rad:M:units}{122.2\mHz}
\setsymbol{LFI:knee:frequency:LFI18:Rad:S:units}{17.7\mHz}
\setsymbol{LFI:knee:frequency:LFI19:Rad:S:units}{22.0\mHz}
\setsymbol{LFI:knee:frequency:LFI20:Rad:S:units}{8.7\mHz}
\setsymbol{LFI:knee:frequency:LFI21:Rad:S:units}{25.9\mHz}
\setsymbol{LFI:knee:frequency:LFI22:Rad:S:units}{15.8\mHz}
\setsymbol{LFI:knee:frequency:LFI23:Rad:S:units}{129.8\mHz}
\setsymbol{LFI:knee:frequency:LFI24:Rad:S:units}{100.9\mHz}
\setsymbol{LFI:knee:frequency:LFI25:Rad:S:units}{38.9\mHz}
\setsymbol{LFI:knee:frequency:LFI26:Rad:S:units}{58.9\mHz}
\setsymbol{LFI:knee:frequency:LFI27:Rad:S:units}{104.4\mHz}
\setsymbol{LFI:knee:frequency:LFI28:Rad:S:units}{40.7\mHz}

\setsymbol{LFI:knee:frequency:LFI18:Rad:M}{16.3}
\setsymbol{LFI:knee:frequency:LFI19:Rad:M}{15.1}
\setsymbol{LFI:knee:frequency:LFI20:Rad:M}{18.7}
\setsymbol{LFI:knee:frequency:LFI21:Rad:M}{37.2}
\setsymbol{LFI:knee:frequency:LFI22:Rad:M}{12.7}
\setsymbol{LFI:knee:frequency:LFI23:Rad:M}{34.6}
\setsymbol{LFI:knee:frequency:LFI24:Rad:M}{46.2}
\setsymbol{LFI:knee:frequency:LFI25:Rad:M}{24.9}
\setsymbol{LFI:knee:frequency:LFI26:Rad:M}{67.6}
\setsymbol{LFI:knee:frequency:LFI27:Rad:M}{187.4}
\setsymbol{LFI:knee:frequency:LFI28:Rad:M}{122.2}
\setsymbol{LFI:knee:frequency:LFI18:Rad:S}{17.7}
\setsymbol{LFI:knee:frequency:LFI19:Rad:S}{22.0}
\setsymbol{LFI:knee:frequency:LFI20:Rad:S}{8.7}
\setsymbol{LFI:knee:frequency:LFI21:Rad:S}{25.9}
\setsymbol{LFI:knee:frequency:LFI22:Rad:S}{15.8}
\setsymbol{LFI:knee:frequency:LFI23:Rad:S}{129.8}
\setsymbol{LFI:knee:frequency:LFI24:Rad:S}{100.9}
\setsymbol{LFI:knee:frequency:LFI25:Rad:S}{38.9}
\setsymbol{LFI:knee:frequency:LFI26:Rad:S}{58.9}
\setsymbol{LFI:knee:frequency:LFI27:Rad:S}{104.4}
\setsymbol{LFI:knee:frequency:LFI28:Rad:S}{40.7}

\setsymbol{LFI:slope:70GHz:units}{$-1.03$\mHz}
\setsymbol{LFI:slope:44GHz:units}{$-0.89$\mHz}
\setsymbol{LFI:slope:30GHz:units}{$-0.87$\mHz}

\setsymbol{LFI:slope:70GHz}{$-1.03$}
\setsymbol{LFI:slope:44GHz}{$-0.89$}
\setsymbol{LFI:slope:30GHz}{$-0.87$}

\setsymbol{LFI:slope:LFI18:Rad:M:units}{$-1.04$\mHz}
\setsymbol{LFI:slope:LFI19:Rad:M:units}{$-1.09$\mHz}
\setsymbol{LFI:slope:LFI20:Rad:M:units}{$-0.69$\mHz}
\setsymbol{LFI:slope:LFI21:Rad:M:units}{$-1.56$\mHz}
\setsymbol{LFI:slope:LFI22:Rad:M:units}{$-1.01$\mHz}
\setsymbol{LFI:slope:LFI23:Rad:M:units}{$-0.96$\mHz}
\setsymbol{LFI:slope:LFI24:Rad:M:units}{$-0.83$\mHz}
\setsymbol{LFI:slope:LFI25:Rad:M:units}{$-0.91$\mHz}
\setsymbol{LFI:slope:LFI26:Rad:M:units}{$-0.95$\mHz}
\setsymbol{LFI:slope:LFI27:Rad:M:units}{$-0.87$\mHz}
\setsymbol{LFI:slope:LFI28:Rad:M:units}{$-0.88$\mHz}
\setsymbol{LFI:slope:LFI18:Rad:S:units}{$-1.15$\mHz}
\setsymbol{LFI:slope:LFI19:Rad:S:units}{$-1.00$\mHz}
\setsymbol{LFI:slope:LFI20:Rad:S:units}{$-0.95$\mHz}
\setsymbol{LFI:slope:LFI21:Rad:S:units}{$-0.92$\mHz}
\setsymbol{LFI:slope:LFI22:Rad:S:units}{$-1.01$\mHz}
\setsymbol{LFI:slope:LFI23:Rad:S:units}{$-0.95$\mHz}
\setsymbol{LFI:slope:LFI24:Rad:S:units}{$-0.73$\mHz}
\setsymbol{LFI:slope:LFI25:Rad:S:units}{$-1.16$\mHz}
\setsymbol{LFI:slope:LFI26:Rad:S:units}{$-0.79$\mHz}
\setsymbol{LFI:slope:LFI27:Rad:S:units}{$-0.82$\mHz}
\setsymbol{LFI:slope:LFI28:Rad:S:units}{$-0.91$\mHz}

\setsymbol{LFI:slope:LFI18:Rad:M}{$-1.04$}
\setsymbol{LFI:slope:LFI19:Rad:M}{$-1.09$}
\setsymbol{LFI:slope:LFI20:Rad:M}{$-0.69$}
\setsymbol{LFI:slope:LFI21:Rad:M}{$-1.56$}
\setsymbol{LFI:slope:LFI22:Rad:M}{$-1.01$}
\setsymbol{LFI:slope:LFI23:Rad:M}{$-0.96$}
\setsymbol{LFI:slope:LFI24:Rad:M}{$-0.83$}
\setsymbol{LFI:slope:LFI25:Rad:M}{$-0.91$}
\setsymbol{LFI:slope:LFI26:Rad:M}{$-0.95$}
\setsymbol{LFI:slope:LFI27:Rad:M}{$-0.87$}
\setsymbol{LFI:slope:LFI28:Rad:M}{$-0.88$}
\setsymbol{LFI:slope:LFI18:Rad:S}{$-1.15$}
\setsymbol{LFI:slope:LFI19:Rad:S}{$-1.00$}
\setsymbol{LFI:slope:LFI20:Rad:S}{$-0.95$}
\setsymbol{LFI:slope:LFI21:Rad:S}{$-0.92$}
\setsymbol{LFI:slope:LFI22:Rad:S}{$-1.01$}
\setsymbol{LFI:slope:LFI23:Rad:S}{$-0.95$}
\setsymbol{LFI:slope:LFI24:Rad:S}{$-0.73$}
\setsymbol{LFI:slope:LFI25:Rad:S}{$-1.16$}
\setsymbol{LFI:slope:LFI26:Rad:S}{$-0.79$}
\setsymbol{LFI:slope:LFI27:Rad:S}{$-0.82$}
\setsymbol{LFI:slope:LFI28:Rad:S}{$-0.91$}

\setsymbol{LFI:FWHM:70GHz:units}{13\parcm01}
\setsymbol{LFI:FWHM:44GHz:units}{27\parcm92}
\setsymbol{LFI:FWHM:30GHz:units}{32\parcm65}

\setsymbol{LFI:FWHM:70GHz}{13.01}
\setsymbol{LFI:FWHM:44GHz}{27.92}
\setsymbol{LFI:FWHM:30GHz}{32.65}

\setsymbol{LFI:FWHM:LFI18:units}{13\parcm39}
\setsymbol{LFI:FWHM:LFI19:units}{13\parcm01}
\setsymbol{LFI:FWHM:LFI20:units}{12\parcm75}
\setsymbol{LFI:FWHM:LFI21:units}{12\parcm74}
\setsymbol{LFI:FWHM:LFI22:units}{12\parcm87}
\setsymbol{LFI:FWHM:LFI23:units}{13\parcm27}
\setsymbol{LFI:FWHM:LFI24:units}{22\parcm98}
\setsymbol{LFI:FWHM:LFI25:units}{30\parcm46}
\setsymbol{LFI:FWHM:LFI26:units}{30\parcm31}
\setsymbol{LFI:FWHM:LFI27:units}{32\parcm65}
\setsymbol{LFI:FWHM:LFI28:units}{32\parcm66}

\setsymbol{LFI:FWHM:LFI18}{13.39}
\setsymbol{LFI:FWHM:LFI19}{13.01}
\setsymbol{LFI:FWHM:LFI20}{12.75}
\setsymbol{LFI:FWHM:LFI21}{12.74}
\setsymbol{LFI:FWHM:LFI22}{12.87}
\setsymbol{LFI:FWHM:LFI23}{13.27}
\setsymbol{LFI:FWHM:LFI24}{22.98}
\setsymbol{LFI:FWHM:LFI25}{30.46}
\setsymbol{LFI:FWHM:LFI26}{30.31}
\setsymbol{LFI:FWHM:LFI27}{32.65}
\setsymbol{LFI:FWHM:LFI28}{32.66}

\setsymbol{LFI:FWHM:uncertainty:LFI18:units}{0.170\arcm}
\setsymbol{LFI:FWHM:uncertainty:LFI19:units}{0.174\arcm}
\setsymbol{LFI:FWHM:uncertainty:LFI20:units}{0.170\arcm}
\setsymbol{LFI:FWHM:uncertainty:LFI21:units}{0.156\arcm}
\setsymbol{LFI:FWHM:uncertainty:LFI22:units}{0.164\arcm}
\setsymbol{LFI:FWHM:uncertainty:LFI23:units}{0.171\arcm}
\setsymbol{LFI:FWHM:uncertainty:LFI24:units}{0.652\arcm}
\setsymbol{LFI:FWHM:uncertainty:LFI25:units}{1.075\arcm}
\setsymbol{LFI:FWHM:uncertainty:LFI26:units}{1.131\arcm}
\setsymbol{LFI:FWHM:uncertainty:LFI27:units}{1.266\arcm}
\setsymbol{LFI:FWHM:uncertainty:LFI28:units}{1.287\arcm}

\setsymbol{LFI:FWHM:uncertainty:LFI18}{0.170}
\setsymbol{LFI:FWHM:uncertainty:LFI19}{0.174}
\setsymbol{LFI:FWHM:uncertainty:LFI20}{0.170}
\setsymbol{LFI:FWHM:uncertainty:LFI21}{0.156}
\setsymbol{LFI:FWHM:uncertainty:LFI22}{0.164}
\setsymbol{LFI:FWHM:uncertainty:LFI23}{0.171}
\setsymbol{LFI:FWHM:uncertainty:LFI24}{0.652}
\setsymbol{LFI:FWHM:uncertainty:LFI25}{1.075}
\setsymbol{LFI:FWHM:uncertainty:LFI26}{1.131}
\setsymbol{LFI:FWHM:uncertainty:LFI27}{1.266}
\setsymbol{LFI:FWHM:uncertainty:LFI28}{1.287}

\setsymbol{HFI:center:frequency:100GHz:units}{100\,GHz}
\setsymbol{HFI:center:frequency:143GHz:units}{143\,GHz}
\setsymbol{HFI:center:frequency:217GHz:units}{217\,GHz}
\setsymbol{HFI:center:frequency:353GHz:units}{353\,GHz}
\setsymbol{HFI:center:frequency:545GHz:units}{545\,GHz}
\setsymbol{HFI:center:frequency:857GHz:units}{857\,GHz}

\setsymbol{HFI:center:frequency:100GHz}{100}
\setsymbol{HFI:center:frequency:143GHz}{143}
\setsymbol{HFI:center:frequency:217GHz}{217}
\setsymbol{HFI:center:frequency:353GHz}{353}
\setsymbol{HFI:center:frequency:545GHz}{545}
\setsymbol{HFI:center:frequency:857GHz}{857}

\setsymbol{HFI:Ndetectors:100GHz}{8}
\setsymbol{HFI:Ndetectors:143GHz}{11}
\setsymbol{HFI:Ndetectors:217GHz}{12}
\setsymbol{HFI:Ndetectors:353GHz}{12}
\setsymbol{HFI:Ndetectors:545GHz}{3}
\setsymbol{HFI:Ndetectors:857GHz}{4}

\setsymbol{HFI:FWHM:Maps:100GHz:units}{9\parcm88}
\setsymbol{HFI:FWHM:Maps:143GHz:units}{7\parcm18}
\setsymbol{HFI:FWHM:Maps:217GHz:units}{4\parcm87}
\setsymbol{HFI:FWHM:Maps:353GHz:units}{4\parcm65}
\setsymbol{HFI:FWHM:Maps:545GHz:units}{4\parcm72}
\setsymbol{HFI:FWHM:Maps:857GHz:units}{4\parcm39}
\setsymbol{HFI:FWHM:Maps:100GHz}{9.88}
\setsymbol{HFI:FWHM:Maps:143GHz}{7.18}
\setsymbol{HFI:FWHM:Maps:217GHz}{4.87}
\setsymbol{HFI:FWHM:Maps:353GHz}{4.65}
\setsymbol{HFI:FWHM:Maps:545GHz}{4.72}
\setsymbol{HFI:FWHM:Maps:857GHz}{4.39}

\setsymbol{HFI:beam:ellipticity:Maps:100GHz}{1.15}
\setsymbol{HFI:beam:ellipticity:Maps:143GHz}{1.01}
\setsymbol{HFI:beam:ellipticity:Maps:217GHz}{1.06}
\setsymbol{HFI:beam:ellipticity:Maps:353GHz}{1.05}
\setsymbol{HFI:beam:ellipticity:Maps:545GHz}{1.14}
\setsymbol{HFI:beam:ellipticity:Maps:857GHz}{1.19}

\setsymbol{HFI:FWHM:Mars:100GHz:units}{9\parcm37}
\setsymbol{HFI:FWHM:Mars:143GHz:units}{7\parcm04}
\setsymbol{HFI:FWHM:Mars:217GHz:units}{4\parcm68}
\setsymbol{HFI:FWHM:Mars:353GHz:units}{4\parcm43}
\setsymbol{HFI:FWHM:Mars:545GHz:units}{3\parcm80}
\setsymbol{HFI:FWHM:Mars:857GHz:units}{3\parcm67}

\setsymbol{HFI:FWHM:Mars:100GHz}{9.37}
\setsymbol{HFI:FWHM:Mars:143GHz}{7.04}
\setsymbol{HFI:FWHM:Mars:217GHz}{4.68}
\setsymbol{HFI:FWHM:Mars:353GHz}{4.43}
\setsymbol{HFI:FWHM:Mars:545GHz}{3.80}
\setsymbol{HFI:FWHM:Mars:857GHz}{3.67}

\setsymbol{HFI:beam:ellipticity:Mars:100GHz}{1.18}
\setsymbol{HFI:beam:ellipticity:Mars:143GHz}{1.03}
\setsymbol{HFI:beam:ellipticity:Mars:217GHz}{1.14}
\setsymbol{HFI:beam:ellipticity:Mars:353GHz}{1.09}
\setsymbol{HFI:beam:ellipticity:Mars:545GHz}{1.25}
\setsymbol{HFI:beam:ellipticity:Mars:857GHz}{1.03}

\setsymbol{HFI:CMB:relative:calibration:100GHz}{$\lsim 1\%$}
\setsymbol{HFI:CMB:relative:calibration:143GHz}{$\lsim 1\%$}
\setsymbol{HFI:CMB:relative:calibration:217GHz}{$\lsim 1\%$}
\setsymbol{HFI:CMB:relative:calibration:353GHz}{$\lsim 1\%$}
\setsymbol{HFI:CMB:relative:calibration:545GHz}{}
\setsymbol{HFI:CMB:relative:calibration:857GHz}{}

\setsymbol{HFI:CMB:absolute:calibration:100GHz}{$\lsim 2\%$}
\setsymbol{HFI:CMB:absolute:calibration:143GHz}{$\lsim 2\%$}
\setsymbol{HFI:CMB:absolute:calibration:217GHz}{$\lsim 2\%$}
\setsymbol{HFI:CMB:absolute:calibration:353GHz}{$\lsim 2\%$}
\setsymbol{HFI:CMB:absolute:calibration:545GHz}{}
\setsymbol{HFI:CMB:absolute:calibration:857GHz}{}

\setsymbol{HFI:FIRAS:gain:calibration:accuracy:statistical:100GHz}{}
\setsymbol{HFI:FIRAS:gain:calibration:accuracy:statistical:143GHz}{}
\setsymbol{HFI:FIRAS:gain:calibration:accuracy:statistical:217GHz}{}
\setsymbol{HFI:FIRAS:gain:calibration:accuracy:statistical:353GHz}{2.5\%}
\setsymbol{HFI:FIRAS:gain:calibration:accuracy:statistical:545GHz}{1\%}
\setsymbol{HFI:FIRAS:gain:calibration:accuracy:statistical:857GHz}{0.5\%}

\setsymbol{HFI:FIRAS:gain:calibration:accuracy:systematic:100GHz}{}
\setsymbol{HFI:FIRAS:gain:calibration:accuracy:systematic:143GHz}{}
\setsymbol{HFI:FIRAS:gain:calibration:accuracy:systematic:217GHz}{}
\setsymbol{HFI:FIRAS:gain:calibration:accuracy:systematic:353GHz}{}
\setsymbol{HFI:FIRAS:gain:calibration:accuracy:systematic:545GHz}{7\%}
\setsymbol{HFI:FIRAS:gain:calibration:accuracy:systematic:857GHz}{7\%}

\setsymbol{HFI:FIRAS:zero:point:accuracy:100GHz:units}{0.8\MJysr}
\setsymbol{HFI:FIRAS:zero:point:accuracy:143GHz:units}{}
\setsymbol{HFI:FIRAS:zero:point:accuracy:217GHz:units}{}
\setsymbol{HFI:FIRAS:zero:point:accuracy:353GHz:units}{1.4\MJysr}
\setsymbol{HFI:FIRAS:zero:point:accuracy:545GHz:units}{2.2\MJysr}
\setsymbol{HFI:FIRAS:zero:point:accuracy:857GHz:units}{1.7\MJysr}

\setsymbol{HFI:FIRAS:zero:point:accuracy:100GHz}{0.8}
\setsymbol{HFI:FIRAS:zero:point:accuracy:143GHz}{}
\setsymbol{HFI:FIRAS:zero:point:accuracy:217GHz}{}
\setsymbol{HFI:FIRAS:zero:point:accuracy:353GHz}{1.4}
\setsymbol{HFI:FIRAS:zero:point:accuracy:545GHz}{2.2}
\setsymbol{HFI:FIRAS:zero:point:accuracy:857GHz}{1.7}

\setsymbol{HFI:unit:conversion:100GHz:units}{0.2415\MJysrmK}
\setsymbol{HFI:unit:conversion:143GHz:units}{0.3694\MJysrmK}
\setsymbol{HFI:unit:conversion:217GHz:units}{0.4811\MJysrmK}
\setsymbol{HFI:unit:conversion:353GHz:units}{0.2883\MJysrmK}
\setsymbol{HFI:unit:conversion:545GHz:units}{0.05826\MJysrmK}
\setsymbol{HFI:unit:conversion:857GHz:units}{0.002238\MJysrmK}

\setsymbol{HFI:unit:conversion:100GHz}{0.2415}
\setsymbol{HFI:unit:conversion:143GHz}{0.3694}
\setsymbol{HFI:unit:conversion:217GHz}{0.4811}
\setsymbol{HFI:unit:conversion:353GHz}{0.2883}
\setsymbol{HFI:unit:conversion:545GHz}{0.05826}
\setsymbol{HFI:unit:conversion:857GHz}{0.002238}

\setsymbol{HFI:colour:correction:alpha=-2:V1.01:100GHz}{0.9893}
\setsymbol{HFI:colour:correction:alpha=-2:V1.01:143GHz}{0.9759}
\setsymbol{HFI:colour:correction:alpha=-2:V1.01:217GHz}{1.0007}
\setsymbol{HFI:colour:correction:alpha=-2:V1.01:353GHz}{1.0028}
\setsymbol{HFI:colour:correction:alpha=-2:V1.01:545GHz}{1.0019}
\setsymbol{HFI:colour:correction:alpha=-2:V1.01:857GHz}{0.9889}

\setsymbol{HFI:colour:correction:alpha=0:V1.01:100GHz}{1.0008}
\setsymbol{HFI:colour:correction:alpha=0:V1.01:143GHz}{1.0148}
\setsymbol{HFI:colour:correction:alpha=0:V1.01:217GHz}{0.9909}
\setsymbol{HFI:colour:correction:alpha=0:V1.01:353GHz}{0.9888}
\setsymbol{HFI:colour:correction:alpha=0:V1.01:545GHz}{0.9878}
\setsymbol{HFI:colour:correction:alpha=0:V1.01:857GHz}{1.0014}

\providecommand{\sorthelp}[1]{}

\newcommand{\be}{\begin{equation}}
\newcommand{\ee}{\end{equation}}
\newcommand{\bp}{\begin{figure}[!ht]}
\newcommand{\ep}{\end{figure}}
\newcommand{\bpm}{\begin{figure*}[!ht]}
\newcommand{\epm}{\end{figure*}}

\newcommand{\healpix} {\texttt {HEALPix}}
\newcommand{\planck} {\textit{Planck}}
\newcommand{\WMAP}{\textit{WMAP}}

\def\arcm{\ifmmode {^{\scriptstyle\prime}}
          \else $^{\scriptstyle\prime}$\fi}

\def\fsky{\ifmmode f_{\rm sky}\else $f_{\rm sky}$\fi}
\def\Tsky{\ifmmode T_{\rm sky}\else $T_{\rm sky}$\fi}
\def\Tskyest{\ifmmode \widetilde T_{\rm sky}\else $\widetilde T_{\rm sky}$\fi}
\def\phat{\ifmmode \hat p\else $\hat p$\fi}
\def\xhat{\ifmmode \hat x\else $\hat x$\fi}
\def\Vout{\ifmmode V_{\rm out}\else $V_{\rm out}$\fi}
\def\Dw{\ifmmode D_{\rm W}\else $D_{\rm W}$\fi}
\def\Gest{\ifmmode \widetilde G\else $\widetilde G$\fi}
\def\Bmain{\ifmmode B_{\rm main}\else $B_{\rm main}$\fi}
\def\Bside{\ifmmode B_{\rm side}\else $B_{\rm side}$\fi}
\def\Bpencil{\ifmmode B_{\rm pencil}\else $B_{\rm pencil}$\fi}

\def\Pmain{\ifmmode P_{\rm main}\else $P_{\rm main}$\fi}
\def\Pnominal{\ifmmode P_{\rm nominal}\else $P_{\rm nominal}$\fi}
\def\Pside{\ifmmode P_{\rm side}\else $P_{\rm side}$\fi}
\def\Ppencil{\ifmmode P_{\rm pencil}\else $P_{\rm pencil}$\fi}

\def\fsl{\ifmmode f_{\rm sl}\else $f_{\rm sl}$\fi}
\def\Psl{\ifmmode P_{\rm sl}\else $P_{\rm sl}$\fi}
\def\phiD{\ifmmode \phi_{\rm D}\else $\phi_{\rm D}$\fi}
\def\phisky{\ifmmode \phi_{\rm sky}\else $\phi_{\rm sky}$\fi}
\def\microK{\ifmmode \,\mu{\rm K}\else $\,\mu{\rm K}$\fi}

\begin{document}

\all2013resultspapers

\title{\planck\ 2013 results. XXXI. Consistency of the \Planck\ data}

\author{\small
Planck Collaboration:
P.~A.~R.~Ade\inst{77}
\and
M.~Arnaud\inst{66}
\and
M.~Ashdown\inst{63, 6}
\and
J.~Aumont\inst{53}
\and
C.~Baccigalupi\inst{76}
\and
A.~J.~Banday\inst{83, 9}
\and
R.~B.~Barreiro\inst{60}
\and
E.~Battaner\inst{84, 85}
\and
K.~Benabed\inst{54, 82}
\and
A.~Benoit-L\'{e}vy\inst{21, 54, 82}
% ...
}
\institute{\small
APC, AstroParticule et Cosmologie, Universit\'{e} Paris Diderot, CNRS/IN2P3, CEA/lrfu, Observatoire de Paris, Sorbonne Paris Cit\'{e}, 10, rue Alice Domon et L\'{e}onie Duquet, 75205 Paris Cedex 13, France\goodbreak
\and
Aalto University Mets\"{a}hovi Radio Observatory and Dept of Radio Science and Engineering, P.O. Box 13000, FI-00076 AALTO, Finland\goodbreak
\and
African Institute for Mathematical Sciences, 6-8 Melrose Road, Muizenberg, Cape Town, South Africa\goodbreak
\and
Agenzia Spaziale Italiana Science Data Center, Via del Politecnico snc, 00133, Roma, Italy\goodbreak
\and
Agenzia Spaziale Italiana, Viale Liegi 26, Roma, Italy\goodbreak
\and
Astrophysics Group, Cavendish Laboratory, University of Cambridge, J J Thomson Avenue, Cambridge CB3 0HE, U.K.\goodbreak
\and
Astrophysics \& Cosmology Research Unit, School of Mathematics, Statistics \& Computer Science, University of KwaZulu-Natal, Westville Campus, Private Bag X54001, Durban 4000, South Africa\goodbreak
\and
CITA, University of Toronto, 60 St. George St., Toronto, ON M5S 3H8, Canada\goodbreak
\and
CNRS, IRAP, 9 Av. colonel Roche, BP 44346, F-31028 Toulouse cedex 4, France\goodbreak
\and
California Institute of Technology, Pasadena, California, U.S.A.\goodbreak
\and
Centre for Theoretical Cosmology, DAMTP, University of Cambridge, Wilberforce Road, Cambridge CB3 0WA, U.K.\goodbreak
\and
Computational Cosmology Center, Lawrence Berkeley National Laboratory, Berkeley, California, U.S.A.\goodbreak
\and
DSM/Irfu/SPP, CEA-Saclay, F-91191 Gif-sur-Yvette Cedex, France\goodbreak
\and
DTU Space, National Space Institute, Technical University of Denmark, Elektrovej 327, DK-2800 Kgs. Lyngby, Denmark\goodbreak
\and
D\'{e}partement de Physique Th\'{e}orique, Universit\'{e} de Gen\`{e}ve, 24, Quai E. Ansermet,1211 Gen\`{e}ve 4, Switzerland\goodbreak
\and
Departamento de F\'{\i}sica, Universidad de Oviedo, Avda. Calvo Sotelo s/n, Oviedo, Spain\goodbreak
\and
Department of Astrophysics/IMAPP, Radboud University Nijmegen, P.O. Box 9010, 6500 GL Nijmegen, The Netherlands\goodbreak
\and
Department of Electrical Engineering and Computer Sciences, University of California, Berkeley, California, U.S.A.\goodbreak
\and
Department of Physics \& Astronomy, University of British Columbia, 6224 Agricultural Road, Vancouver, British Columbia, Canada\goodbreak
\and
Department of Physics and Astronomy, Dana and David Dornsife College of Letter, Arts and Sciences, University of Southern California, Los Angeles, CA 90089, U.S.A.\goodbreak
\and
Department of Physics and Astronomy, University College London, London WC1E 6BT, U.K.\goodbreak
\and
Department of Physics, Florida State University, Keen Physics Building, 77 Chieftan Way, Tallahassee, Florida, U.S.A.\goodbreak
\and
Department of Physics, Gustaf H\"{a}llstr\"{o}min katu 2a, University of Helsinki, Helsinki, Finland\goodbreak
\and
Department of Physics, Princeton University, Princeton, New Jersey, U.S.A.\goodbreak
\and
Department of Physics, University of California, Santa Barbara, California, U.S.A.\goodbreak
\and
Department of Physics, University of Illinois at Urbana-Champaign, 1110 West Green Street, Urbana, Illinois, U.S.A.\goodbreak
\and
Dipartimento di Fisica e Astronomia G. Galilei, Universit\`{a} degli Studi di Padova, via Marzolo 8, 35131 Padova, Italy\goodbreak
\and
Dipartimento di Fisica e Scienze della Terra, Universit\`{a} di Ferrara, Via Saragat 1, 44122 Ferrara, Italy\goodbreak
\and
Dipartimento di Fisica, Universit\`{a} La Sapienza, P. le A. Moro 2, Roma, Italy\goodbreak
\and
Dipartimento di Fisica, Universit\`{a} degli Studi di Milano, Via Celoria, 16, Milano, Italy\goodbreak
\and
Dipartimento di Fisica, Universit\`{a} degli Studi di Trieste, via A. Valerio 2, Trieste, Italy\goodbreak
\and
Dipartimento di Fisica, Universit\`{a} di Roma Tor Vergata, Via della Ricerca Scientifica, 1, Roma, Italy\goodbreak
\and
Discovery Center, Niels Bohr Institute, Blegdamsvej 17, Copenhagen, Denmark\goodbreak
\and
Dpto. Astrof\'{i}sica, Universidad de La Laguna (ULL), E-38206 La Laguna, Tenerife, Spain\goodbreak
\and
European Space Agency, ESAC, Planck Science Office, Camino bajo del Castillo, s/n, Urbanizaci\'{o}n Villafranca del Castillo, Villanueva de la Ca\~{n}ada, Madrid, Spain\goodbreak
\and
European Space Agency, ESTEC, Keplerlaan 1, 2201 AZ Noordwijk, The Netherlands\goodbreak
\and
Finnish Centre for Astronomy with ESO (FINCA), University of Turku, V\"{a}is\"{a}l\"{a}ntie 20, FIN-21500, Piikki\"{o}, Finland\goodbreak
\and
Haverford College Astronomy Department, 370 Lancaster Avenue, Haverford, Pennsylvania, U.S.A.\goodbreak
\and
Helsinki Institute of Physics, Gustaf H\"{a}llstr\"{o}min katu 2, University of Helsinki, Helsinki, Finland\goodbreak
\and
INAF - Osservatorio Astronomico di Padova, Vicolo dell'Osservatorio 5, Padova, Italy\goodbreak
\and
INAF - Osservatorio Astronomico di Roma, via di Frascati 33, Monte Porzio Catone, Italy\goodbreak
\and
INAF - Osservatorio Astronomico di Trieste, Via G.B. Tiepolo 11, Trieste, Italy\goodbreak
\and
INAF/IASF Bologna, Via Gobetti 101, Bologna, Italy\goodbreak
\and
INAF/IASF Milano, Via E. Bassini 15, Milano, Italy\goodbreak
\and
INFN, Sezione di Bologna, Via Irnerio 46, I-40126, Bologna, Italy\goodbreak
\and
INFN, Sezione di Roma 1, Universit\`{a} di Roma Sapienza, Piazzale Aldo Moro 2, 00185, Roma, Italy\goodbreak
\and
INFN/National Institute for Nuclear Physics, Via Valerio 2, I-34127 Trieste, Italy\goodbreak
\and
IPAG: Institut de Plan\'{e}tologie et d'Astrophysique de Grenoble, Universit\'{e} Joseph Fourier, Grenoble 1 / CNRS-INSU, UMR 5274, Grenoble, F-38041, France\goodbreak
\and
IUCAA, Post Bag 4, Ganeshkhind, Pune University Campus, Pune 411 007, India\goodbreak
\and
Imperial College London, Astrophysics group, Blackett Laboratory, Prince Consort Road, London, SW7 2AZ, U.K.\goodbreak
\and
Infrared Processing and Analysis Center, California Institute of Technology, Pasadena, CA 91125, U.S.A.\goodbreak
\and
Institut Universitaire de France, 103, bd Saint-Michel, 75005, Paris, France\goodbreak
\and
Institut d'Astrophysique Spatiale, CNRS (UMR8617) Universit\'{e} Paris-Sud 11, B\^{a}timent 121, Orsay, France\goodbreak
\and
Institut d'Astrophysique de Paris, CNRS (UMR7095), 98 bis Boulevard Arago, F-75014, Paris, France\goodbreak
\and
Institute for Space Sciences, Bucharest-Magurale, Romania\goodbreak
\and
Institute of Astronomy, University of Cambridge, Madingley Road, Cambridge CB3 0HA, U.K.\goodbreak
\and
Institute of Theoretical Astrophysics, University of Oslo, Blindern, Oslo, Norway\goodbreak
\and
Instituto Nacional de Astrof\'{\i}sica, \'{O}ptica y Electr\'{o}nica (INAOE), Apartado Postal 51 y 216, 72000 Puebla, M\'{e}xico\goodbreak
\and
Instituto de Astrof\'{\i}sica de Canarias, C/V\'{\i}a L\'{a}ctea s/n, La Laguna, Tenerife, Spain\goodbreak
\and
Instituto de F\'{\i}sica de Cantabria (CSIC-Universidad de Cantabria), Avda. de los Castros s/n, Santander, Spain\goodbreak
\and
Jet Propulsion Laboratory, California Institute of Technology, 4800 Oak Grove Drive, Pasadena, California, U.S.A.\goodbreak
\and
Jodrell Bank Centre for Astrophysics, Alan Turing Building, School of Physics and Astronomy, The University of Manchester, Oxford Road, Manchester, M13 9PL, U.K.\goodbreak
\and
Kavli Institute for Cosmology Cambridge, Madingley Road, Cambridge, CB3 0HA, U.K.\goodbreak
\and
LAL, Universit\'{e} Paris-Sud, CNRS/IN2P3, Orsay, France\goodbreak
\and
LERMA, CNRS, Observatoire de Paris, 61 Avenue de l'Observatoire, Paris, France\goodbreak
\and
Laboratoire AIM, IRFU/Service d'Astrophysique - CEA/DSM - CNRS - Universit\'{e} Paris Diderot, B\^{a}t. 709, CEA-Saclay, F-91191 Gif-sur-Yvette Cedex, France\goodbreak
\and
Laboratoire Traitement et Communication de l'Information, CNRS (UMR 5141) and T\'{e}l\'{e}com ParisTech, 46 rue Barrault F-75634 Paris Cedex 13, France\goodbreak
\and
Laboratoire de Physique Subatomique et de Cosmologie, Universit\'{e} Joseph Fourier Grenoble I, CNRS/IN2P3, Institut National Polytechnique de Grenoble, 53 rue des Martyrs, 38026 Grenoble cedex, France\goodbreak
\and
Laboratoire de Physique Th\'{e}orique, Universit\'{e} Paris-Sud 11 \& CNRS, B\^{a}timent 210, 91405 Orsay, France\goodbreak
\and
Max-Planck-Institut f\"{u}r Astrophysik, Karl-Schwarzschild-Str. 1, 85741 Garching, Germany\goodbreak
\and
McGill Physics, Ernest Rutherford Physics Building, McGill University, 3600 rue University, Montr\'{e}al, QC, H3A 2T8, Canada\goodbreak
\and
National University of Ireland, Department of Experimental Physics, Maynooth, Co. Kildare, Ireland\goodbreak
\and
Niels Bohr Institute, Blegdamsvej 17, Copenhagen, Denmark\goodbreak
\and
Observational Cosmology, Mail Stop 367-17, California Institute of Technology, Pasadena, CA, 91125, U.S.A.\goodbreak
\and
SB-ITP-LPPC, EPFL, CH-1015, Lausanne, Switzerland\goodbreak
\and
SISSA, Astrophysics Sector, via Bonomea 265, 34136, Trieste, Italy\goodbreak
\and
School of Physics and Astronomy, Cardiff University, Queens Buildings, The Parade, Cardiff, CF24 3AA, U.K.\goodbreak
\and
School of Physics and Astronomy, University of Nottingham, Nottingham NG7 2RD, U.K.\goodbreak
\and
Space Sciences Laboratory, University of California, Berkeley, California, U.S.A.\goodbreak
\and
Special Astrophysical Observatory, Russian Academy of Sciences, Nizhnij Arkhyz, Zelenchukskiy region, Karachai-Cherkessian Republic, 369167, Russia\goodbreak
\and
Theory Division, PH-TH, CERN, CH-1211, Geneva 23, Switzerland\goodbreak
\and
UPMC Univ Paris 06, UMR7095, 98 bis Boulevard Arago, F-75014, Paris, France\goodbreak
\and
Universit\'{e} de Toulouse, UPS-OMP, IRAP, F-31028 Toulouse cedex 4, France\goodbreak
\and
University of Granada, Departamento de F\'{\i}sica Te\'{o}rica y del Cosmos, Facultad de Ciencias, Granada, Spain\goodbreak
\and
University of Granada, Instituto Carlos I de F\'{\i}sica Te\'{o}rica y Computacional, Granada, Spain\goodbreak
% ...
}

\authorrunning{Planck Collaboration}
\titlerunning{Planck 2013 results. XXXI.}

\abstract {The \Planck\ design and scanning strategy provide many levels of redundancy that can be exploited to provide tests of internal consistency.  One of the most important is the comparison of the 70\,GHz (amplifier) and 100\,GHz (bolometer) channels.  Based on different instrument technologies, with feeds located differently in the focal plane, analysed independently by different teams using different software, and near the minimum of diffuse foreground emission, these channels are in effect two different experiments.  The 143\,GHz channel has the lowest noise level on \Planck, and is near the minimum of unresolved foreground emission.  In this paper, we analyse the level of consistency achieved in the 2013 \Planck\ data.  We  concentrate on comparisons between the 70, 100, and 143\,GHz channel maps and power spectra, particularly over the angular scales of the first and second acoustic peaks, on maps masked for diffuse Galactic emission and for strong unresolved sources.  Difference maps covering angular scales from 8\deg\ to 15\arcm\ are consistent with noise, and show no evidence of cosmic microwave background structure.  Including small but important corrections for unresolved-source residuals, we demonstrate agreement (measured by deviation of the ratio from unity) between 70 and 100\,GHz power spectra averaged over $70\le\ell\le390$ at the 0.8\,\% level, and agreement between 143 and 100\,GHz power spectra of 0.4\,\% over the same $\ell$ range.  These values are within and consistent with the overall uncertainties in calibration given in the \Planck\ 2013 results.  We also present results based on the 2013 likelihood analysis showing consistency at the 0.35\,\% between the 100, 143, and 217\,GHz power spectra.  We analyse calibration procedures and beams to determine what fraction of these differences can be accounted for by known  approximations or systematic errors that could be controlled even better in the future, reducing uncertainties still further.  Several possible small improvements are described.  Subsequent analysis of the beams quantifies the importance of asymmetry in the near sidelobes, which was not fully accounted for initially, affecting the 70/100 ratio.  Correcting for this, the 70, 100, and 143\,GHz power spectra agree to 0.4\,\% over the first two acoustic peaks.  The likelihood analysis that produced the 2013 cosmological parameters incorporated uncertainties larger than this.  We show explicitly that correction of the missing near sidelobe power in the HFI channels would result in shifts in the posterior distributions of parameters of less than $0.3\sigma$ except for $A_{\rm s}$, the amplitude of the primordial curvature perturbations at 0.05\,Mpc\mo, which changes by about $1\sigma$.  We extend these comparisons to include the sky maps from the complete nine-year mission of the {\it Wilkinson Microwave Anisotropy Probe\/} (\WMAP\/), and find a roughly  2\,\% difference between the \Planck\ and \WMAP\ power spectra in the region of the first acoustic peak.  
}
\keywords{Cosmology: observations --- Cosmic background radiation
 --- Instrumentation: detectors} 
\maketitle

\section{Introduction}
\label{sec:introduction}

This paper, one of a set associated with the 2013 release of data from the \Planck\footnote{\Planck\ (\url{http://www.esa.int/Planck}) is a project of the European Space Agency (ESA) with instruments provided by two scientific consortia funded by ESA member states (in particular the lead countries France and Italy), with contributions from NASA (USA) and telescope reflectors provided by a collaboration between ESA and a scientific consortium led and funded by Denmark.} mission \citep{planck2013-p01}, describes aspects of the internal consistency of the \Planck\ data in the 2013 release not addressed in the other papers.  The \Planck\ design and scanning strategy provide many levels of redundancy, which can be exploited to provide tests of consistency (\citealt{planck2013-p01}), most of which are carried out routinely in the \Planck\ data processing pipelines (\citealt{planck2013-p02}; \citealt{planck2013-p03}; \citealt{planck2013-p08}; and \citealt{planck2013-p11}). One of the most important consistency tests for \Planck\ is the comparison of the LFI and HFI channels, and indeed this was a key feature of its original experimental concept. Based on different instrument technologies, with feeds located differently in the focal plane, and analysed independently by different teams, these two instruments provide a powerful mutual assessment and test of systematic errors.  This paper focuses on comparison of the LFI and HFI channels closest in frequency to each other and to the diffuse foreground minimum, namely the 70\,GHz (LFI) and 100\,GHz (HFI) channels\footnote{The frequency at which extragalactic foregrounds are at a minimum level depends on angular scale, shifting from around 65\,GHz at low $\ell$ to 143\,GHz at $\ell \approx 200$.}, together with the 143\,GHz HFI channel, which has the greatest sensitivity to the cosmic microwave background (CMB) of all the \Planck\ channels.

Quantitative comparisons involving different frequencies must take into account the effects of frequency-dependent foregrounds, both diffuse and unresolved.  \Planck\ processing for the 2013 results proceeds along two main lines, depending on the scientific purpose.  For non-Gaussianity and higher-order statistics, and for the $\ell<50$ likelihood, diffuse foregrounds are separated at map level \citep{planck2013-p06}.  Only the strongest unresolved sources, however, can be identified and masked from the maps, and the effects of residual unresolved foregrounds must be dealt with statistically.  They therefore require corrections later in processing (e.g., \citealt{planck2013-p09}, \citealt{planck2013-p09a}).  For power spectra, the $\ell\ge50$ likelihood, and parameters, both diffuse and unresolved source residuals are handled in the power spectra with a combination of masking and fitting of a parametric foreground model \citep{planck2013-p11}.  

In this paper, we compare \Planck\ channels for consistency in two different ways.  First, in Sects.~\ref{sec:planckskymaps} and \ref{sec:PSestimation}, we compare frequency maps from the \Planck\ 2013 data release -- available from the \Planck\ Legacy Archive (PLA)\footnote{\url{http://archives.esac.esa.int/pla2}} and referred to hereafter as PLA maps -- and power spectra calculated from them, looking first at the effects of noise and foregrounds (both diffuse and unresolved), and then, in Sect.~\ref{sec:calibration-and-transfer-function}, at calibration and beam effects.  This comparison based on publicly released maps ties effects in the data directly to characteristics of the instruments and their determination.  This examination has provided important insights into our calibration and beam determination procedures even since the 2013 results were first released publicly in March 2013, confirming the validity of the 2013 cosmological results, resolving some issues that had been contributing to the uncertainties, and suggesting future improvements that will reduce uncertainties further.

Second, in Sect.~\ref{sec:PSlate}, we compare power spectra again, this time from the ``detector set'' data at 100, 143, and 217\,GHz (see Table~1 in \citealt{planck2013-p08}) used in the likelihood analysis described in \citet{planck2013-p08}, but extending that analysis to include 70\,GHz as well.  These detector-set/likelihood comparisons give a measure of the agreement between frequencies in the data used to generate the \Planck\ 2013 cosmological parameter results in \citet{planck2013-p11}.  Taking into account differences in the data and processing, the same level of consistency is seen as in the comparison based on PLA frequency maps in Sect.~\ref{sec:PSestimation}.  We then show that the small changes in beam window functions discussed in Sect.~\ref{sec:calibration-and-transfer-function} have no significant effect on the 2013 parameter results other than the overall amplitude of the primordial curvature perturbations at 0.05\,Mpc\mo, $A_{\rm s}$.  

After having established consistency within the \Planck\ data, specifically agreement between 70, 100, and 143\,GHz over the first acoustic peak to better than 0.5\,\% in the power spectrum, in Sect.~\ref{sec:wmap} we compare \Planck\ with \WMAP, specifically the \WMAP\/9 release\footnote{Available from the LAMBDA site: \url{http://lambda.gsfc.nasa.gov}}.  The absolute calibration of the \Planck\ 2013 results is based on the ``solar dipole'' (i.e., the motion of the Solar System barycentre with respect to the CMB) determined by \WMAP\/7 \citep{hinshaw2009}, whose uncertainty leads to a calibration error of 0.25\,\% \citep{planck2013-p02b}.   For the \Planck\ channels considered in this paper, the overall calibration uncertainty is 0.6\,\% in the 70\,GHz maps and 0.5\,\% in the 100 and 143\,GHz maps (1.2\,\% and 1.0\,\%, respectively, in the power spectra; \citealt{planck2013-p01}, Table~6).  When comparing \Planck\ and \WMAP\ calibrated maps, however, one should remove from these uncertainties in the \Planck\ maps the 0.25\,\% contribution from the \WMAP\ dipole, since it was the reference calibrator for both LFI and HFI.  In the planned 2014 release, the \Planck\ absolute calibration will be based on the ``orbital dipole'' (i.e., the modulation of the solar dipole due to the Earth's orbital motion around the Sun), bypassing uncertainties in the solar dipole.  

Throughout this paper we refer to frequency bands by their nominal designations of 30, 44, 70, 100, 143, 217, 353, 545, and 857\,GHz for \planck\ and 23, 33, 41, 61, and 94\,GHz for \WMAP;  however, we take bandpasses into account in all calculations.  The actual weighted central frequencies determined by convolution of the bandpass response with a CMB spectrum are 28.4, 44.1, 70.4, 100.0, 143.0, 217.0, 353.0, 545.0, and 857.0\,GHz for \Planck, and 22.8, 33.2, 41.0, 61.4, and 94.0\,GHz for \WMAP.  These correspond to the effective frequencies for CMB emission.  For emission with different spectra, the effective frequency is slightly shifted.

The maps discussed in this paper are structured according to the \healpix\footnote{See \url{http://healpix.jpl.nasa.gov}} scheme \citep{gorski05} displayed in Mollweide projections in Galactic coordinates.

\begin{figure}
\centerline{\includegraphics[width=88mm]{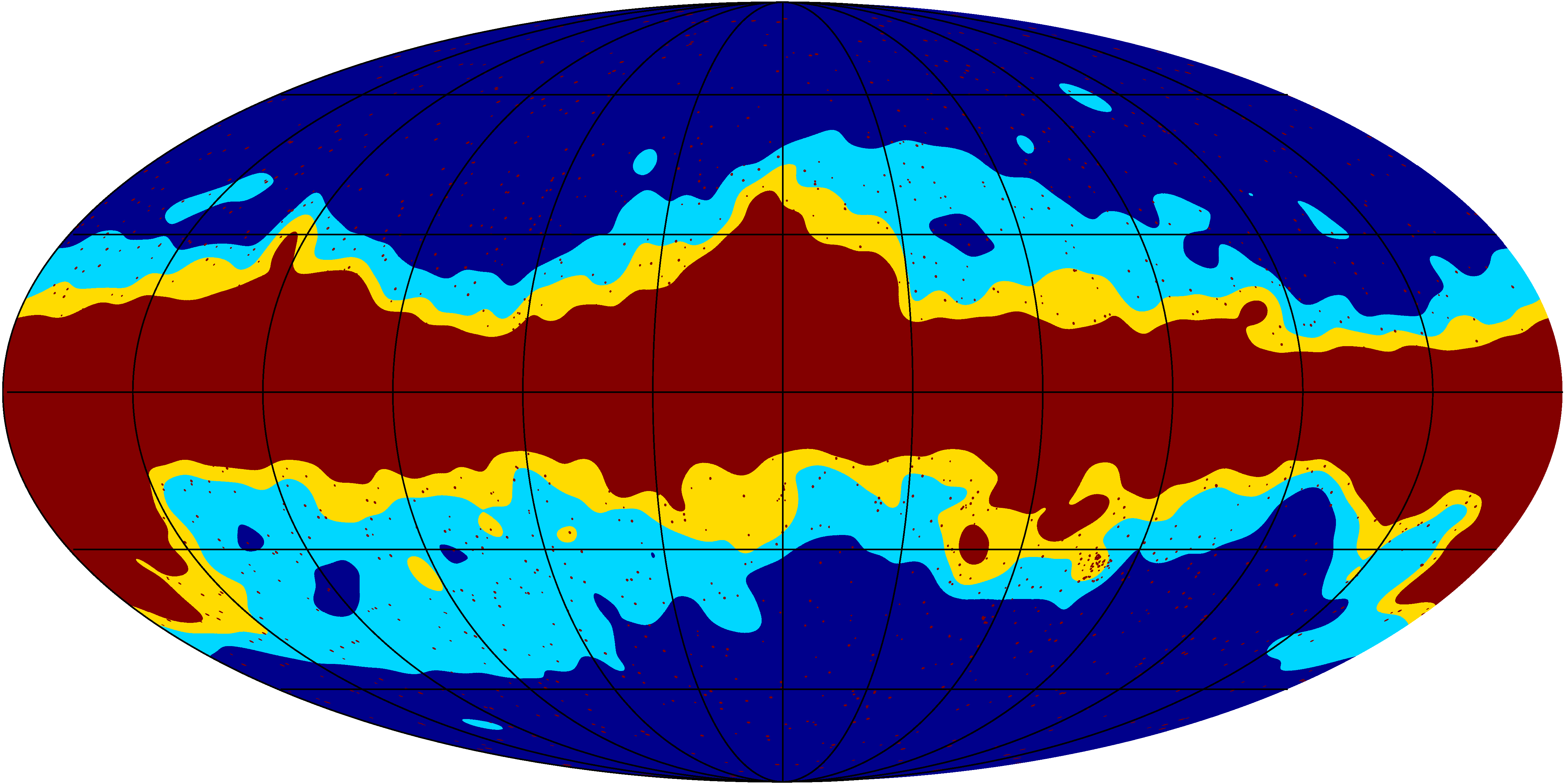}}
\caption{Sky masks used for spectral analysis of the \planck\ 70, 100, and 143\,GHz maps. The light blue, yellow, and red masks leave observable sky fractions $\fsky$ of 39.7\,\%, 59.6\,\%, and 69.4\,\%, respectively, and are named GAL040, GAL060, and GAL070 in the \hbox{PLA}.  These masks are  extended by exclusion of unresolved sources in the PCCS 70, 100, and 143\,GHz source lists above the $5\,\sigma$ flux density cuts.}
\label{fig:Planck_masks}
\end{figure}

\section{Comparison of frequency maps}
\label{sec:planckskymaps}

The \planck\ 2013 data release includes maps based on 15.5~months of data, as well as maps of subsets of the data that enable tests of data quality and systematic errors.  Examples include (see \citealt{planck2013-p01} for complete descriptions) single survey maps and half-ring difference maps, made by splitting the data from each pointing period of the satellite into halves,  making separate sky maps from the two halves, and taking the difference of the two maps.  Half-ring maps are particularly useful in characterizing the noise, and also enable signal estimation based on cross-spectra, with significant noise reduction compared to auto-spectra. 

The 100 and 143\,GHz maps are released at \healpix\ resolution $N_{\rm side} = 2048$, with $N_{\rm pix}=12 \times N_{\rm side}^2 \approx 5\times 10^7$ pixels of approximately 1\parcm7.  Although the LFI maps are generally released at $N_{\rm side}=1024$, the 70\,GHz maps are also released at $N_{\rm side}=2048$. All mapmaking steps except map binning at the given pixel resolution are the same for the two resolutions.  In this paper we use the 70\,GHz maps made at  $N_{\rm side} = 2048$ for comparison with the 100 and 143\,GHz sky maps.

\subsection{Sky masks}
\label{sec:sky-masks}

Comparison of maps at different frequencies over the full sky is quite revealing of foregrounds, as will be seen.  For most purposes in this paper we need to mask regions of strong foreground emission.  We do this using the publicly-released\footnote{Available from the \Planck\ Legacy Archive: \url{http://archives.esac.esa.int/pla2}} Galactic masks GAL040, GAL060, and GAL070, shown in Fig.~\ref{fig:Planck_masks}.  These leave unmasked $\fsky = 39.7$\,\%, 59.6\,\%, and 69.4\,\%  of the full sky.  We mask unresolved (``point'') sources detected above $5\sigma$ in the 70, 100, and 143\,GHz channels, as described in the \Planck\ Catalogue of Compact Sources (PCCS; \citealt{planck2013-p05}).  The point source masks are circular holes centred on detected sources with diameter 2.25 times the FWHM beamsize of the frequency channel in question.  The masks are unapodized, as the effect of apodization on large angular scales is primarily to improve the accuracy of covariance matrices.

\subsection{Monopole/dipole removal}
\label{sec:monopole-dipole-removal}

The \planck\ data have an undetermined absolute zero level, and the \planck\ maps contain low-amplitude offsets generated in the process of mapmaking, as well as small residual dipoles that remain after removal of the kinematic dipole anisotropy.  We remove the $\ell = 0$ and $\ell=1$ modes from the maps using $\chi^2$-minimization and the GAL040 mask, extended where applicable to a constant latitude of $\pm 45\deg$.  Diffuse Galactic emission at both low and high frequencies is still present even at high latitude, so this first step can leave residual offsets that become visible at the few microkelvin level in the difference maps smoothed to 8\deg\ shown in Appendix~\ref{sec:mapappendix}.  In those cases, a small offset adjustment, typically no more than a few microkelvin, is made to keep the mean value very close to zero in patches of sky visually clear of foregrounds.

\subsection{\textit{Comparisons}}
\label{sec:30-353_map}

\begin{figure*}
\centerline{\includegraphics[width=180mm]{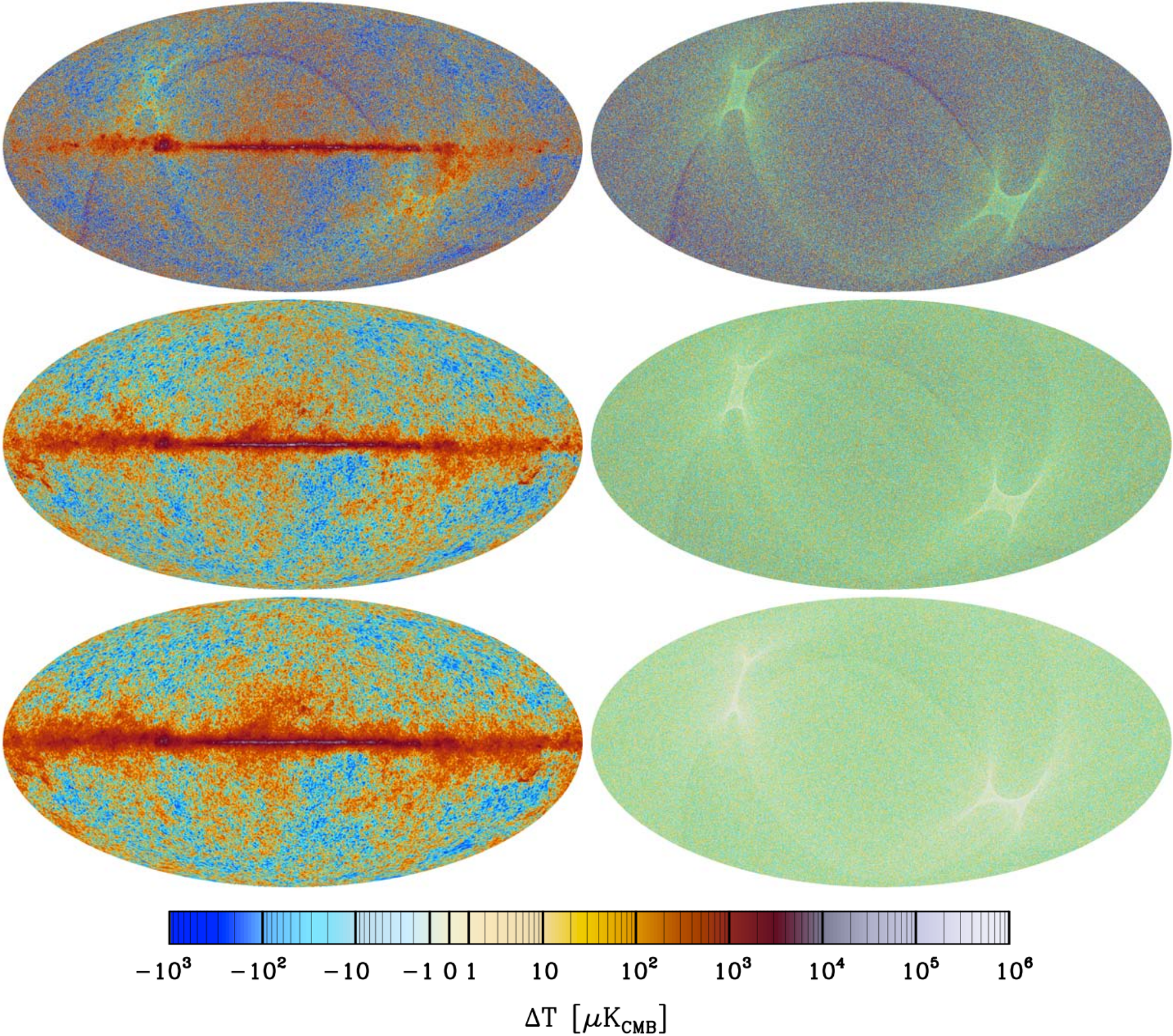}}
\caption{Sky maps used in the analysis of \Planck\ data consistency.  \textit{Top row}: 70\,GHz. \textit{Middle row}: 100\,GHz. \textit{Bottom row}: 143\,GHz.  \textit{Left column}: signal maps. \textit{Right column}: noise maps derived from half-ring differences.  All maps are $N_{\rm side}=2048$.  These are the publicly-released maps corrected for monopole and dipole terms as described in the text.  The impression of overall colour differences between the maps is due to the interaction between noise, the colour scale, and display resolution.  For example, the larger positive and negative swings between pixels in the 70\,GHz noise map pick up darker reds and blues farther from zero.  Smaller swings around zero in the 100 and 143\,GHz noise maps result in pastel yellows and blues in adjacent pixels, which when displayed at less than full-pixel resolution give an overall impression of green, a colour not used in the colour bar.}
\label{fig:Maps1}
\end{figure*}

\begin{figure*}
\centerline{\includegraphics[width=130mm]{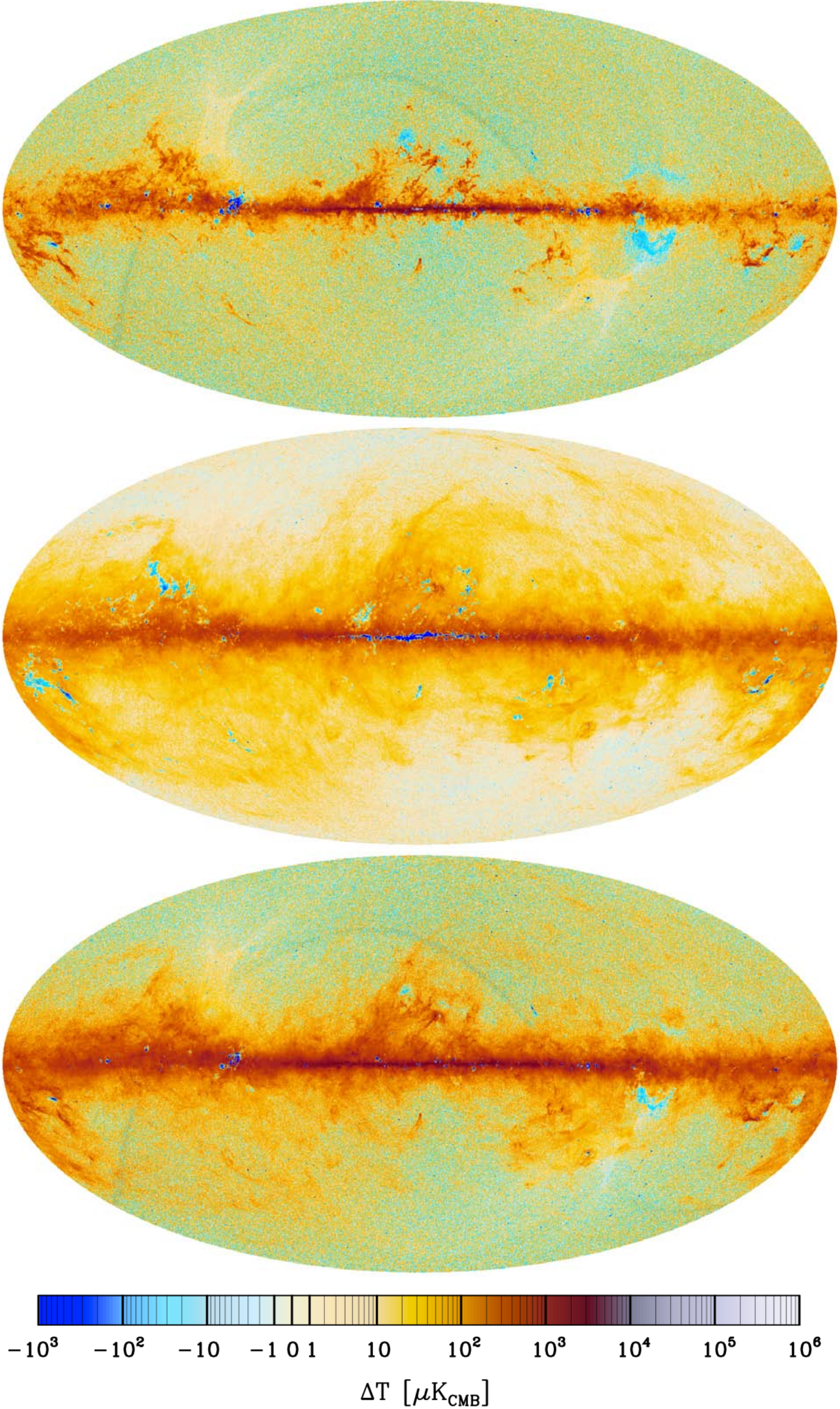}}
\caption{Difference maps.  \textit{Top}: 100\,GHz minus 70\,GHz. \textit{Middle}: 143\,GHz minus 100\,GHz.  \textit{Bottom}: 143\,GHz minus 70\,GHz.  All sky maps are smoothed to angular resolution ${\rm FWHM} =15\arcm$ by a filter that accounts for the difference between the effective beam response at each frequency and a Gaussian of FWHM 15\arcm.  These maps illustrate clearly the difference in the noise level of the individual maps, excellent overall nulling of the CMB anisotropy signal, and frequency-dependent foregrounds.  The $100-70$ difference shows predominantly CO ($J\,{=}\,0\,{\rightarrow}\,1$) emission (positive) and free-free emission (negative).  The $143-100$ difference shows dust emission (positive) and CO emission (negative).  The $143-70$ difference shows dust emission (positive) and free-free emission (negative).  The darker stripe in the top and bottom maps is due to reduced integration time in the 70\,GHz channel in the first days of observation (see \citealt{planck2013-p02}, Sect.~9.5).}
\label{fig:Maps2}
\end{figure*}

Figure~\ref{fig:Maps1} shows the monopole- and dipole-removed maps at 70, 100, and 143\,GHz, along with the corresponding half-ring difference maps.  Figure~\ref{fig:Maps2} shows the difference maps between these three frequencies.  The strong frequency-dependence of foregrounds is obvious.  Equally obvious, and the essential point of the comparison, is the nearly complete nulling of the CMB anisotropies.  This shows that these three channels on \Planck\ are measuring the same CMB sky.

\begin{figure*}
\includegraphics[width=93mm]{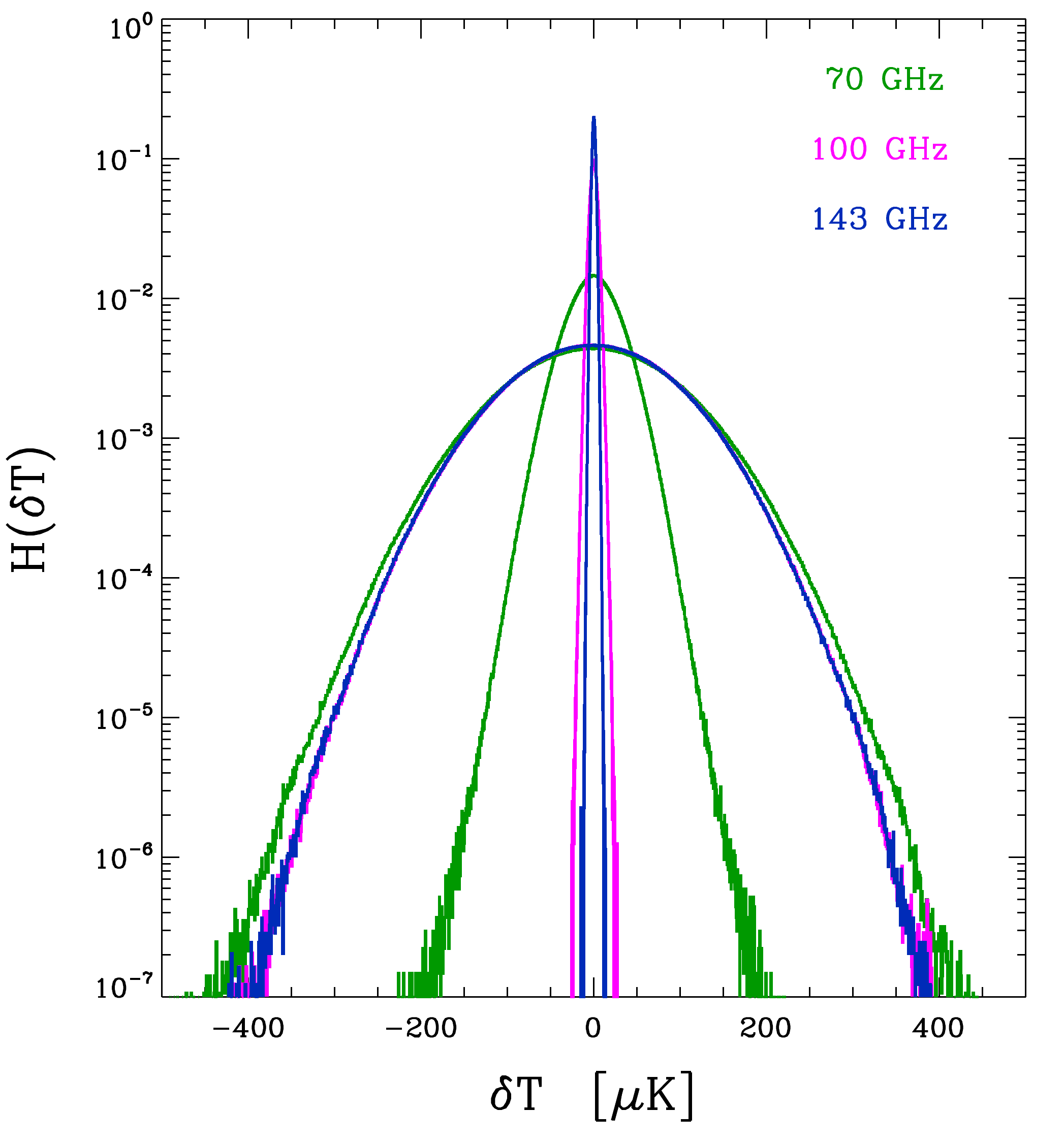}
\includegraphics[width=93mm]{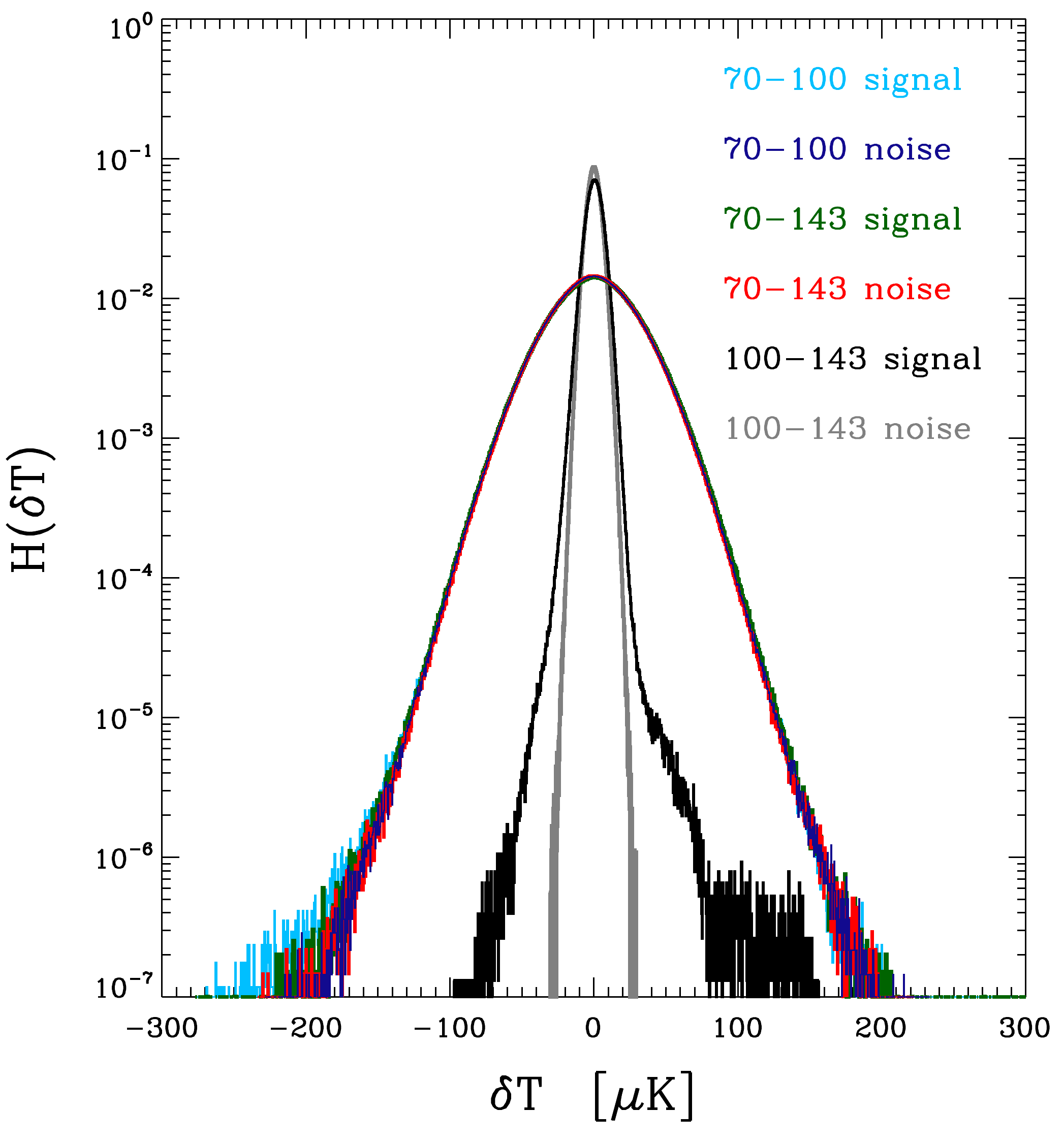}
\caption{Signal and noise for the frequency maps of Fig.~\ref{fig:Maps1} ({\it left panel\/}) and the difference maps of Fig.~\ref{fig:Maps2} ({\it right panel\/}), with the 59.6\,\% mask in all cases.  The broader, signal+noise curves are nearly Gaussian due to the dominant CMB anisotropies.  The 70\,GHz curve is broader than the 100 and 143\,GHz curves because of the higher noise level, but is still signal-dominated for $|dT/T| \gsim 50\muK$.  The narrower noise curves, derived from the half-ring difference maps, are not Gaussian because of the scanning-induced spatial dependence of pixel noise in \planck\ maps.  The considerably higher noise level of the 70\,GHz map is again apparent.  The histograms of the difference maps show noise domination near the peak of each pair of curves (the signal+noise and noise curves overlap).  The pairs involving 70\,GHz are wider and dominated by the 70\,GHz noise, but the wings at low pixel counts show the signature of foregrounds that exceed the noise levels, primarily dust and CO emission in the negative wing, and free-free and synchrotron emission in the positive wing.  In the low-noise 100 minus 143\,GHz pairs, the signal, due mostly to dust emission in the negative wing and to free-free and CO residuals in the positive wing, stands out clearly from the noise.} 
\label{fig:stats_histograms}
\end{figure*}

For a quantitative comparison, we calculate root mean square (rms) values of unmasked regions of the frequency and difference maps shown in Figs.~\ref{fig:Maps1} and \ref{fig:Maps2}, for the three masks shown in Fig.~\ref{fig:Planck_masks}.  To avoid spurious values caused on small scales by the differing angular resolution of the three frequencies, and on large scales by diffuse foregrounds, we first smooth the maps to a common resolution of 15\arcm.  We then smooth them further to 8\deg\ resolution, and subtract the 8\deg\ maps from the 15\arcm\ maps.  This leaves maps that can be directly compared for structure on angular scales from 8\deg\ to 15\arcm.  We calculate rms values for half-ring sum  (``Freq.'') and half-ring difference (``Diff.'') maps at 70, 100, and 143\,GHz, and for the frequency-difference maps 70\,GHz\,$-$\,100\,GHz, 70\,GHz\,$-$\,143\,GHz, and 100\,GHz\,$-$\,143\,GHz.  The rms values are given in Table~\ref{tab:rms}.  The maps are shown in Fig.~\ref{fig:stats_maps}.  Histograms of the frequency maps and difference maps are shown in Fig.~\ref{fig:stats_histograms}.

\begin{table*}
\caption{Rms values of the unmasked regions of the frequency and difference maps shown in Fig.~\ref{fig:stats_maps}, and for the $f_{\rm sky} = 69.4$\,\%, 59.6\,\%, and 39.7\,\% masks shown in Fig.~\ref{fig:Planck_masks}.  Diagonal blocks give the rms values for half-ring sums (``Freq.'') and differences (``Diff.'') of single-frequency maps.  Off-diagonal blocks give the same quantities for frequency-{\it difference\/} maps (70\,GHz\,$-$\,100\,GHz, 70\,GHz\,$-$\,143\,GHz, and 100\,GHz\,$-$\,143\,GHz.  As described in the text, the maps are smoothed to a common resolution of 15\arcm, somewhat lower than the resolution of the 70\,GHz maps.  In addition, structure on scales larger than 8\deg\ is determined and removed from all maps to avoid introducing biases from residual monopoles and dipoles, so that only structure from 15\arcm\ to 8\deg\ in angular scale is included in these calculations.}
\label{tab:rms}
\vskip -3.5mm
\tiny
\setbox\tablebox=\vbox{
\newdimen\digitwidth
\setbox0=\hbox{\rm 0}
\digitwidth=\wd0
\catcode`*=\active
\def*{\kern\digitwidth}
\newdimen\signwidth
\setbox0=\hbox{+}
\signwidth=\wd0
\catcode`!=\active
\def!{\kern\signwidth}
\halign{    \hfil#\hfil\tabskip=1.0em& 
    \hbox to 0.7in{#\leaderfil}\tabskip=2em&
    \hfil#\hfil\tabskip=1em& 
    \hfil#\hfil\tabskip=2em& 
    \hfil#\hfil\tabskip=1em& 
    \hfil#\hfil\tabskip=2em& 
    \hfil#\hfil\tabskip=1em& 
    \hfil#\hfil\tabskip=0pt\cr
\noalign{\doubleline}
\omit&\omit&\multispan6\hfil R{\sc MS} [\muK]\hfil\cr
\noalign{\vskip -3pt}
\omit&\omit&\multispan6\hrulefill\cr
\omit&\omit&\multispan2\hfil 70\,GHz\hfil&\multispan2\hfil 100\,GHz\hfil&\multispan2\hfil 143\,GHz\hfil\cr
\noalign{\vskip -3pt}
\omit&\omit&\multispan2\hrulefill&\multispan2\hrulefill&\multispan2\hrulefill\cr
\noalign{\vskip 3pt}
\omit\hfil$\nu$\hfil&\omit\hfil $f_{\rm sky}$\hfil&Freq.&Diff.&Freq.&Diff.&Freq.&Diff.\cr
\noalign{\vskip 3pt\hrule\vskip 5pt}
&        39.7\,\%&90.44&28.62&29.01&28.93&29.00&28.69\cr
*70\,GHz&59.6\,\%&90.09&29.48&29.66&29.47&29.77&29.22\cr
&        69.4\,\%&90.12&29.46&29.79&29.36&30.03&29.12\cr
\noalign{\vskip 7pt}
&        39.7\,\%&\dots&\dots&85.63&*4.27&*5.49&*4.76\cr
100\,GHz&59.6\,\%&\dots&\dots&85.05&*4.38&*6.09&*4.83\cr
&        69.4\,\%&\dots&\dots&85.16&*4.39&*6.76&*4.81\cr
\noalign{\vskip 7pt}
&        39.7\,\%&\dots&\dots&\dots&\dots&85.70&*2.11\cr
143\,GHz&59.6\,\%&\dots&\dots&\dots&\dots&85.23&*2.17\cr
&        69.4\,\%&\dots&\dots&\dots&\dots&85.45&*2.18\cr
\noalign{\vskip 5pt\hrule\vskip 3pt}}}
\endPlancktablewide
\end{table*}

Except for obvious foreground structures and noise, the difference maps lie close to zero, showing the excellent agreement between the three \Planck\ frequencies for the CMB anisotropies.

The map comparisons give a comprehensive view of consistency between 70, 100, and 143\,GHz, but the two-dimensional nature of the comparisons makes it somewhat difficult to grasp the key similarities.  To make this easier, we turn now to comparisons at the power spectrum level.

\section{Comparison of power spectra from 2013 results frequency maps}
\label{sec:PSestimation}

Power spectra of the unmasked regions of the maps are estimated as follows.

\begin{itemize}

\item Starting from half-ring maps (section~5.1 of \citealt{planck2013-p01}),  cross-spectra are computed on the masked, incomplete sky using the {\tt HEALPix} routine ${\tt anafast}$ with $\ell = 0, ~1$ removal.  These are so-called ``pseudo-spectra.''

\item The {\tt MASTER} spectral coupling kernel \citep{Hietal02}, which describes spectral mode coupling on an incomplete sky, is calculated based on the mask used.  The pseudo-spectra from the previous step are converted to $4\pi$-equivalent amplitude using the inverse of the {\tt MASTER} kernel.

\item Beam and pixel smoothing effects are removed from the spectra by dividing out the appropriate beam and pixel window functions.  Beam response functions in $\ell$ space are required.  We use the effective beam window functions derived using {\tt FEBeCoP} \citep{mitra2010}.

\end{itemize}

\subsection{Spectral analysis of signals and noise}
\label{sec:spectra_first_look}

\begin{figure*}
\includegraphics[width=180mm]{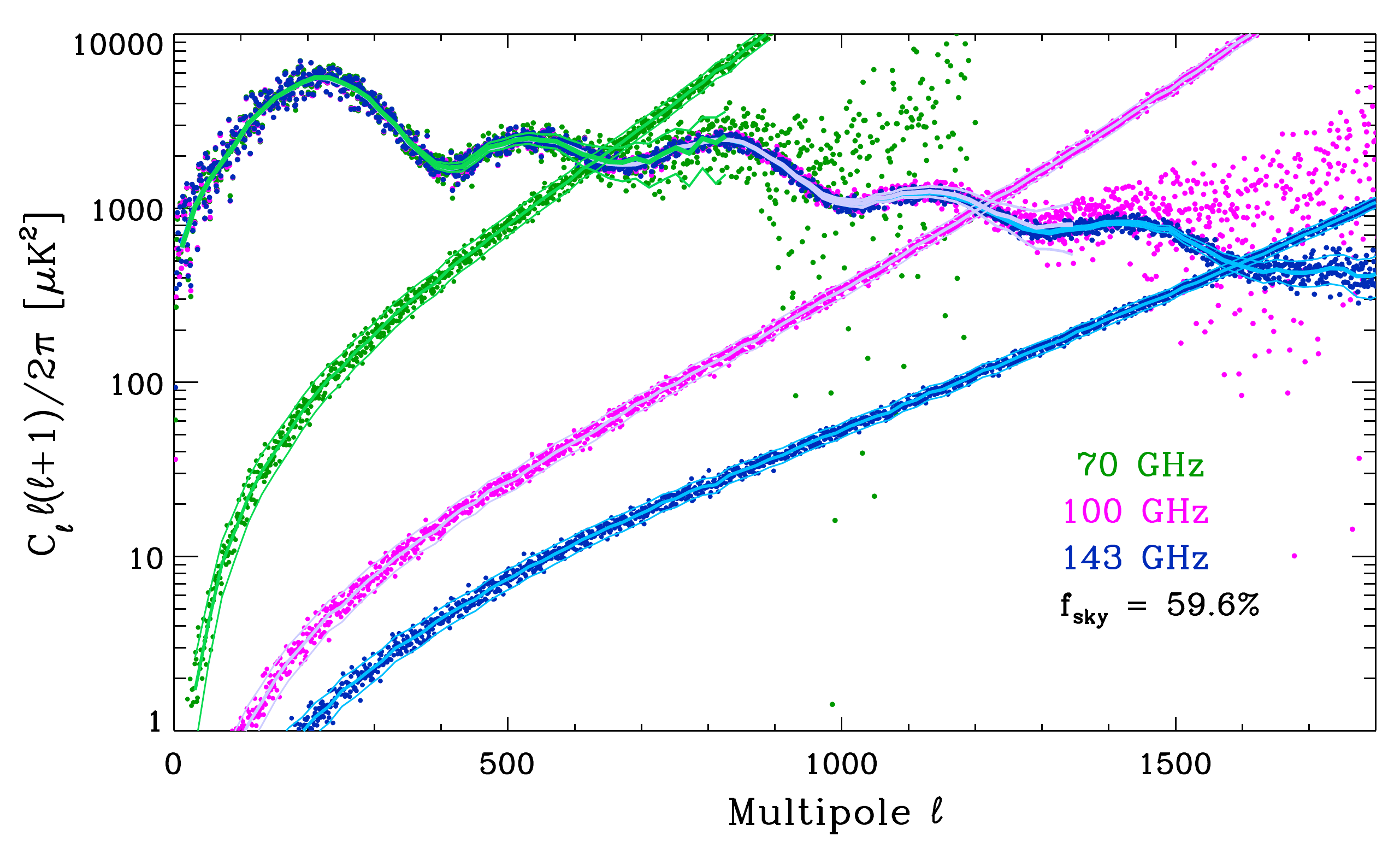}
\caption{\Planck\ 70, 100, and 143\,GHz CMB anisotropy power spectra computed for the GAL060 mask. Mask- and beam-deconvolved cross-spectra of the half-ring maps show the signal; auto-correlation spectra of the half-ring difference maps show the noise.
Points show single multipoles up to $\ell = 1200$ for 70\,GHz and $\ell = 1700$ for 100 and 143\,GHz.  Heavy solid lines show $\Delta \ell = 20$ boxcar averages.  The S/N near the first peak ($\ell =220$) is approximately 80, 1900, and 6000 for 70, 100, and 143\,GHz, respectively.  Noise power is calculated according to the large-$\ell$ approximation, i.e., as a  $\chi^2_{2\ell +1}$ distribution with mean $C_{\ell}$ and rms $C_{\ell}\left[f_{\rm sky}(2\ell +1)/2\right]^{-1/2}$.  Pairs of thin lines mark $\pm3\,\sigma$ bands of noise power around the noise spectra.  We translate this statistical spread of noise power $C_{\ell}$s into the signal spectra estimated via half-ring map cross-spectra.  Under the simplifying assumption that each $C_{\ell}$ of the noise in the cross-spectrum at high-$\ell$ is distributed as a sum of $(2\ell +1)$ products of independent Gaussian deviates, each with variance $2 C_{\ell}^{\rm noise}$ derived from the half-ring difference maps, the Gaussianized high-$\ell$ noise in the cross-spectra has zero mean and rms of $2 C_{\ell}^{\rm noise}\left[f_{\rm sky}(2\ell+1)\right]^{-1/2}$.  Pairs of thin lines mark $\pm 1\,\sigma$ bands of noise around the boxcar-averaged cross-spectra.}
\label{fig:spectra1_planck_70_100_143}
\end{figure*}

Figure~\ref{fig:spectra1_planck_70_100_143} shows the signal (half-ring map cross-spectra), and noise (half-ring difference map auto-spectra) of the 70, 100, and 143\,GHz channels.  As stated earlier, the 70--100\,GHz channel comparison quantifies the cross-instrument consistency of \planck.

This description of the statistics of noise contributions to the empirical cross-spectra derived from the \planck\ sky maps sets up the analysis of inter-frequency consistency of \planck\ data.  The pure instrumental noise contribution to the empirical cross-spectra is very small over a large $\ell$-range for the HFI channels, and at 70\,GHz over the $\ell$-range of the first peak in the spectrum, where we now focus our analysis.  Cosmic variance is irrelevant for our discussion because we are assessing inter-frequency data consistency, and the instruments observe the same CMB anisotropy.  Any possible departures from complete consistency of the measurements must be accounted for by frequency-dependent foreground emission, accurate accounting of systematic effects, or (at a very low level) residual noise.

\subsection{Spectral consistency}
\label{sec:estimating_spectral_consistency}

\begin{figure*}
\centerline{
\includegraphics[width=6.2cm]{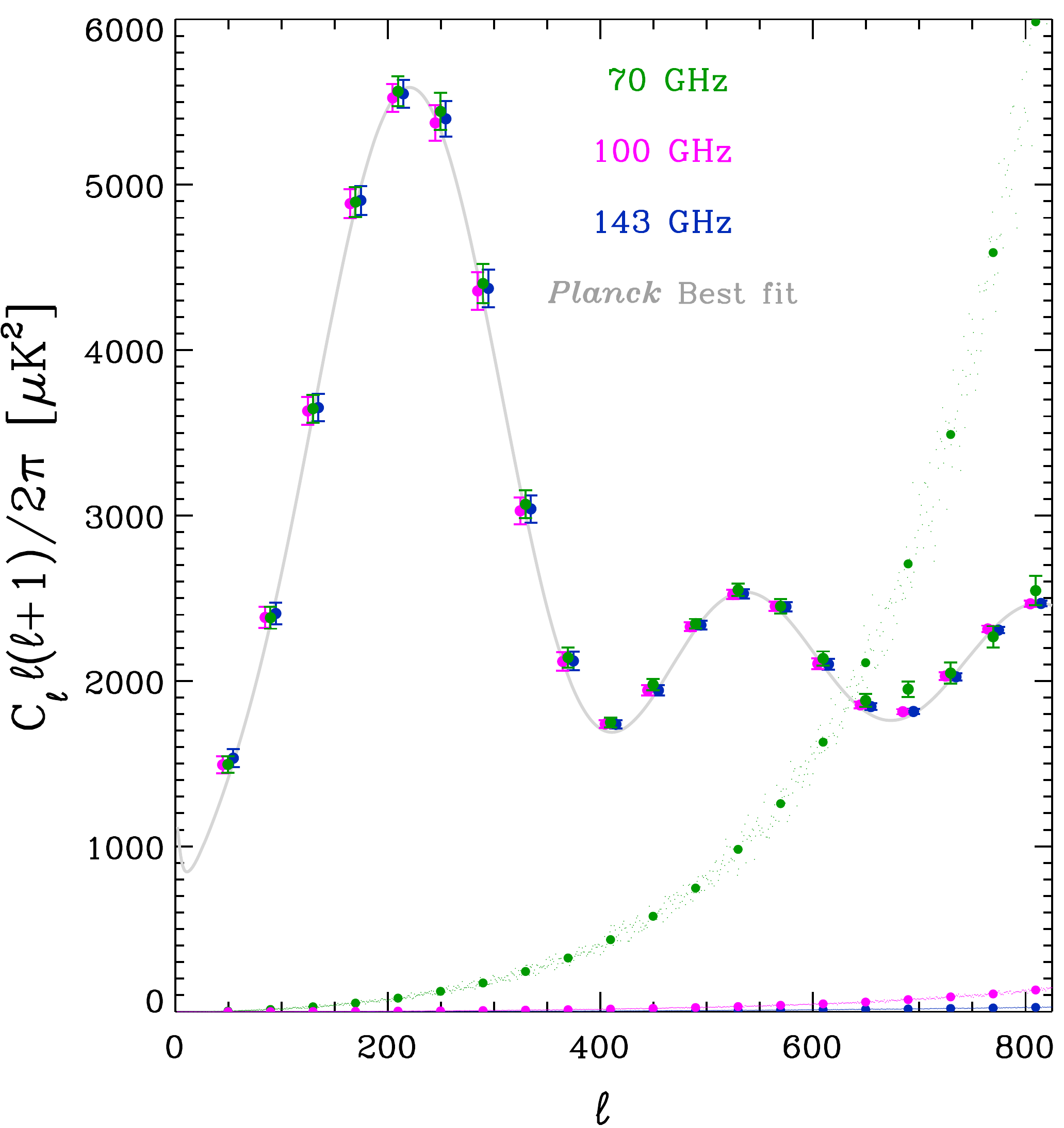}
\includegraphics[width=6.2cm]{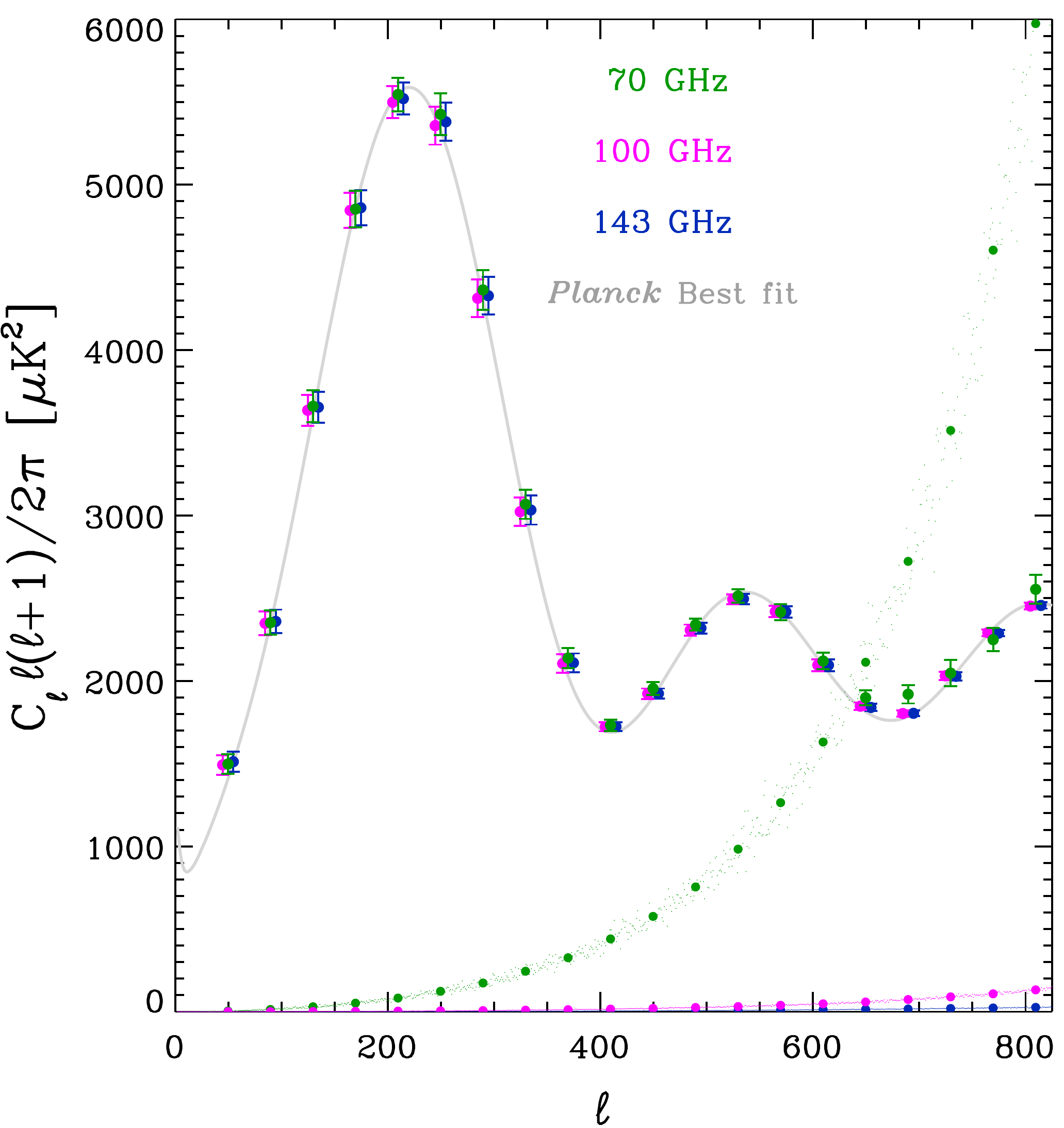}
\includegraphics[width=6.2cm]{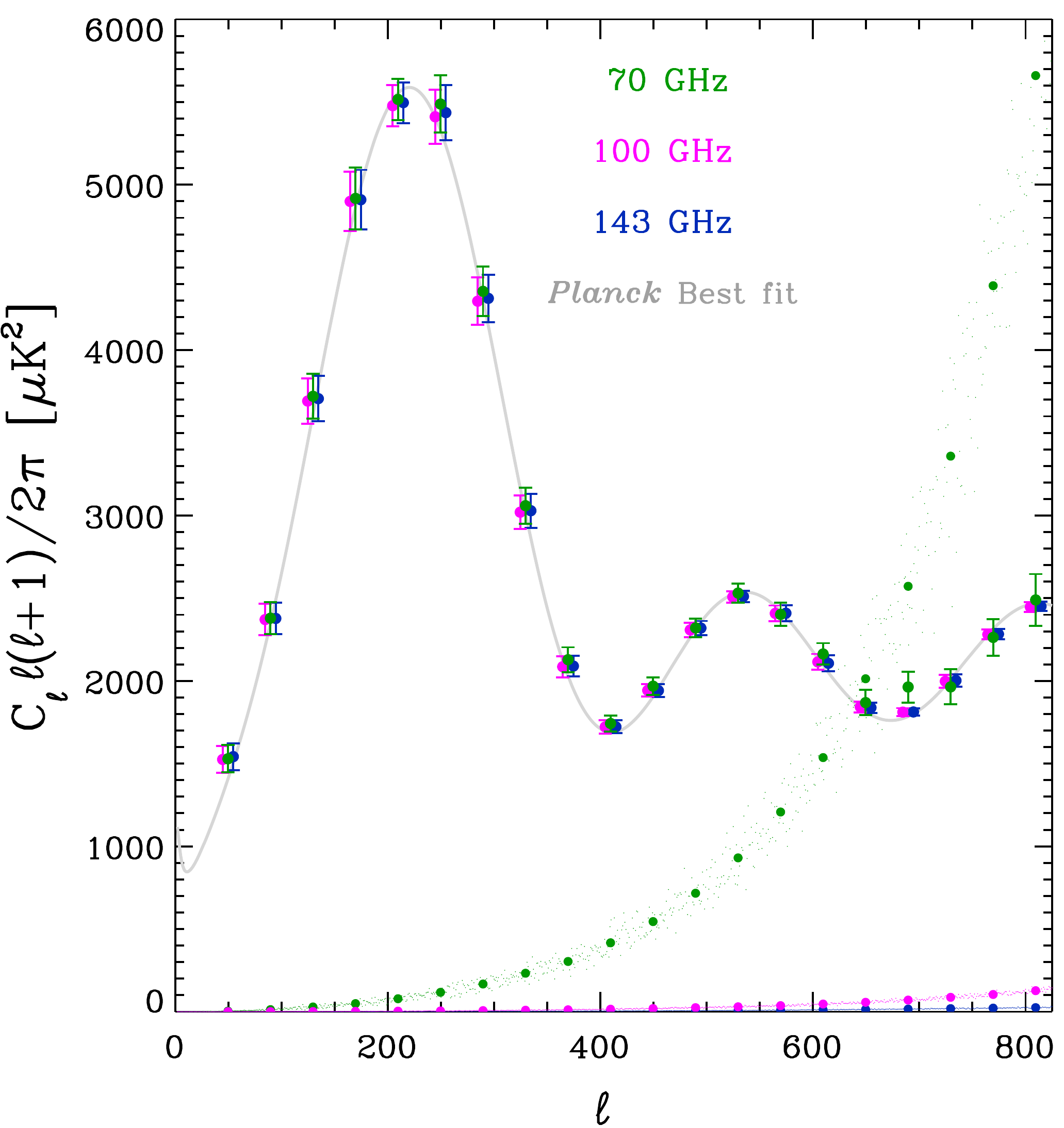}
}
\centerline{
\includegraphics[width=6.2cm]{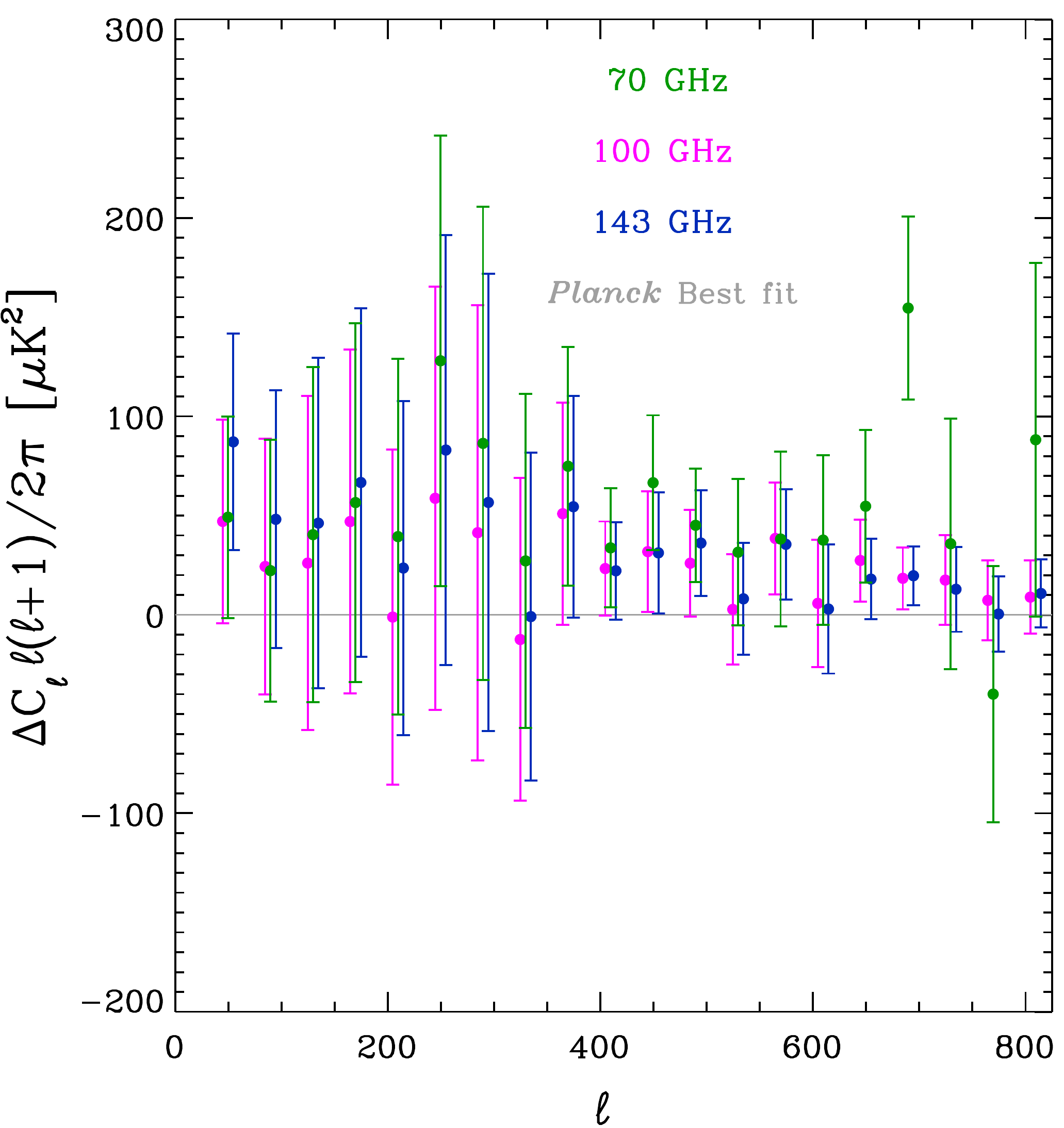}
\includegraphics[width=6.2cm]{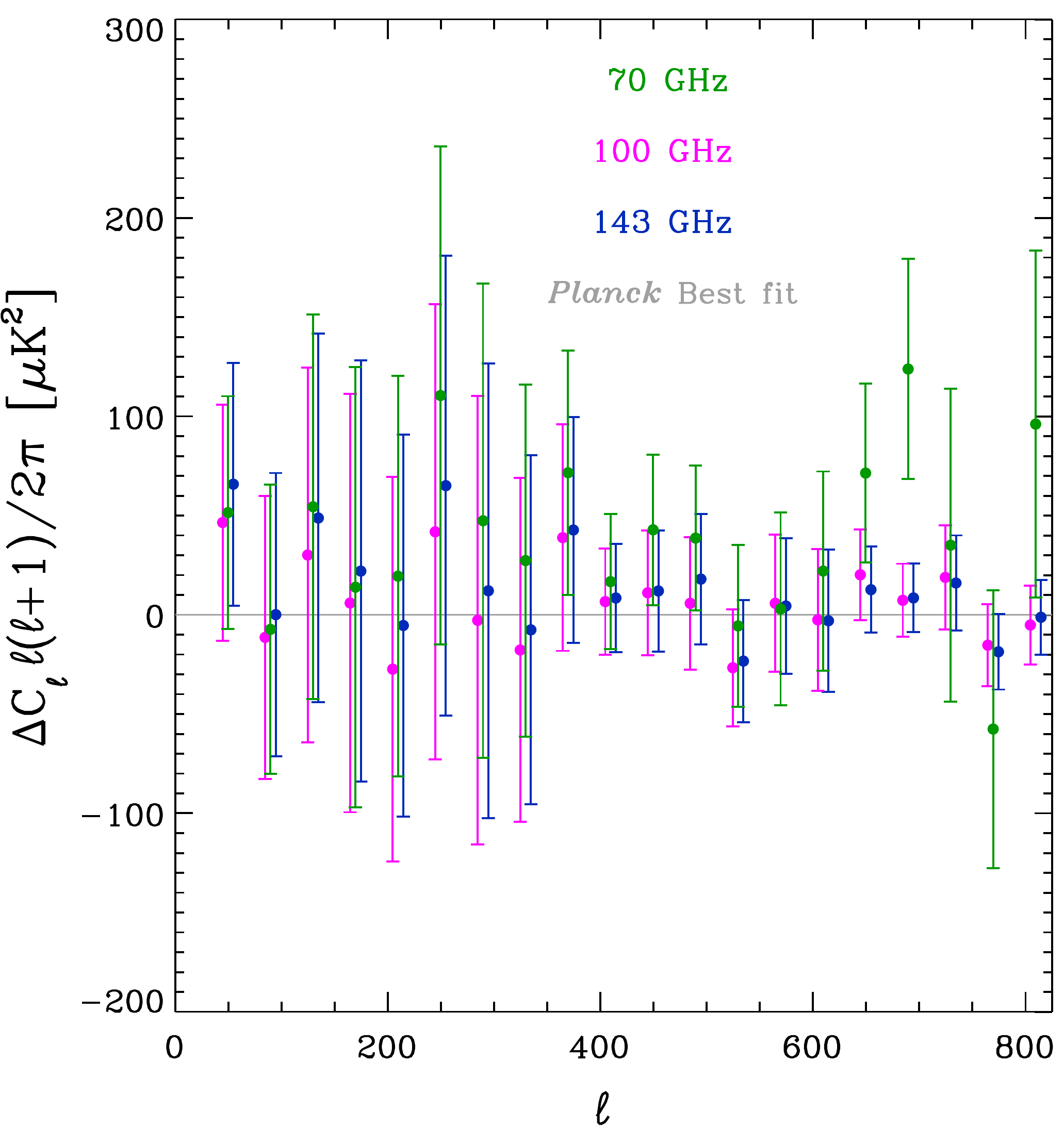}
\includegraphics[width=6.2cm]{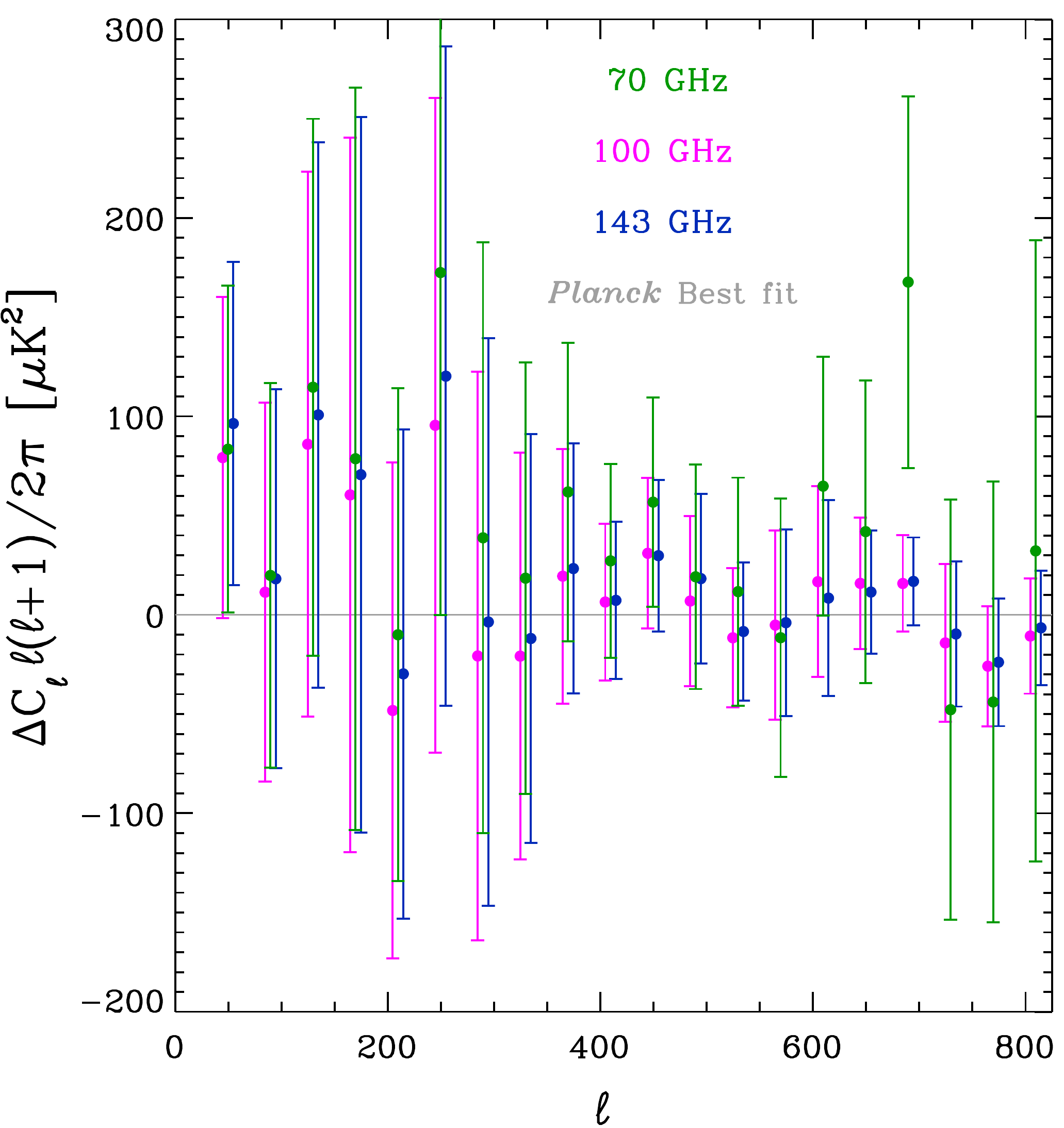}
}
\centerline{
\includegraphics[width=6.2cm]{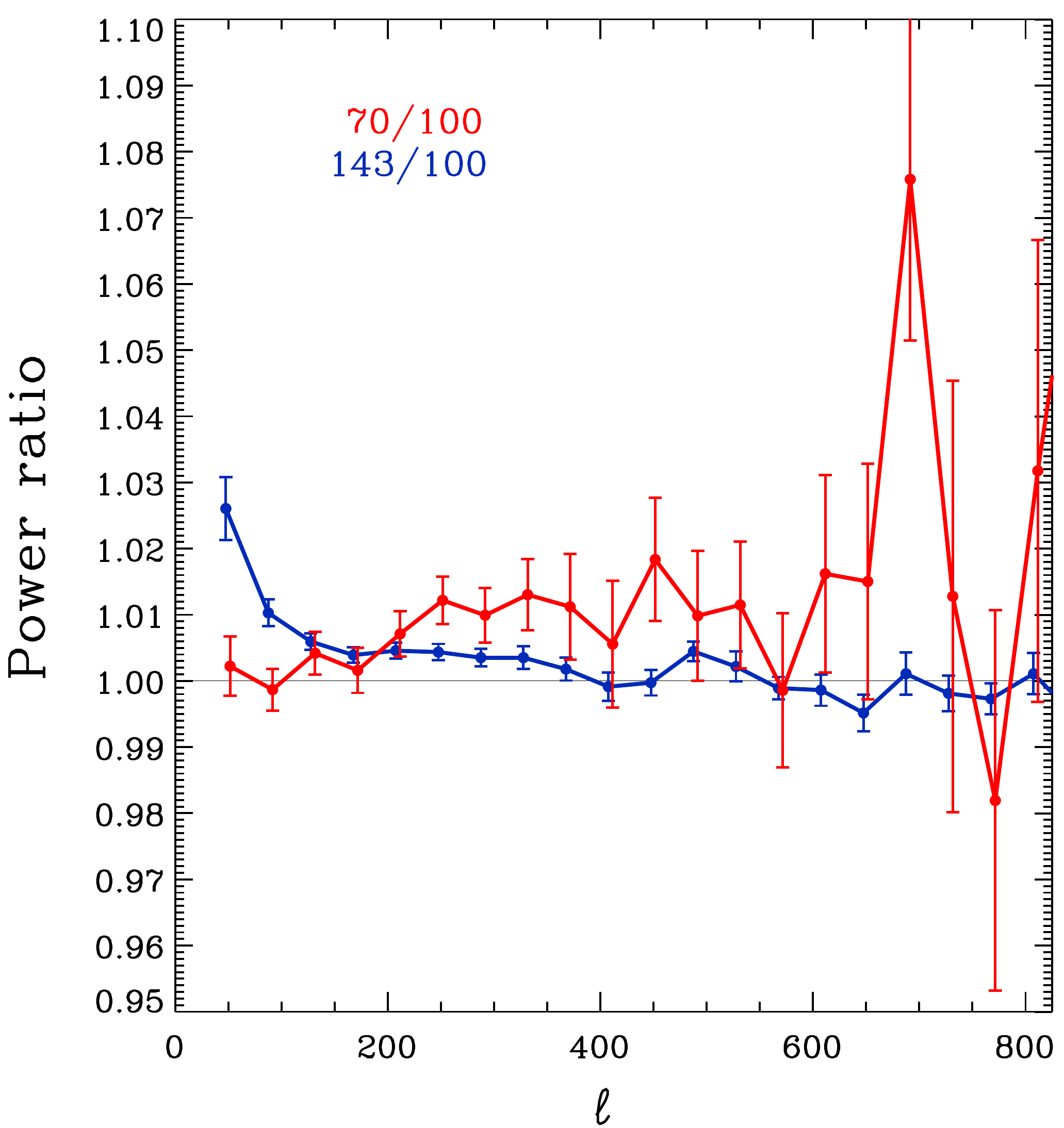}
\includegraphics[width=6.2cm]{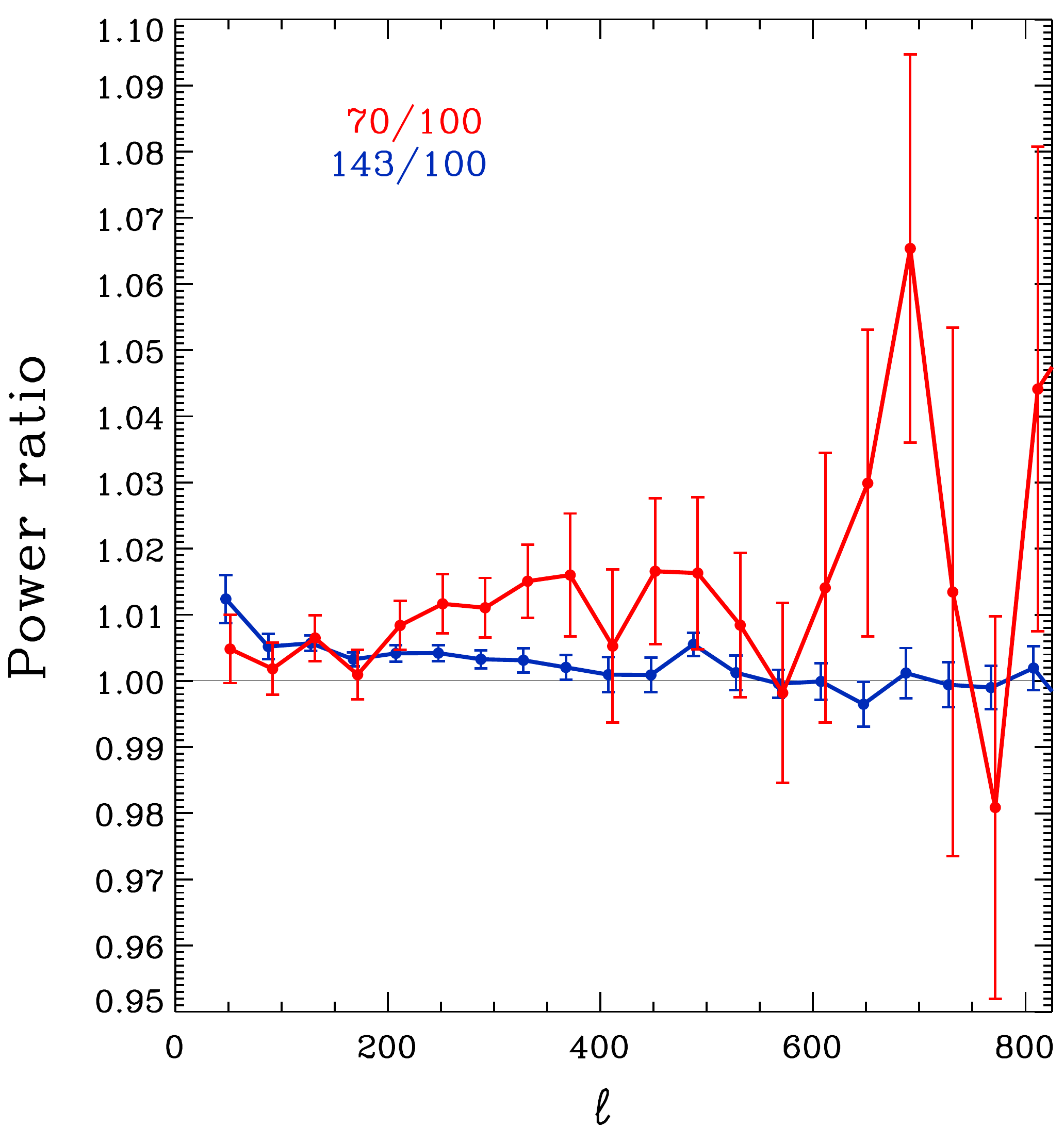}
\includegraphics[width=6.2cm]{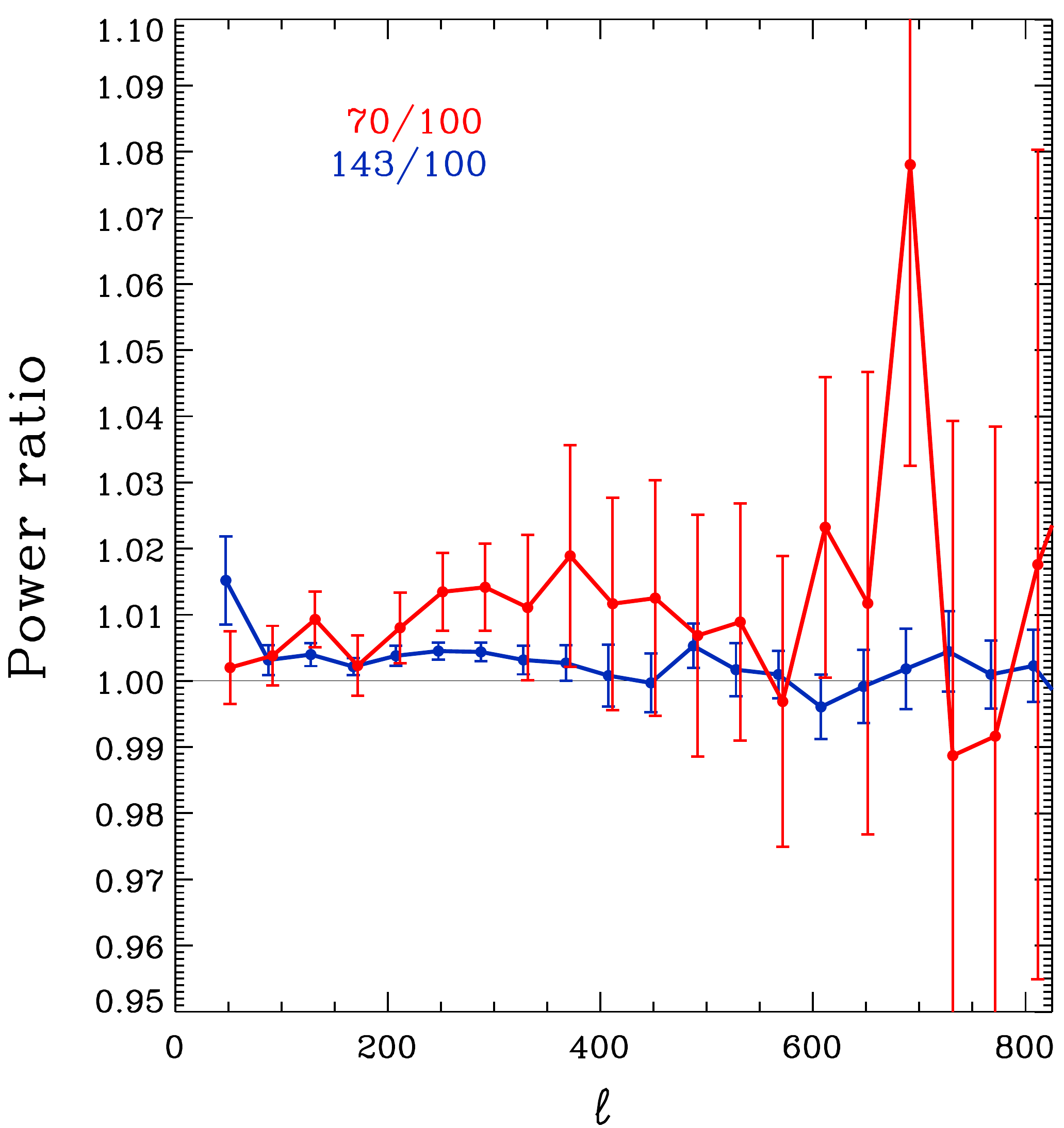}
}
\vskip 0.33in
\caption{Spectral analysis of the \planck\ 70, 100, and 143\,GHz maps.  Columns show results computed using the three sky masks in Fig.~1, with, from left to right, $\fsky = 69.4\,\%$, 59.6\,\%, and 39.7\,\%.  \textit{Top row}: CMB anisotropy spectra binned over a range of multipoles $\Delta \ell = 40$, for $\ell \ge 30$, with $(2\ell +1)$-weighting applied within the bin.  Error bars are computed as a measure of the {\rm rms}-power within each bin, and hence comprise both the measurement inaccuracy and cosmic variance. The grey curve is the best-fit \planck\ 6-parameter $\Lambda$CDM model from \citet{planck2013-p11}. Noise spectra computed from the half-ring-difference maps are shown: for the 70\,GHz channel, the ${\rm S/N} \approx 1$ at $\ell \approx 650$.  \textit{Middle row}: Residuals of the same power spectra with respect to the \planck\ best-fit model.  \textit{Bottom row}: Power ratios for the 70 vs.~100\,GHz and 143 vs.~100\,GHz channels of \planck.  The ratios are calculated $\ell$ by $\ell$, then binned.  The error bars show the standard error of the mean for the bin.  The effect of diffuse foregrounds is clearly seen in the  changes in the 143/100 ratio with sky fraction at $\ell \approx 100$.  Bin-to-bin variations in the exact values of the ratios with sky fraction emphasize the importance of making such comparisons precisely.}
\label{fig:planck_spectra_70_100_143}
\end{figure*}

Figure~\ref{fig:planck_spectra_70_100_143} shows spectra of the 70, 100, and 143\,GHz maps for the three sky masks, differences with respect to the \Planck\ 2013 best-fit model, and ratios of different frequencies.  In the $\ell$-range of the first peak and below, the 143/100 ratio shows the effects of residual diffuse foreground emission outside the masks.  The largest mask reduces the detected amplitude, but does not remove it completely.  The frequency dependence of the ratios conforms to what is well known, namely, that diffuse foreground emission is at a minimum between 70 and 100\,GHz.  The 143/100 pair is more affected by diffuse foregrounds than the 70/100 pair, as the dust emission gets brighter at 143\,GHz.  The effects of residual unresolved foregrounds in Fig.~\ref{fig:planck_spectra_70_100_143} are discussed in the next section.

\begin{figure}
\includegraphics[width=88mm]{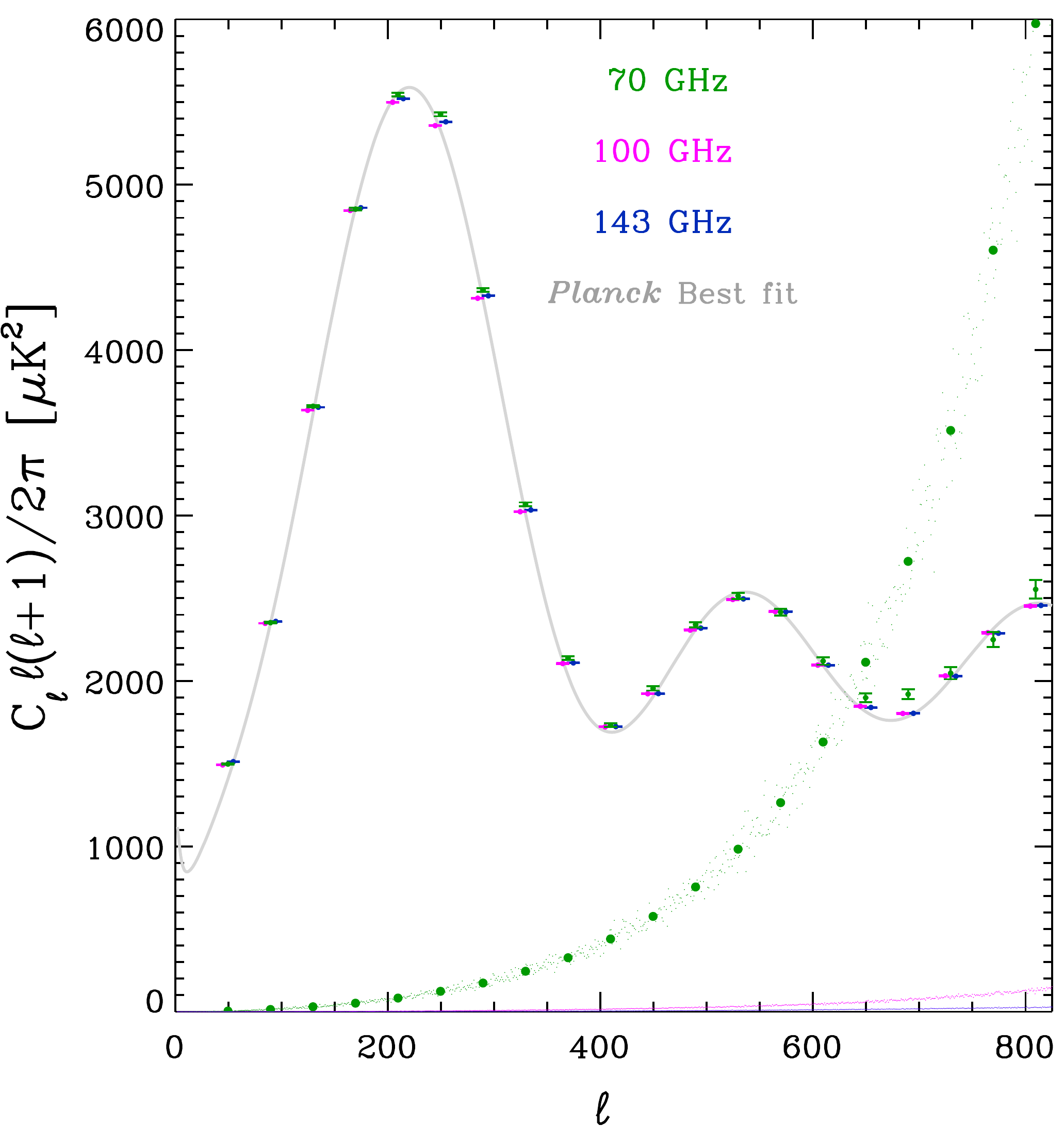}
\includegraphics[width=88mm]{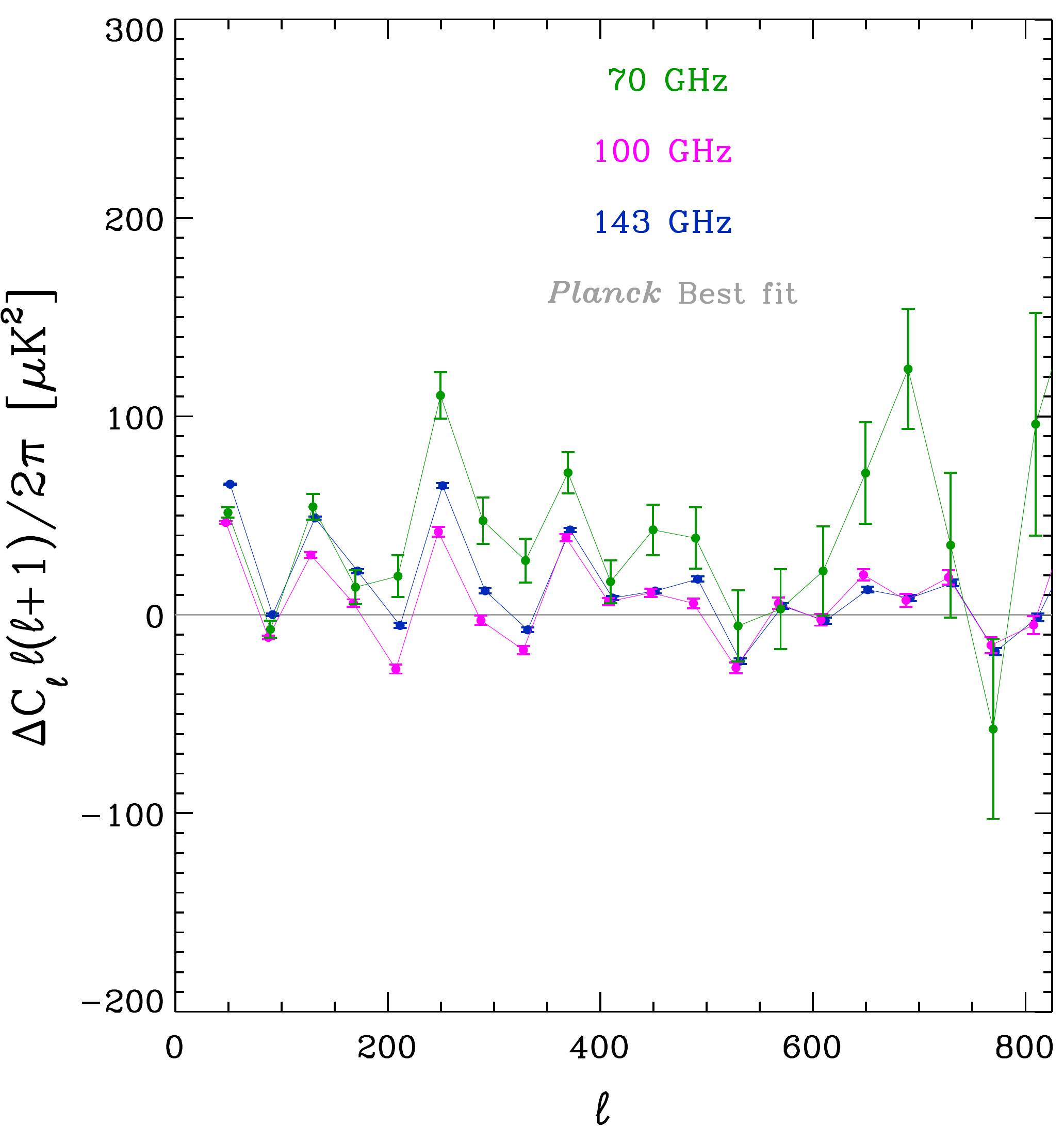}
\caption{Same as the top two panels of the middle column of Fig.~\ref{fig:planck_spectra_70_100_143}, but \textit{without} inclusion of signal cosmic variance in the uncertainties.  Both signal $\times$ noise and noise $\times$ noise terms are included.} 
\label{fig:planck_spectra_without_cosmic_variance_uncertainties}
\end{figure}

Near the first acoustic peak, measurements in the three \planck\ channels agree to better than one percent of the CMB signal, and to much better than their uncertainties, which are dominated by the effects of cosmic/sample variance (see Fig.~\ref{fig:planck_spectra_70_100_143}).  Inclusion of cosmic/sample variance is essential for making inferences about the underlying statistical processes of the Universe; however, since the receivers at all frequencies are observing a single realization of the CMB, cosmic variance is irrelevant in the comparison of the measurements themselves.  Figure~\ref{fig:planck_spectra_without_cosmic_variance_uncertainties} is the same as the top two middle panels of Fig.~\ref{fig:planck_spectra_70_100_143} (i.e., over 60\,\% of the sky), but {\it without\/} inclusion of cosmic/sample variance in the uncertainties.  As can be seen, cosmic/sample variance completely dominates the measurement uncertainties up to multipoles of 400, after which noise dominates.

\subsection{Residual unresolved sources}
\label{sec:residualsources}

Figure~\ref{fig:planck_spectra_70_100_143} shows that while diffuse foregrounds are significant for low multipoles, they are much less important on smaller angular scales.  To see clearly the intrinsic consistency between frequencies, however, we must remove the effects of unresolved sources.  

Discrete extragalactic foregrounds comprise synchrotron radio sources, Sunyaev-Zeldovich (SZ) emission in clusters, and dust emission in galaxies.  These have complicated behaviour in $\ell$ and $\nu$.  All have a Poisson part, but the SZ and cosmic infrared background sources also have a correlated part.  These are the dominant foregrounds (for a 39.7\,\% Galactic mask) for $\ell\ga200$.  For frequencies in the range 70--143\,GHz and multipoles in the range 50--200, they stay below 0.2\,\%.  The minimum in unresolved foregrounds remains at 143\,GHz, with less than 2\,\% contamination up to $\ell=1000$.

\begin{figure}
\includegraphics[width=88mm]{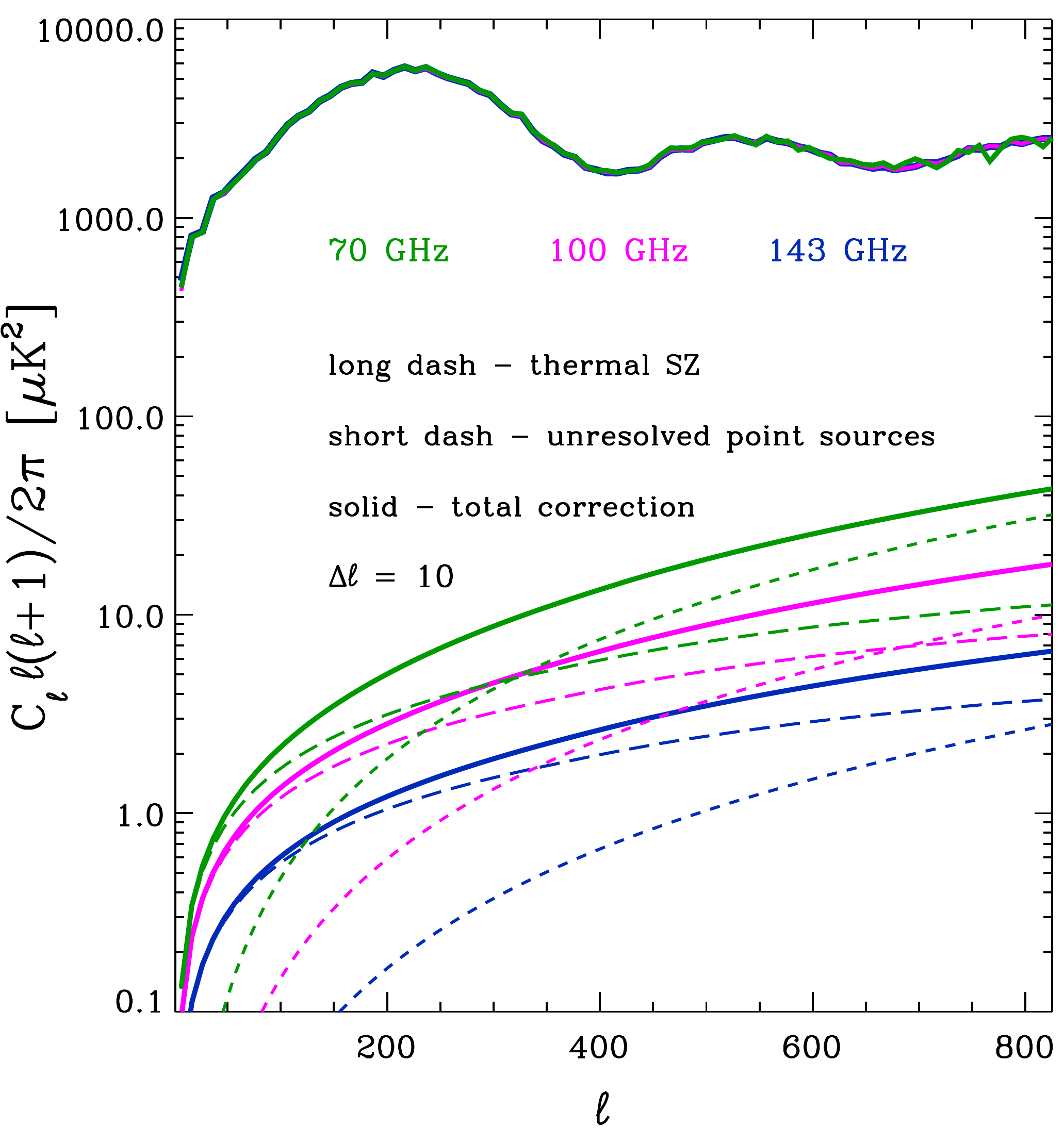}
\caption{Estimates of the residual thermal SZ and unresolved radio and infrared source residuals that must be removed.} 
\label{fig:sourcecorrections}
\end{figure}

\begin{figure}
\includegraphics[width=88mm]{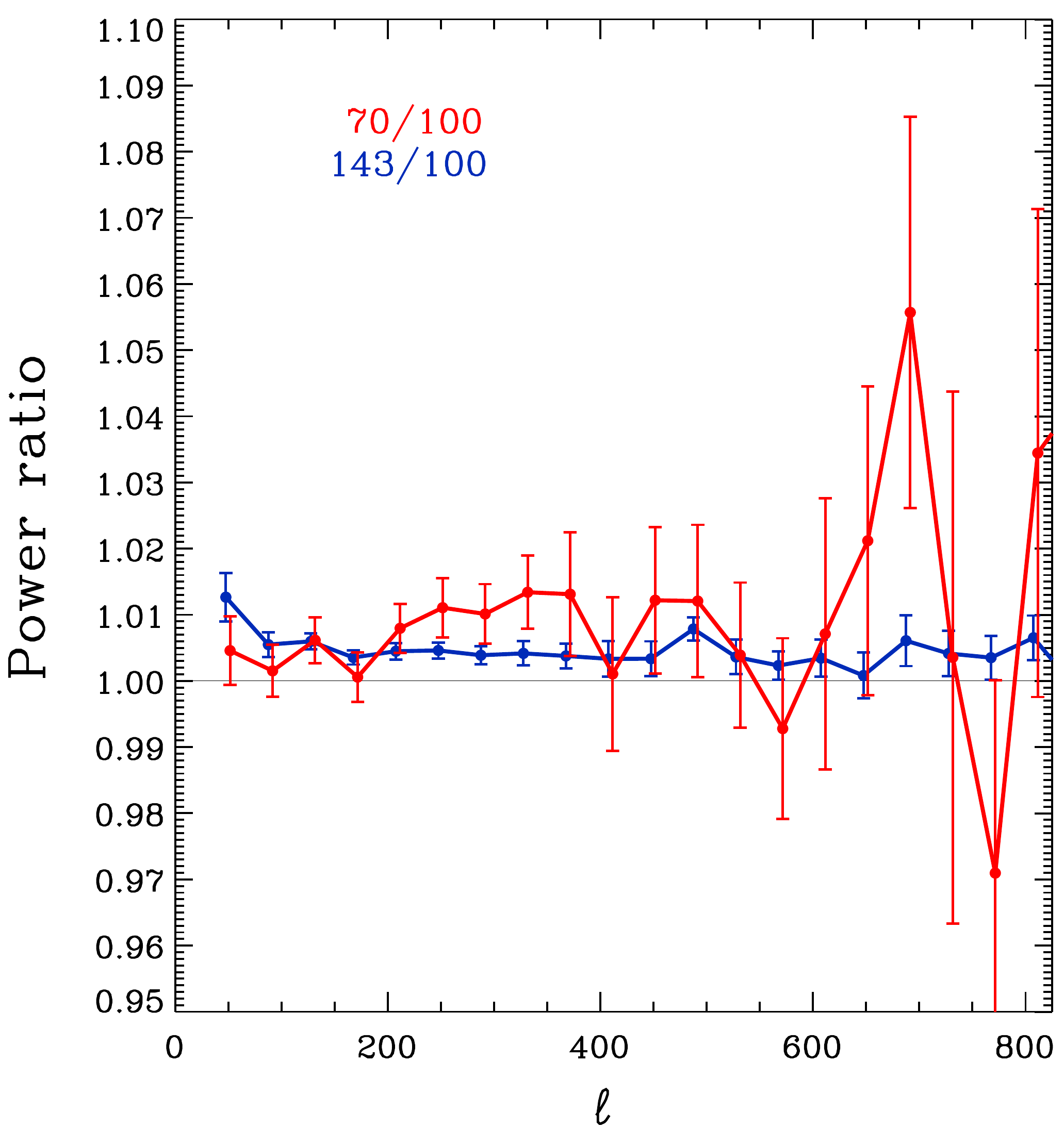}
\caption{Same as the bottom middle panel of Fig.~\ref{fig:planck_spectra_70_100_143}, but corrected for differences in unresolved-source residuals (see text).  We have not tried to account for uncertainties in the foreground correction itself; however, since the correction is small, the effect on the uncertainties would be small.} 
\label{fig:discretecorrected}
\end{figure}

Discrete sources detected above $5\,\sigma$ in the PCCS \citep{planck2013-p05} are individually masked, as described in Sect.~\ref{sec:planckskymaps}.  Corrections for residual unresolved radio sources are determined by fitting the differential Euclidean-normalized number counts $S^{5/2} dN/dS$ in ${\rm Jy}^{1.5}$\,sr\mo at each frequency with a double power law plus Euclidean term:
\begin{equation}
S^{5/2}\,dN/dS = {A_{\rm f} S^{5/2} \over \left[ (S/S_1)^{b_{\rm f1}} + (S/S_2)^{b_{\rm f2}}\right] + A_{\rm E} \, (1 - e^{-S/S_{\rm E}})},
\end{equation}
where $A_{\rm f}$ is the amplitude at faint flux density levels, $S_1$ is the first faint flux density level, $b_{\rm f1}$ is the exponent of the first power law at faint flux densities, $S_2$ is the second faint flux density level, $b_{\rm f2}$ is the exponent of the second power law at faint flux densities, $A_{\rm E}$ is the amplitude of the Euclidean part, i.e., at large flux density, and $S_{\rm E}$ is the flux density level for the Euclidean part ($\ga 1$\,Jy).  These are then integrated from a cutoff flux density corresponding to the $5\,\sigma$ selection limit in the PCCS at 143\,GHz, and the equivalent levels for a radio source with $S\propto \nu^{-0.7}$ at 100 and 70\,GHz.  Thermal SZ and CIB fluctuations are fitted as part of likelihood function determination described in \citet{planck2013-p08}; the values found there are used here.  Figure~\ref{fig:sourcecorrections} shows the level of these corrections, while Fig.~\ref{fig:discretecorrected} shows the ratios of power spectra after the corrections are made.

\subsection{Assessment}
\label{subsec:assessment}

The 70/100 and 143/100 ratios in Fig.~\ref{fig:discretecorrected}, for 59.6\,\% of the sky, averaged over the range $70\le\ell\le390$ where the 70\,GHz signal-to-noise ratio (S/N) is high, are 1.0080 and 1.0045, respectively.  Over the range $70\le\ell\le830$, the ratios are 1.0094 and 1.0043, respectively. 
Table~\ref{tab:ratios} collects these ratios and following ones for easy comparison.

\begin{table*}[t]
\caption{Summary of ratios of \Planck\ 70, 100, and 143\,GHz power spectra appearing in this paper.}
\label{tab:ratios}
\vskip -3.5mm
\tiny
\setbox\tablebox=\vbox{
\newdimen\digitwidth
\setbox0=\hbox{\rm 0}
\digitwidth=\wd0
\catcode`*=\active
\def*{\kern\digitwidth}
\newdimen\signwidth
\setbox0=\hbox{+}
\signwidth=\wd0
\catcode`!=\active
\def!{\kern\signwidth}
\halign{\hbox to 1.6in{#\leaderfil}\tabskip=2em&
         #\hfil\tabskip=3em& 
    \hfil#\hfil\tabskip=3em& 
    \hfil#\hfil\tabskip=3em& 
    \hfil#\hfil\tabskip=1em& 
    \hfil#\hfil\tabskip=0pt\cr
\noalign{\doubleline}
\omit&&&&\multispan2\hfil Spectrum Ratios\hfil\cr
\noalign{\vskip -3pt}
\omit&&&&\multispan2\hrulefill\cr
\noalign{\vskip 3pt}
\omit\hfil Location\hfil&\omit\hfil Features\hfil&$\fsky$&$\ell$ Range&70/100&143/100\cr
\noalign{\vskip 3pt\hrule\vskip 5pt}
Sect.~3.4, Fig.~6, bottom centre&No corrections&          59.6\,\%&$*70\le\ell\le390$&1.0089&1.0039*\cr
\omit&                                                           &&$*70\le\ell\le830$&1.0140&1.0020*\cr
\noalign{\vskip 3pt}
Sect.~3.4, Fig.~9&DSR$^{\rm a}$ correction&               59.6\,\%&$*70\le\ell\le390$&1.0080&1.0045*\cr
\omit&                                                           &&$*70\le\ell\le830$&1.0094&1.0043*\cr
\noalign{\vskip 3pt}
Sect.~3.4, {\tt SMICA}&Paper VI, figure~35&         40\,\%$^{\rm d}$&$*50\le\ell\le300$&1.0060&1.0020*\cr
\omit&                                                           &&$300\le\ell\le700$&1.0075&1.0020*\cr
\noalign{\vskip 3pt}
Sect.~4.3, Fig.~12&NS$^{\rm c}$ correction&               59.6\,\%&$*70\le\ell\le390$&1.0052&1.0040*\cr
\omit&                                                           &&$*70\le\ell\le830$&1.0077&1.0020*\cr
\noalign{\vskip 3pt}
Sect.~4.3, Fig.~13&DSR$^{\rm a}$+NS$^{\rm c}$ corrections&59.6\,\%&$*70\le\ell\le390$&1.0043&1.0046*\cr
\omit&                                                           &&$*70\le\ell\le830$&1.0032&1.0043*\cr
\noalign{\vskip 3pt}
Sect.~5,\phantom{.}* Fig.~14&CamSpecLikelihood$^{\rm d}$&                \ldots&\ldots&\dots&1.00058\cr
\noalign{\vskip 5pt\hrule\vskip 3pt}}}
\endPlancktablewide
\tablenote {{\rm a}} Discrete-source residual correction.\par
\tablenote {{\rm b}} The mask used in Paper VI, figure~35 was similar but not identical to the 39.7\,\% mask of Fig.~\ref{fig:Planck_masks}.  The differences do not affect the comparison.\par  
\tablenote {{\rm c}} Near sidelobe correction, 100 and 143\,GHz.\par
\tablenote {{\rm d}} \citet{planck2013-p11}.\par
\end{table*}

Section~7.4 of \citet{planck2013-p03} uses the {\tt SMICA} code to intercalibrate on the common CMB anisotropies themselves, with results given in figure~35 of that paper.  For 40\,\% of the sky, the 70/100 and 143/100 power ratios are 1.006 and 1.002 over the range $50\le\ell\le300$, and 1.0075 and 1.002 over the range $300\le\ell\le700$.  (These {\it gain\/} ratios from figure~35 of \citealt{planck2013-p03} must be squared for comparison with the {\it power\/} ratios discussed in this section and given in Table~\ref{tab:ratios}.)  The {\tt SMICA} equivalent power ratios are systematically about 0.2\,\% closer to unity than those calculated in this section; however, in broad terms the two methods give remarkably similar results.  Moreover, the absolute {\it gain\/} calibration uncertainties given in \citet[Table~8]{planck2013-p02b} and \citet{planck2013-p03f} are 0.62\,\% for 70\,GHz and 0.54\,\% for 100\,GHz and 143\,GHz.  The agreement at the power spectrum level between 70, 100, and 143\,GHz is quite reasonable in terms of these overall uncertainties.  We will return to comparisons of spectra in Sects.~\ref{sec:calibration-and-transfer-function} and \ref{sec:PSlate}.

We are working continuously to refine our understanding of the instrument characteristics, implement more accurate calibration procedures, and understand and control systematic effects better.  All of these will lead to reduced errors and uncertainties in 2014.  In the next section we describe an analysis of beams and calibration procedures that has already been beneficial.

\section{Beams, beam transfer functions, and calibration}
\label{sec:calibration-and-transfer-function}

The residual differences that we see in Sect.~\ref{sec:PSestimation} are small but not negligible.  We now address the question of whether they may be due to beam or calibration errors.  Detailed descriptions and analyses of the LFI and HFI beams and calibration are contained in \citet{planck2013-p02d}, \citet{planck2013-p02b}, \citet{planck2013-p03c}, and \citet{planck2013-p03f}.  In this section we summarize our present understanding of calibration and beam effects for the two instruments, explain the reasons for the approximations that have been made in data processing, provide estimates for the impact of these approximations on the resulting maps and power spectra, and outline plans for changes to be implemented in the 2014 data release.  We will show that the small differences between LFI and HFI at intermediate $\ell$ seen in Fig.~\ref{fig:discretecorrected} are significantly reduced by improvements in our understanding of the near sidelobes in HFI, which affect the window functions in this $\ell$ range.

\subsection{Beam definitions}
\label{subsec:beamdefinitions}

\begin{figure}
\includegraphics[width=88mm]{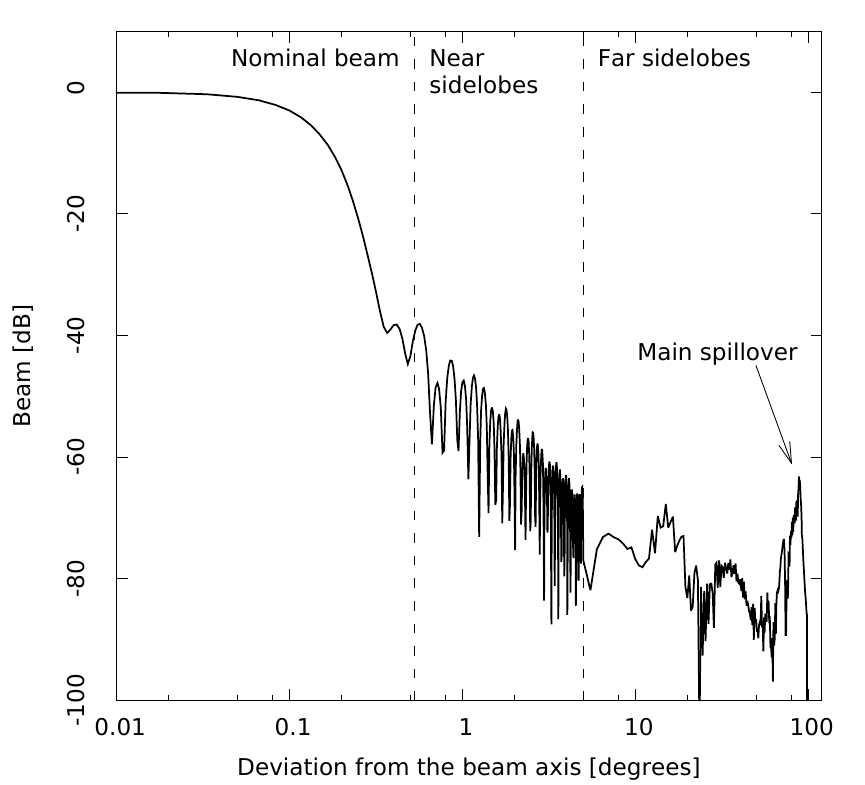}
\caption{Radial slice through a 70\,GHz beam from the {\tt GRASP} model, illustrating the nominal beam, near sidelobe, and far sidelobe regions. The exact choice of angular cutoff for the nominal beam is different for different frequencies. } 
\label{fig:radial_beam_70}
\end{figure}

Calibration of the CMB channels (30 to 353\,GHz) is based on the dipole anisotropies produced by the motion of the Sun relative to the CMB and of the modulation of this dipole by the motion of the spacecraft relative to the Sun (which we refer to as the solar and orbital dipoles, respectively).  For the 2013 data release and all LFI and HFI frequency channels considered in this paper, the time-ordered data have been fit to the solar dipole as measured by \WMAP~\citep{hinshaw2009}.  The present analysis aims to show that the LFI-HFI differences at intermediate $\ell$ seen in Fig.~\ref{fig:discretecorrected} are understood within the present uncertainties due to beams, calibration, and detector noise.  \citet{planck2013-p02d} and \citet{planck2013-p03c} define three regions of the beam response (see Fig.~\ref{fig:radial_beam_70}, Fig.~1 of~\citealt{planck2013-p02d}, and Fig.~5 of~\citealt{tauber2010b}), as follows.

\smallskip\noindent
The nominal beam or main beam is that portion used to create the beam window functions for the 2013 data release. The nominal beam carries most of the beam shape information and more than 99\,\% of the total solid angle, and therefore has most of the information needed for the 2013 cosmological analysis.  The angle from the beam centre to the boundary of the nominal beam varies with frequency and instrument, and is 1\pdeg9, 1\pdeg3, and 0\pdeg9 for 30, 44, and 70\,GHz, respectively, and 0\pdeg5 for 100\,GHz and above.
 
\smallskip\noindent 
The near sidelobes comprise any effective solid angle within 5\deg\ of the centre of the beam that is \emph{not} included in the nominal beam.  The response to the dipole from this region of the beam is very similar to that from the nominal beam, and unaccounted-for near sidelobe response leads to errors in the window function.  

\smallskip\noindent
The far sidelobes comprise the beam response more than 5\deg\ from the centre.  Because of the geometry of the telescope and baffles, the bulk of this solid angle is at large angles from the line of sight, not far from the spin axis, and not in phase with the dipole seen in the nominal beam, and therefore has little effect on the dipole calibration.  However, the secondary mirror spillover, containing typically $1/3$ of the total power in the far sidelobes, is in phase with the dipole, and affects the calibration signal.  The inaccuracy introduced by approximating the optical response with the nominal beam normalized to unity is corrected to first order by our use of a pencil ($\delta$-function) beam to estimate the calibration.

For reference, for the 2013 release we estimated a contribution to the solid angle from near sidelobes of 0.08\,\%, 0.2\,\%, and 0.2\,\%, and from far sidelobes of 0.62\,\%, 0.33\,\%, and 0.31\,\%, for 70, 100, and 143\,GHz,  respectively, of which 0.12\,\%, 0.075\,\%, and 0.055\,\% is from the secondary spillover referred to above. Recent analysis, detailed in Appendix~\ref{sec:LFIHFIassessment}, has resulted in a new estimate for the near sidelobe contribution for 100 and 143\,GHz of $0.30\pm0.2$\,\% and $0.35\pm0.1$\,\%, respectively\footnote{The solid angle statistical errors are 0.53\,\% and 0.14\,\% at 100 and 143\,GHz, respectively.}. The impact of this is described below.

\subsection{Nominal beam approximation}

In the 2013 analysis, both LFI and HFI performed a ``nominal beam'' calibration, i.e., we assumed that the detector response to the dipole can be approximated by the response of the nominal beam alone, which in turn is modelled as a pencil beam (for details see
Appendix~\ref{sec:calibration-convolution-details}).  Clearly, if 100\,\% of the power were contained in the nominal beam, the window function would fully account for beam effects in the reconstructed map and power spectrum.  In reality, however, a fraction of the beam power is missing from the nominal beam and appears in the near and far sidelobes, affecting the map and power spectrum reconstruction in ways that depend on the level of coupling of the sidelobes with the dipole.  Accordingly, a correction factor is applied that has the form (see Eq.~\ref{eq:Tsky_corr--appendix})
\begin{equation}
\Tsky \approx \Tskyest \left(1 - \phisky + \phiD\right),
\end{equation}
where \Tsky\ is the true sky temperature, \Tskyest\ is the sky temperature estimated by the  ``nominal beam" calibration,  $\phiD \equiv (\Pside\ast D)/(\Pnominal\ast D)$ is the coupling of the (near and far) sidelobes with the dipole, and  $\phisky \equiv (\Pside\ast\Tsky)/\Tskyest$  is a small term (of order 0.05\,\%, see Appendix~\ref{sec:LFIHFIassessment}) representing the sidelobe coupling with all-sky sources other than the dipole (mainly CMB anisotropies and Galactic emission). The term \phiD\ is potentially important, since dipole signals contributing to the near sidelobes may bias the dipole calibration. Our current understanding of the value and uncertainty of the scale factors $\eta = \left(1 - \phisky + \phiD\right)$ for LFI and HFI is discussed in detail in Appendix~\ref{sec:LFIHFIassessment}.

\subsection{Key findings}
\label{subsec:keyfindings}

There are two key findings or conclusions from the analyses in Appendices~\ref{sec:calibration-convolution-details} and \ref{sec:LFIHFIassessment}.

\begin{itemize}

\item For LFI, a complete accounting of the corrections using the \emph{current} full $4\pi$ beam model would lead to an adjustment of about 0.1\,\% in the amplitude of the released maps (i.e., 0.2\,\% of the power spectra). At present this is an estimate, and rather than adjusting the maps we include this in our uncertainty.

\item For HFI, recent work on a \emph{hybrid beam profile}, including data from planet measurements and {\tt GRASP}\footnote{Developed by TICRA (Copenhagen, DK) for analysing general reflector antennas (\url{http://www.ticra.it}).} modelling, has led to improvements in the beam window function correction rising from 0 to 0.8, 0.8, 0.5, and 1.2\,\% over the range $\ell=1$ to $\ell= 600$, at 100, 143, 217, and 353\,GHz, respectively.  Uncertainties in these corrections have not been fully characterized, but are dominated by the intercalibration of Mars and Jupiter data and are comparable to the corrections themselves (see Fig.~\ref{fig:HFI_nsl_wfn})
\end{itemize}

\begin{figure}
\includegraphics[width=88mm]{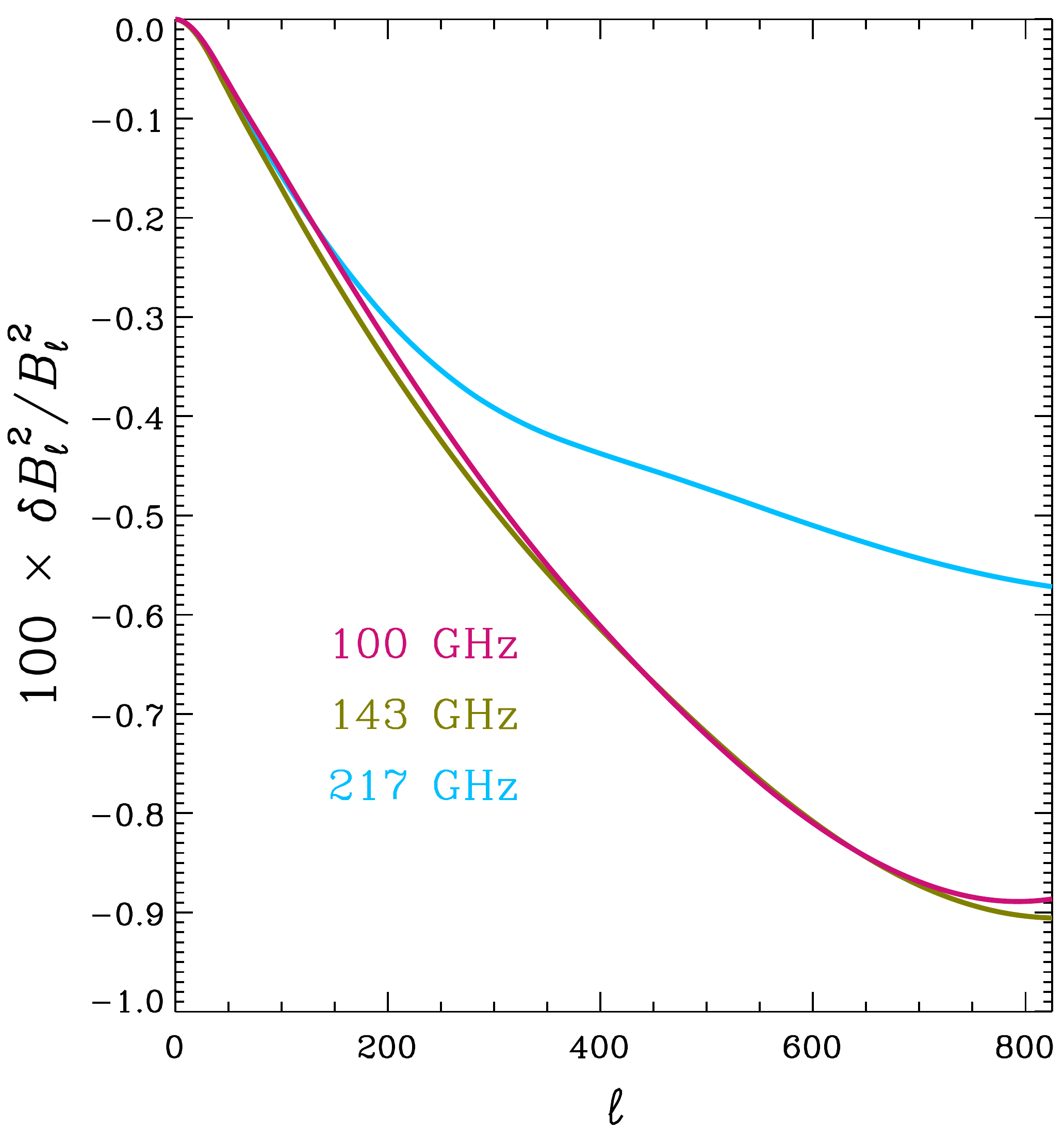}
\caption{Effective beam window function corrections from Fig.~\ref{fig:HFI_nsl_wfn}, which correct for the effect of near-sidelobe power missing in the HFI beams used in the 2013 results (Sect.~\ref{subsubsec:mainandnear}).  Uncertainties are not shown here for clarity, but are shown in Fig.~\ref{fig:HFI_nsl_wfn}, and would be large on the scale of this plot. The 217\,GHz correction is shown for illustration only.}
\label{fig:beamcorrections}
\end{figure}

\begin{figure}
\includegraphics[width=88mm]{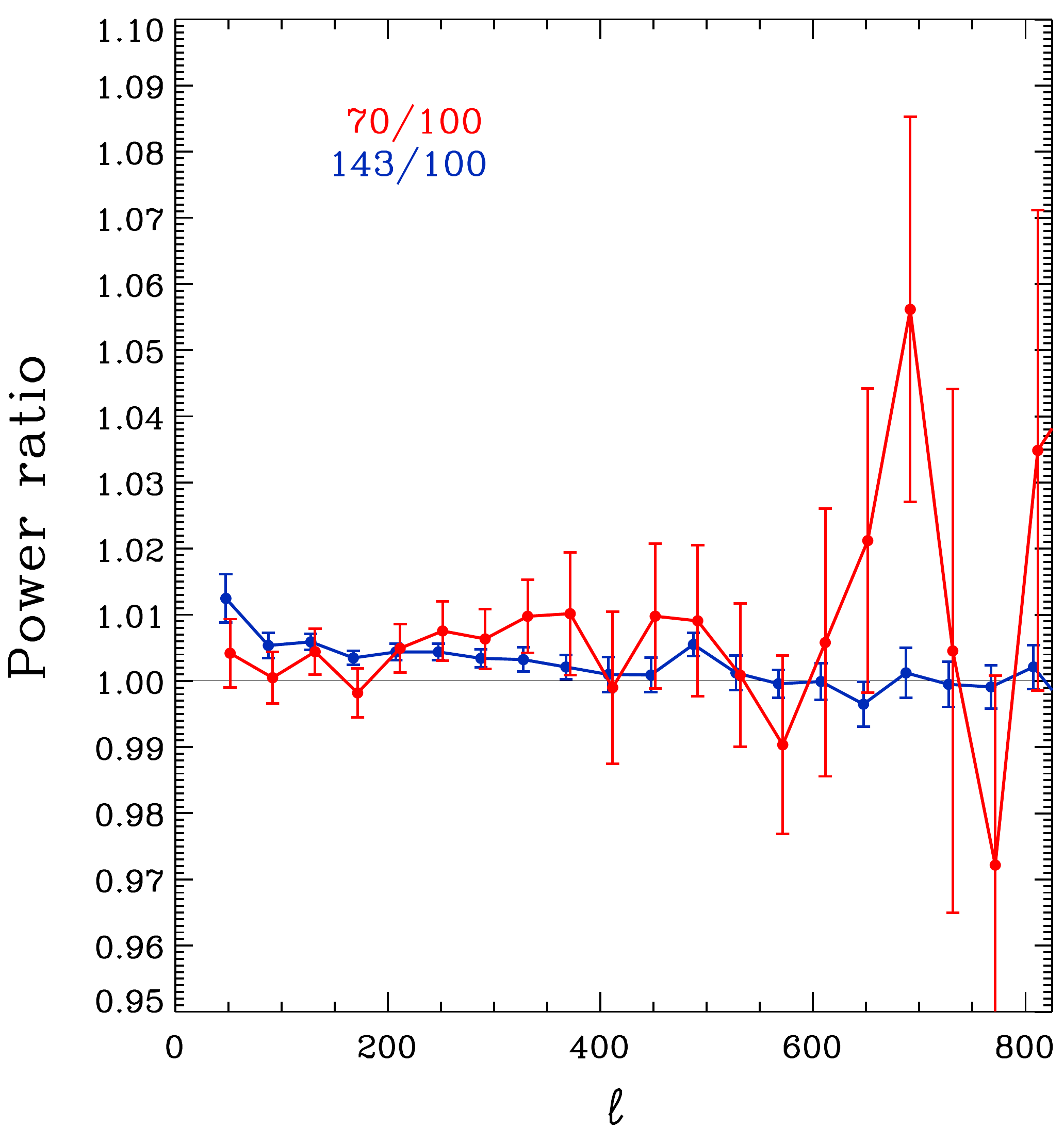}
\caption{Same as the bottom middle panel of Fig.~\ref{fig:planck_spectra_70_100_143}, but corrected for the near-sidelobe power at 100 and 143\,GHz that was not included in the 2013 results.  Since the beam corrections for 100 and 143\,GHz are nearly identical, the ratio 143/100 hardly changes.  The ratio 70/100, however, changes significantly, moving towards unity. Uncertainties in the beam window function corrections are not included.}
\label{fig:beamcorrected}
\end{figure}

Figure~\ref{fig:beamcorrections} shows the corrections to the beam window functions at 100, 143, and 217\,GHz.  Figure~\ref{fig:beamcorrected}  shows the effect of those corrections on the 70/100 and 143/100 power spectrum ratios, uncorrected for unresolved source residuals.  There is almost no effect on the 100/143\,GHz ratio, as the {\it differential} beam window function correction between these two frequencies is small.  The 70/100 ratio, however, is significantly closer to unity.  In Sect.~\ref{sec:PSlate}, we show that such a correction does not materially affect the 2013 cosmology results.

\begin{figure}
\includegraphics[width=88mm]{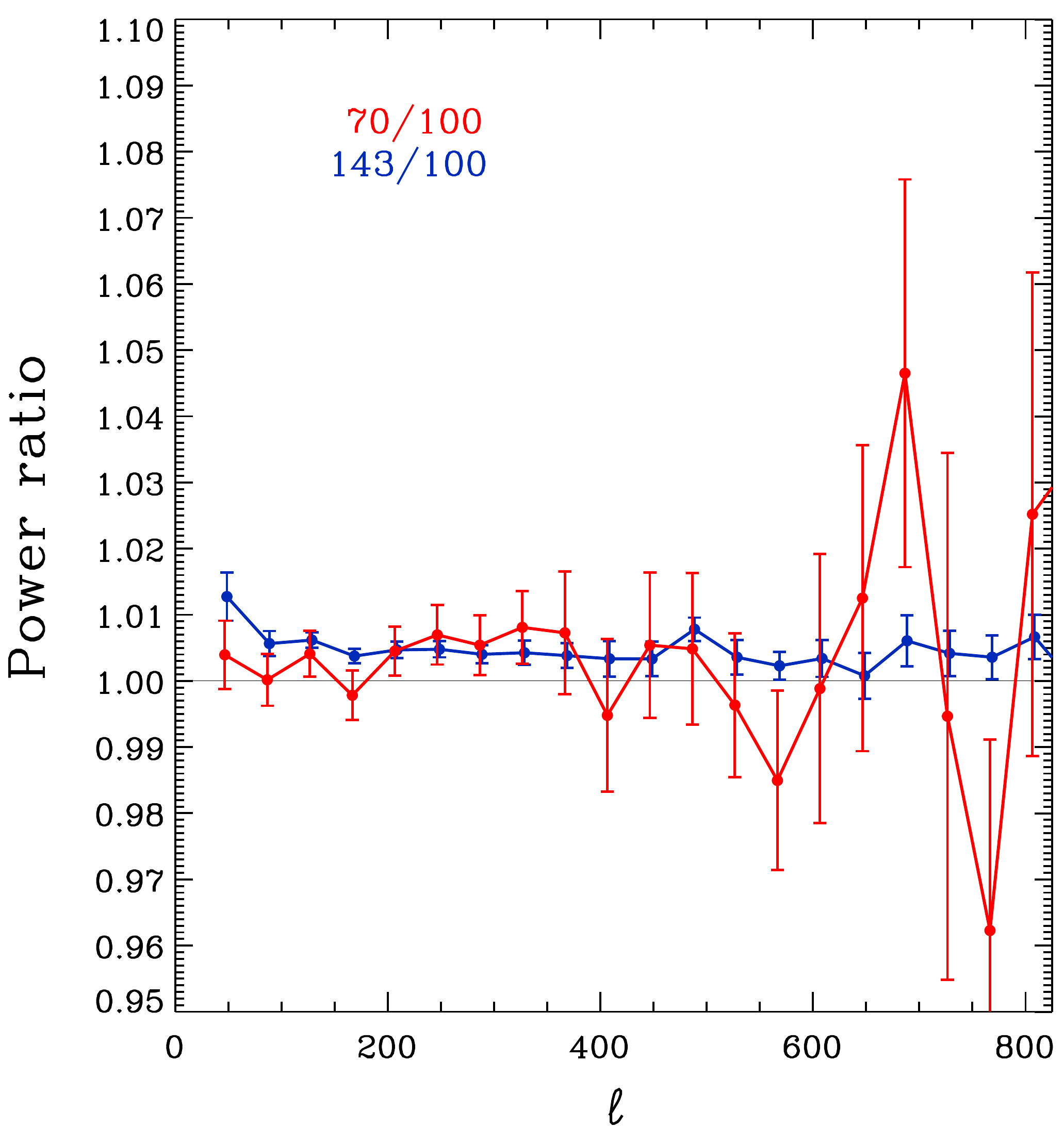}
\caption{Same as the bottom middle panel of Fig.~\ref{fig:planck_spectra_70_100_143}, but corrected for both the near-sidelobe power at 100 and 143\,GHz that was not included in the 2013 results and for unresolved source residuals (Sect.~\ref{sec:residualsources}). Uncertainties in the beam window function corrections are not included.}
\label{fig:corrected}
\end{figure}

Figure~\ref{fig:corrected} shows the power spectrum ratios corrected for both the beam window functions and unresolved source residuals (Sect.~\ref{sec:residualsources}).  The average ratios over the range $70\le\ell\le390$ are 1.0043 and 1.0046 for 70/100 and 143/100, respectively.  For the range $70\le\ell\le830$, they are 1.0032 and 1.0043.

For the 2014 release, we expect internal consistency and uncertainties to further improve as more detailed models of the beam and correction factors are included in the analysis.

\bigskip

We have concentrated in this section on beam effects; however, the transfer function depends also on the residuals of the time transfer function, measured on planets and glitches, and deconvolved in the time-ordered data prior to mapmaking and calibration.  For the HFI channels, the transfer function used for the 2013 cosmological analysis assumes that all remaining effects are contained within a $40\arcm\times40\arcm$ map of a compact scanning beam and corresponding effective beam.  Any residuals from uncorrected time constants longer than 1\,s are left in the maps, and will affect the dipoles and thus the absolute calibration.  This has been investigated since the 2013 data release; time constants in the 1--3\,s range have been identified and shown to be the origin of difficulties encountered with calibration based on the orbital dipole. The 2014 data release will include a correction of these effects, and the absolute calibration will be carried out on the orbital dipole.  A reduction in calibration uncertainties by a factor of a few can be anticipated.

\section{Likelihood analysis}
\label{sec:PSlate}

In the previous section we showed how work since release of the 2013 \Planck\ results has led to an improved understanding of the beams and a small (and well within the stated uncertainties) revision to the near-sidelobe power in the HFI beams, which brings HFI and LFI into even closer agreement. In this section, we show that the revision in the HFI beams has little effect on cosmological parameters.  To do this, we make use of the likelihood and parameter estimation machinery described in \citet{planck2013-p08} and \citet{planck2013-p11}. For both analytical and historical reasons there are differences (e.g., masks, frequencies, multipole ranges) in the analyses in this section and in previous sections; however, as will be seen, the effects of the differences are accounted for straightforwardly, and do not affect the conclusions about parameters.

The \Planck\ 2013 cosmological parameter results given in \citet{planck2013-p11} are determined for $\ell \ge 50$ from 100, 143, and 217\,GHz ``detector set'' data described in \citet[Table~1]{planck2013-p08}, by means of the {\tt CamSpec} likelihood analysis described in the same paper that solves simultaneously for calibration, foreground, and beam parameters.  This approach allows power spectrum comparisons to sub-percent level precision, using only cross-spectra (as in Sect.~\ref{sec:PSestimation}) to avoid the need for accurate subtraction of noise in auto-spectra.  

In this section, we determine the ratios of the 100, 143, and 217\,GHz spectra using this approach, and compare the 143/100 results to those found in Sect.~\ref{sec:PSestimation}.  We show that the apparent difference in the results from the two different approaches is easily accounted for by differences in the sky used, the difference between the detector set data and full frequency channel data, and the use of individual detector recalibration factors in the detector set/likelihood approach.  Having established essentially exact correspondence between the methods, we use the likelihood machinery to estimate the effect on cosmological parameters of the revision in the near-sidelobe power in the HFI beams.

\begin{figure}
\includegraphics[width=88mm]{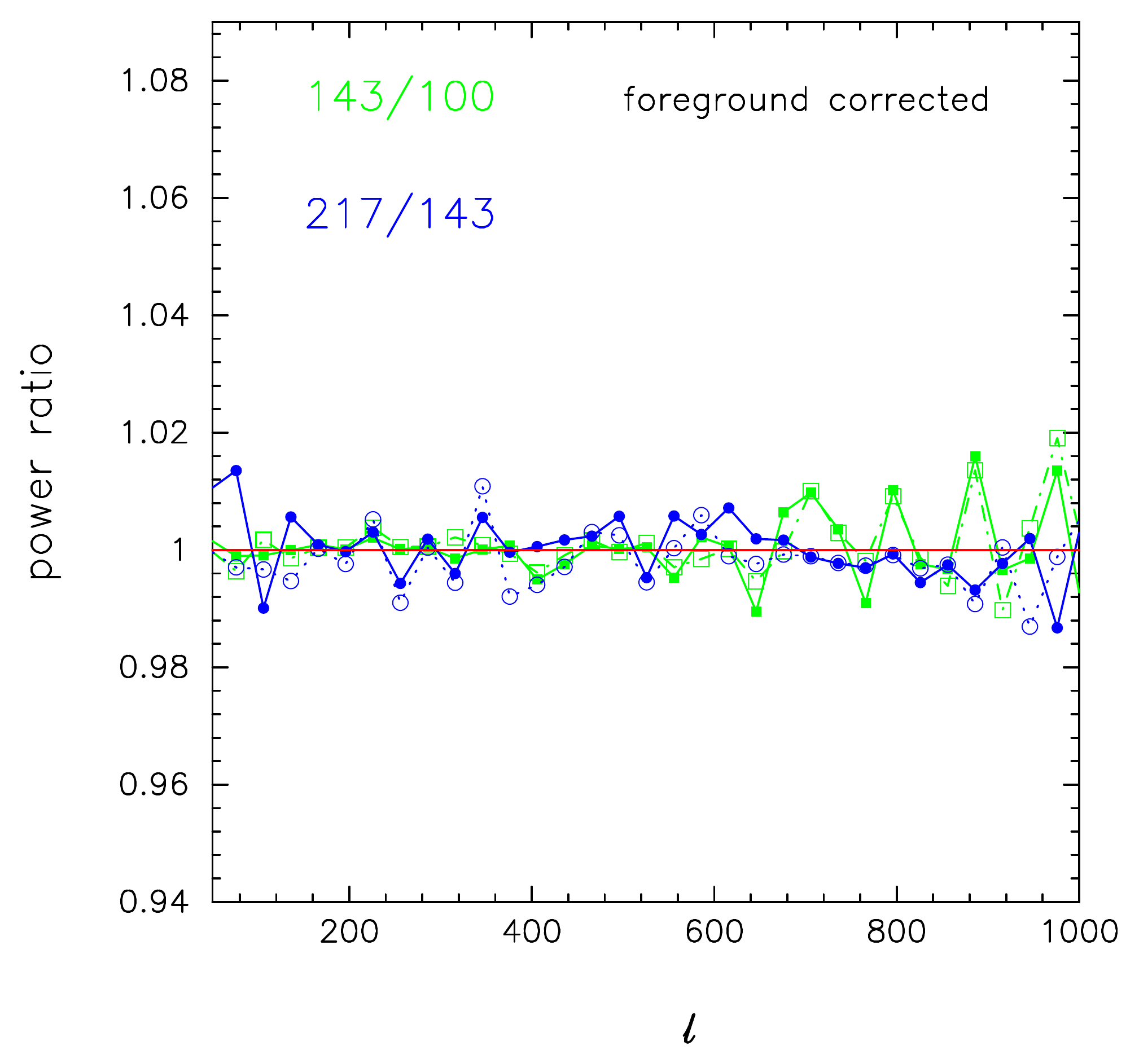}
\caption{Ratios of 100, 143, and 217\,GHz power spectra calculated from detector sets with the likelihood method, including subtraction of the best-fitting foreground model (see text) and correction for the best-fit relative calibration factors for individual detectors.  Solid symbols and lines show ratios for mask G22; open symbols and dotted or dashed lines show ratios for mask G35.  The greater scatter in 217/143 for mask G35 is caused by CMB-foreground cross-correlations.}
\label{fig:HFIratios}
\end{figure}

\begin{figure}
\includegraphics[width=88mm]{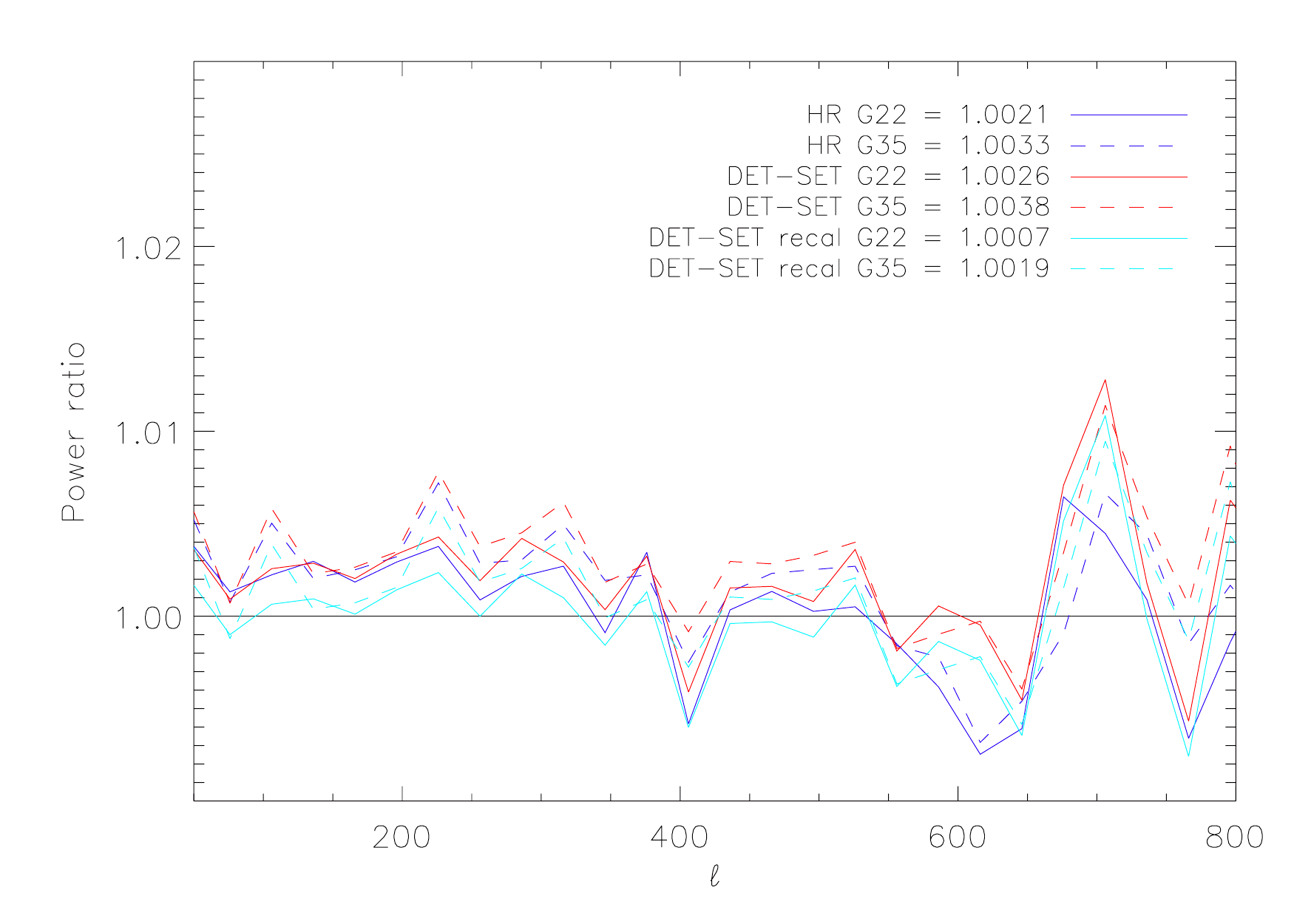}
\caption{Effects on the 143/100 ratio of changes in the mask, choice of detectors, and detector recalibration.  Solid lines indicate ratios calculated with mask G22; dashed lines indicate mask G35.  Use of detector sets gives the cyan curves with recalibration turned on and the red curves with recalibration turned off.  Use of full-frequency half-ring cross spectra, as in Sect.~\ref{sec:PSestimation}, gives the blue curves.  The cyan curves are comparable to the green curves in Fig.~\ref{fig:HFIratios}, which also have intra-frequency calibration and foreground corrections applied.  The blue-dashed curve agrees extremely well with the blue curve in the bottom right panel of Fig.~\ref{fig:planck_spectra_70_100_143}, as it should (see text). }
\label{fig:Marina}
\end{figure}

In \citet{planck2013-p11}, we used mask G45 ($f_{\rm sky} = 0.45$) for $100\times100$\,GHz, and mask G35 ($f_{\rm sky} = 0.37$) for $143\times143$\,GHz and $217\times217$\,GHz to control diffuse foregrounds.  However, here we are interested in precise tests of inter-frequency power spectrum consistency, so (as before) we need to compute spectra using exactly the same masks to cancel the effects of cosmic variance from the primordial \hbox{CMB}.   We have therefore recomputed all of the spectra using mask G22 ($f_{\rm sky} = 0.22$) and mask G35, restricting the sky area to reduce the effects of Galactic dust emission at 143 and 217\,GHz.  The spectra are computed from means of detector set cross-spectra \citet{planck2013-p08}.  For each spectrum, we subtract the best-fitting foreground model from the Planck+WP+high$\ell$ solution for the base six-parameter $\Lambda$CDM model with parameters as tabulated in \citet{planck2013-p11}, and correct for the best-fit relative calibration factors of this solution.  The convention adopted in the CamSpec likelihood fixes the calibration of $143\times143$ to unity, hence calibration factors multiply the $100\times100$ and $217\times217$ power spectra to match the $143\times143$ spectrum. The best fit values of these coefficients are $c_{100}=1.00058$ and $c_{217}=0.9974$ for mask G22, both very close to unity and consistent with the calibration differences between individual detectors at the same frequency (see Table~3 of \citealt{planck2013-p08}).  The results are shown in Fig.~\ref{fig:HFIratios}.

The 143/100 ratio given by the dashed green line can be compared with the bottom right panel of Fig.~\ref{fig:planck_spectra_70_100_143}, which is based on 40\,\% of the sky, nearly the same as mask G35.  As expected, they are not identical; Fig.~\ref{fig:Marina} explains the differences.  In Fig.~\ref{fig:Marina}, pairs of curves in the same colour show the difference between mask G22 and mask G35, as labelled.  
The cyan curves can be compared to the green curves in Fig.~\ref{fig:HFIratios}, which also have inter-frequency calibration and foreground corrections applied.  The red curves show the effect of turning off detector-by-detector intercalibration.  The blue curves show the effect of switching from detector sets to full-frequency half-ring cross spectra (as in Sect.~\ref{sec:PSestimation}).  The progression from solid cyan to dashed blue in Fig.~\ref{fig:Marina} shows the relationship between the PLA map-based results and the detector-set/likelihood results.  As used in the likelihood analysis \citep{planck2013-p08}, the 143/100 ratio is 1.00058 over the full $\ell$ range used in the likelihood analysis, compared to the ratios between 1.0039 and 1.0046 seen in Table~\ref{tab:ratios} over $70\le\ell\le390$ for Figs.~6, 9, 12, or 13.  However, using mask G35 ($\fsky = 0.37$), using half-ring cross-spectra of full-frequency detector sets, and turning off unresolved-source residual and detector-by-detector intercalibration factors, changes the ratio over $60<\ell<390$ to 1.0033, in good agreement with the 1.0039 calculated for the 143/100 comparison in the bottom right panel of Fig.~\ref{fig:planck_spectra_70_100_143}.

This agreement extends to the detailed shapes of the two curves (blue in the bottom right panel of Fig.~\ref{fig:planck_spectra_70_100_143} and blue-dashed in  Fig.~\ref{fig:Marina}) as well.  This is necessarily the case, since they are both cross-spectra of half-ring frequency maps, without corrections for unresolved-source residuals, and using the ``2013'' beams.  The only difference in the data comes from the masks used, which are the GAL040 mask ($f_{\rm sky} = 39.7$\,\%) and mask G35 ($f_{\rm sky} = 37$\,\%), respectively.  This agreement is nevertheless reassuring in showing that the differences in spectral ratios between the PLA map-based approach and the detector set likelihood approach are well-understood, and disappear for common data and masks.

We can now turn to the question of whether the small revision to the HFI beams affects cosmological parameters.  A full revised beam analysis at the detector level that includes the 0.1\,\% power in near sidelobes not taken into account directly in the 100 and 143\,GHz beams in 2013 (Sect.~\ref{sec:calibration-and-transfer-function}) has not yet been completed; however, for an indicative test, we rescaled the averaged cross-spectra appearing in the likelihood by functions corresponding to the new beam shapes for the $\ell$-ranges for which they have been calculated (presently up to $\ell=2000$).  Where necessary the shapes were extrapolated as being flat up to higher $\ell$.  The $143 \times 217$ spectrum was rescaled by the geometric mean of the $143$ and $217$ rescalings.  Then we performed a Monte Carlo Markov Chain (MCMC) analysis for the base $\Lambda$CDM model for the modified ``high-$\ell$'' likelihood with an unmodified low-$\ell$ Planck likelihood and \WMAP\ low-$\ell$ polarized likelihood (``WP'').  To see any change in the beam error behaviour, we choose to sample explicitly over all twenty of the eigenmode amplitudes, rather than sampling over one and marginalizing over the other nineteen, as we did in the parameters paper \citep{planck2013-p11}.

\begin{figure}
\includegraphics[width=88mm]{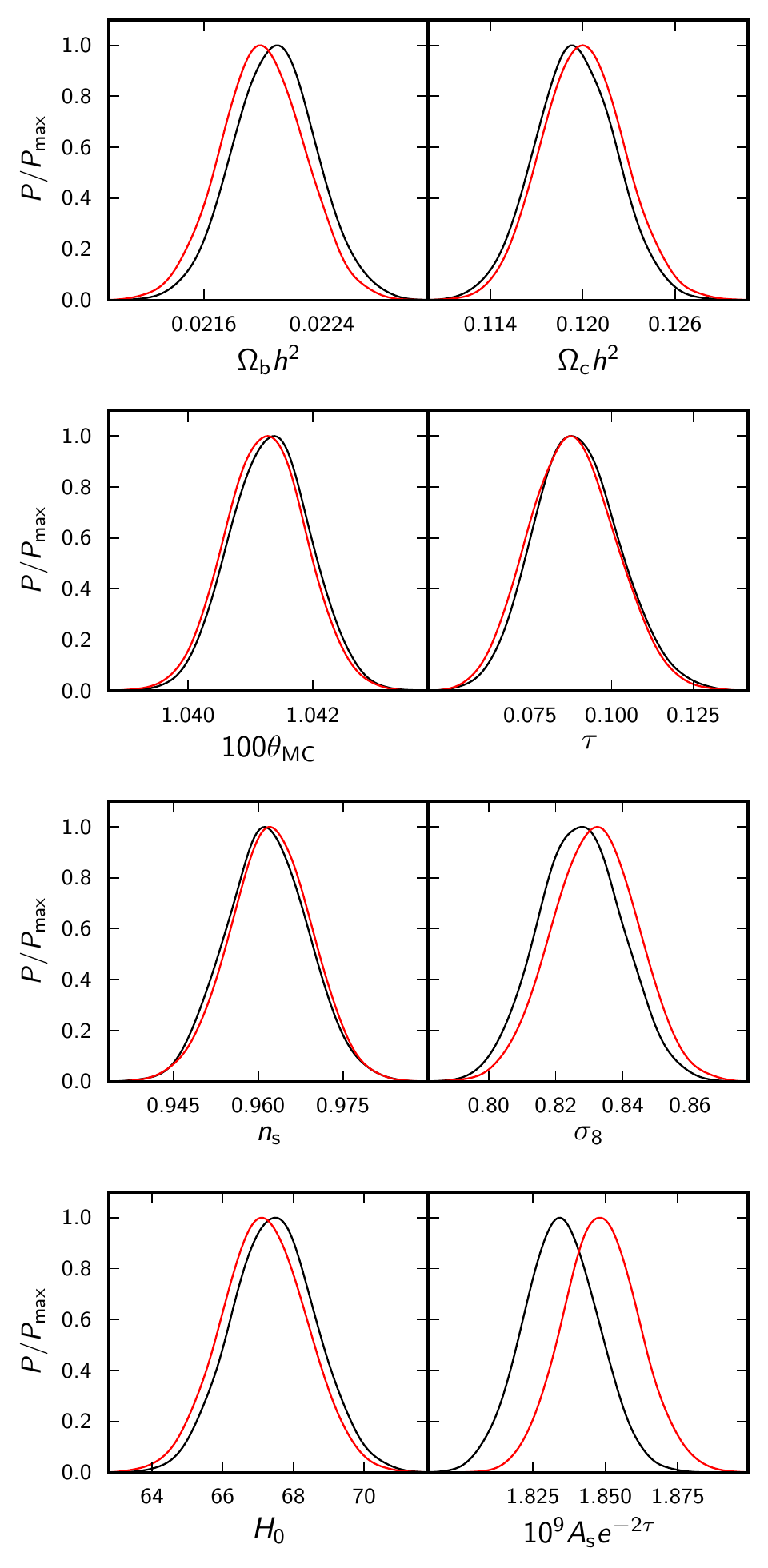}
\caption{Changes in cosmological parameters from the inclusion of the near sidelobe power discussed in the text.  The black curves are the 2013 results for \Planck\ plus the low-$\ell$ \WMAP\ polarization (WP).  The red curves are for \Planck+WP using the revised HFI beams. The shifts in the posteriors are all less than $0.3\sigma$ except for the cosmological amplitude $A_{\rm s}$ and parameters related to it, as expected.}
\label{fig:parameters1}
\end{figure}

\begin{figure*}
\includegraphics[width=180mm]{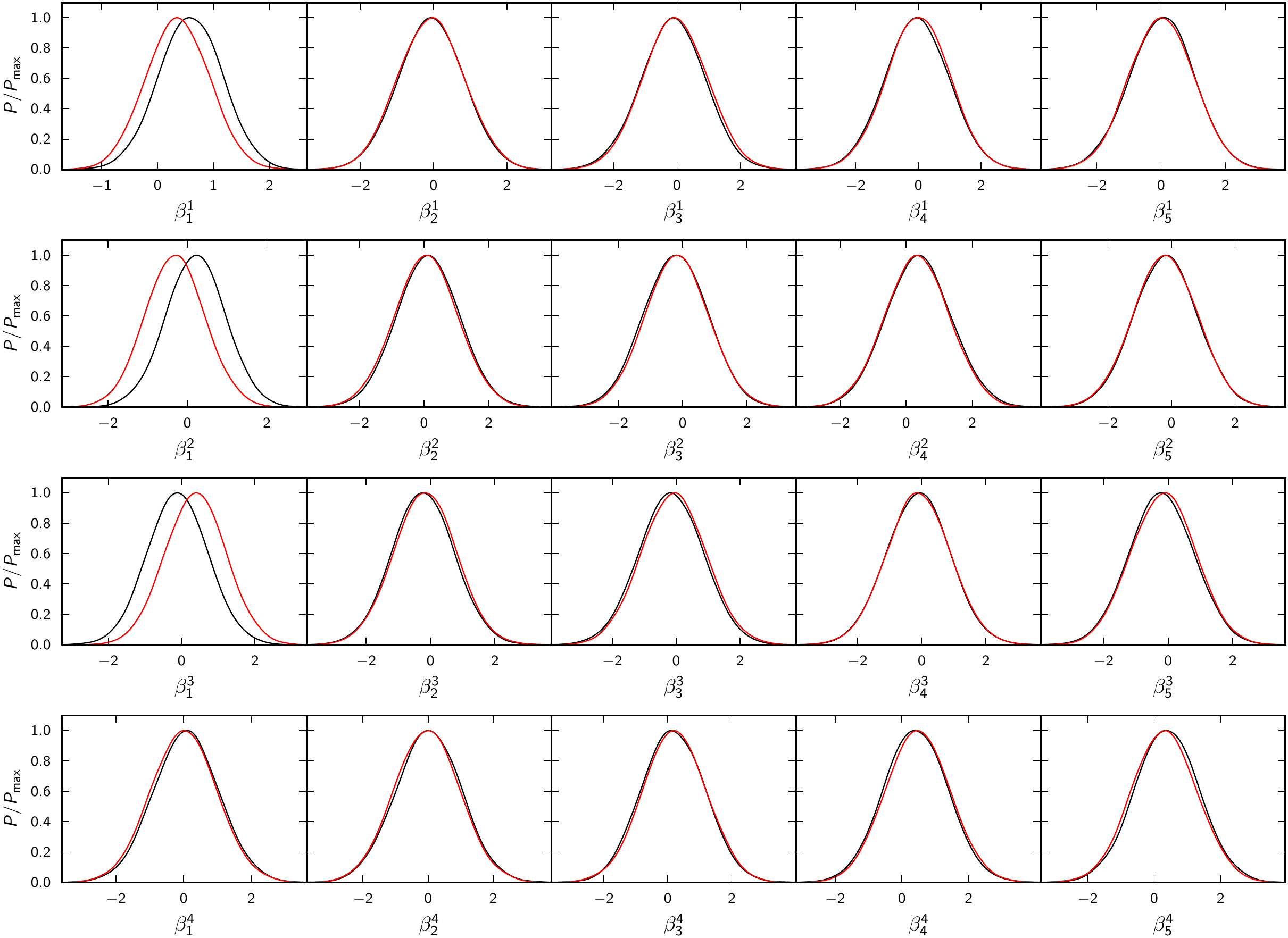}
\caption{Changes in the beam eigenmode coefficients from the inclusion of the near-sidelobe power now established but not included in the processing for the 2013 results.  Black curves are for \Planck+WP; red curves are for \Planck+WP with the revised beams.  The superscript indicates the effective spectum (one to four for $100$, $143$, $217$, and $143\times217$ respectively) while the subscript indicates eigenmode number.}
\label{fig:parameters2}
\end{figure*}

The results are indicated in Figs.~\ref{fig:parameters1} and \ref{fig:parameters2}, showing a selection of cosmological parameters and the beam eigenmode amplitudes, respectively.  As expected, we see a boost in the power spectrum amplitude, resulting in a change to the cosmological amplitude at about the $1\,\sigma$ level.  However, the largest shift in any other cosmological parameter is $0.3\,\sigma$.  The uncertainty in the beam window function is described by a small number of eigenmodes in multipole space and their covariance matrix \citep{planck2013-p03c}.  The posteriors for the first beam eigenmodes for the $100$, $143$, and $217$ effective spectra shift noticeably; others are practically unchanged.  
The beams used here are preliminary and the beam eigenmodes have not been generated self-consistently to match the beam calibration pipeline.
No adjustment was made in the calibration of the low-$\ell$ likelihood.  Nevertheless, from the results presented here, we can anticipate that the 2014 revisions to the beams will affect the overall calibration of the spectra, but will have little other impact on cosmology.

\section{Comparison of \Planck\ and \WMAP}
\label{sec:wmap}

\Planck\ and \WMAP\ have both produced sky maps with excellent large-scale stability, as demonstrated by many null tests both internal to the data and external.  In this section, we compare \Planck\ and \WMAP\ measurements in several different ways.  In Sect.~\ref{subsec:WMAPmapanalysis}, we compare power spectra calculated from 70 and 100\,GHz \Planck\ maps available in the PLA, and from V- and W-band yearly maps in the \WMAP\/9 data release.  In Sec.~\ref{subsec:WMAPlike}, we perform a likelihood analysis similar to that in Sect.~\ref{sec:PSlate} and in \citet{planck2013-p11}, and show that the differences between the map-based and likelihood analyses are well-understood.  In Sect.~\ref{subsec:WMAPassessment} we assess the results in the context of the uncertainties for the two experiments.

\subsection{Map and power spectrum analysis}
\label{subsec:WMAPmapanalysis}

\begin{figure}
\centerline{\includegraphics[width=88mm]{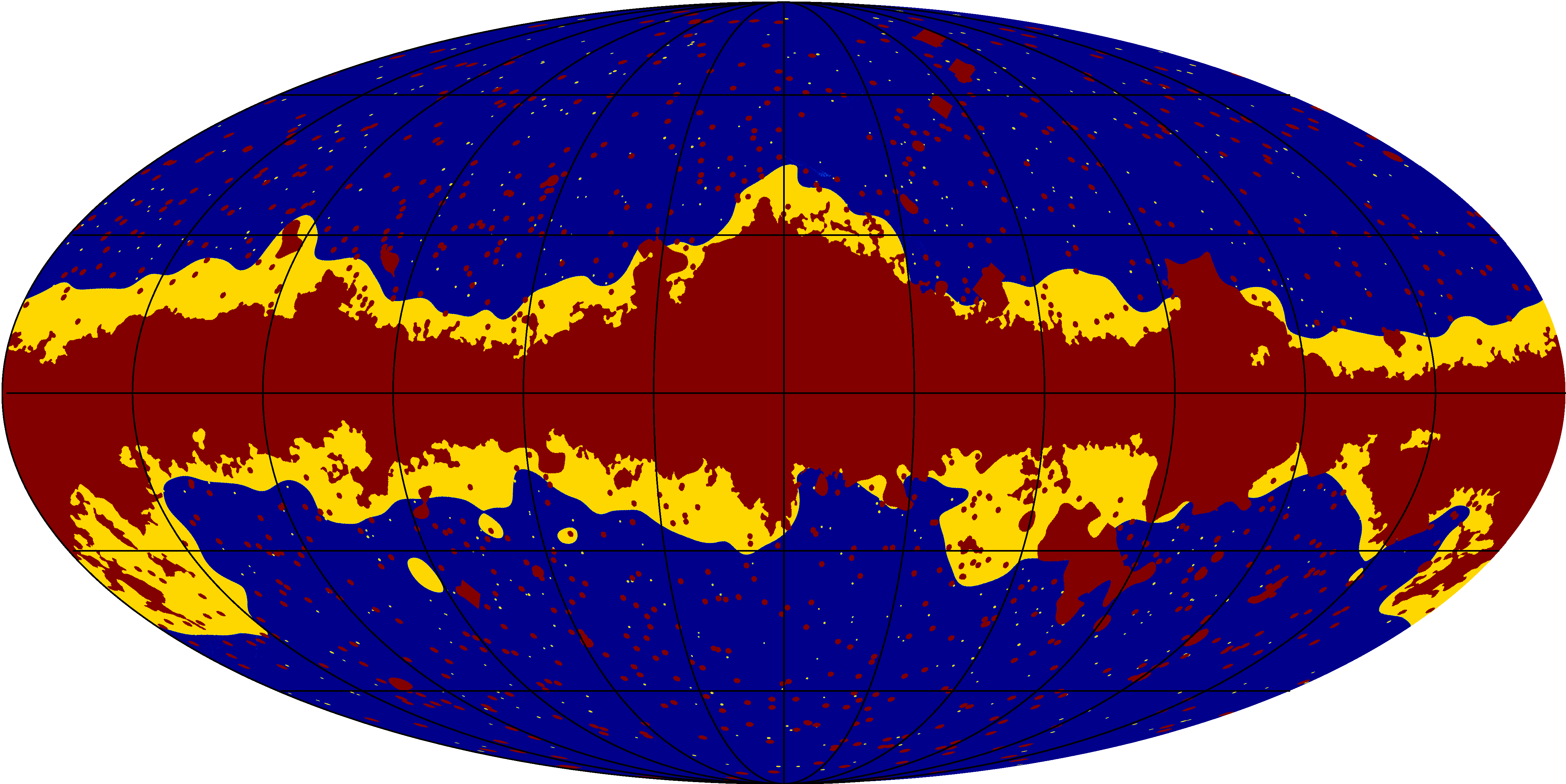}}
\caption{\planck\  $\fsky \approx  60\,\%$ Galactic mask in yellow and \WMAP\ KQ75 at $N_{\rm side}=1024$ in red.  The mask used for comparative spectral analysis of the \planck\ 70, 100, and 143\,GHz, and \WMAP\ nine-year V- and W-band sky maps is the union of the two.  The joint \planck\ 70, 100, and 143\,GHz point source mask is also used, exactly as before.  The final sky fraction is $\fsky = 56.7\,\%$.  The \planck\ mask is degraded to the pixel resolution $N_{\rm side}=1024$, at which the yearly \WMAP\ individual differential assembly maps are available.} 
\label{fig:Planck06Ukq75_mask}
\end{figure}

The {\it WMAP\/}9 data release includes $N_{\rm side} = 1024$ yearly sky maps from individual differential assemblies (DAs), both corrected for foregrounds and uncorrected, as well as $N_{\rm side} = 512$ frequency maps.  \WMAP\ uses somewhat different sky masks than \Planck.  In Sect.~\ref{sec:PSestimation} we emphasized the importance of using exactly the same masks in comparing results.  Accordingly, for \Planck/\WMAP\ comparisons we construct a joint mask, taking the union of the \Planck\ GAL060 mask used in Sects.~\ref{sec:PSestimation} and \ref{sec:calibration-and-transfer-function}, the \WMAP\ KQ85 mask, which imposes larger cuts for radio sources and some galaxy clusters, as required by the poorer angular resolution of \WMAP, and the \Planck\ joint 143, 100, 70\,GHz point source mask.  Fig.~\ref{fig:Planck06Ukq75_mask} shows the mask, which leaves $\fsky=56.7\,\%$ of the sky available for spectral analysis..

We use the same spectrum estimation procedure as in Sect.~\ref{sec:PSestimation}, evaluating the relevant cross-spectra, correcting for the mask with the appropriate kernel, and dividing out the relevant beam response and pixel smoothing functions.  As the mask is different from the one used for Planck-only comparisons, so is the mask-correction kernel.  All maps are analysed using the same mask.

\begin{figure*}
\includegraphics[width=9.0cm]{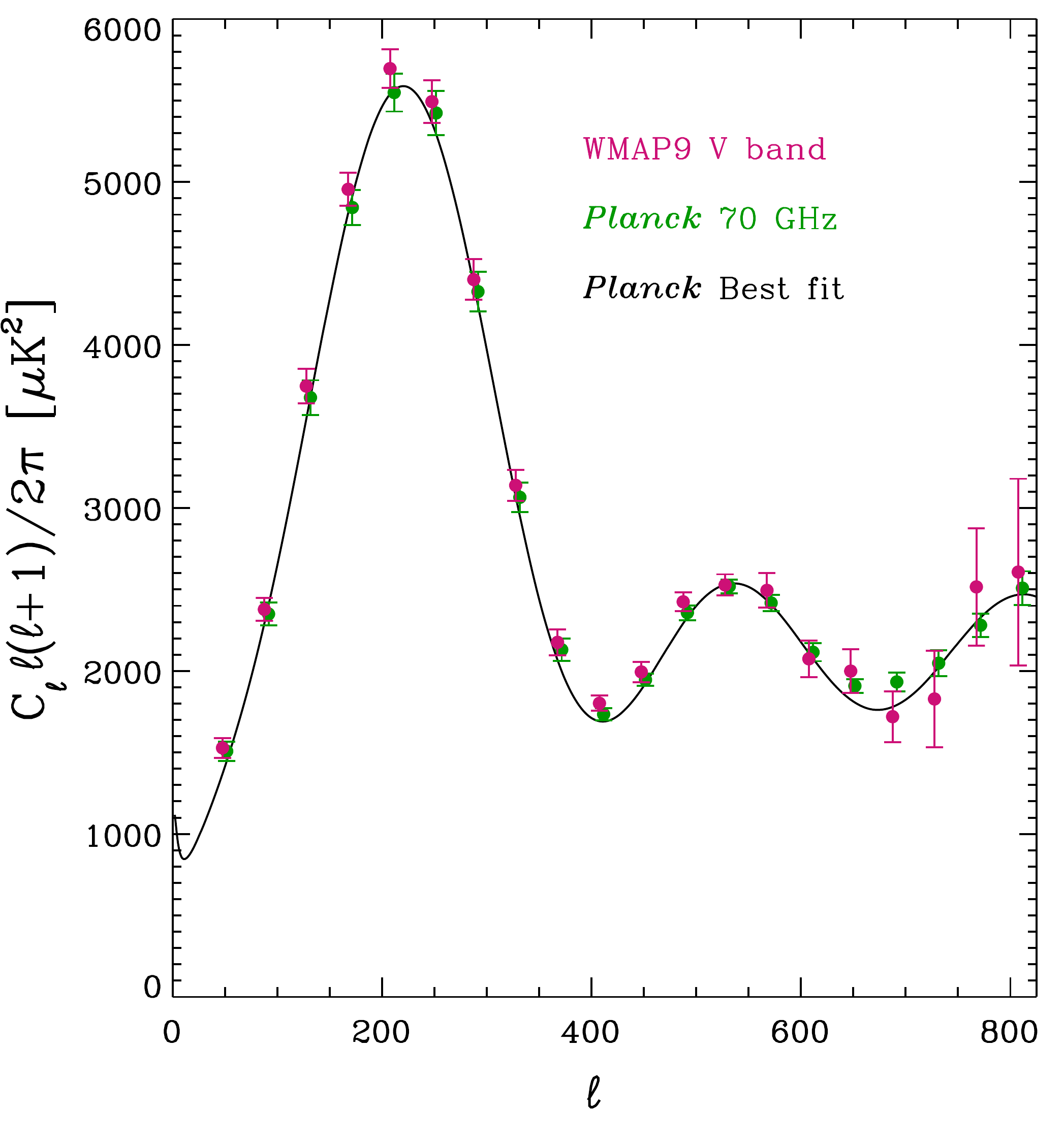}
\includegraphics[width=9.0cm]{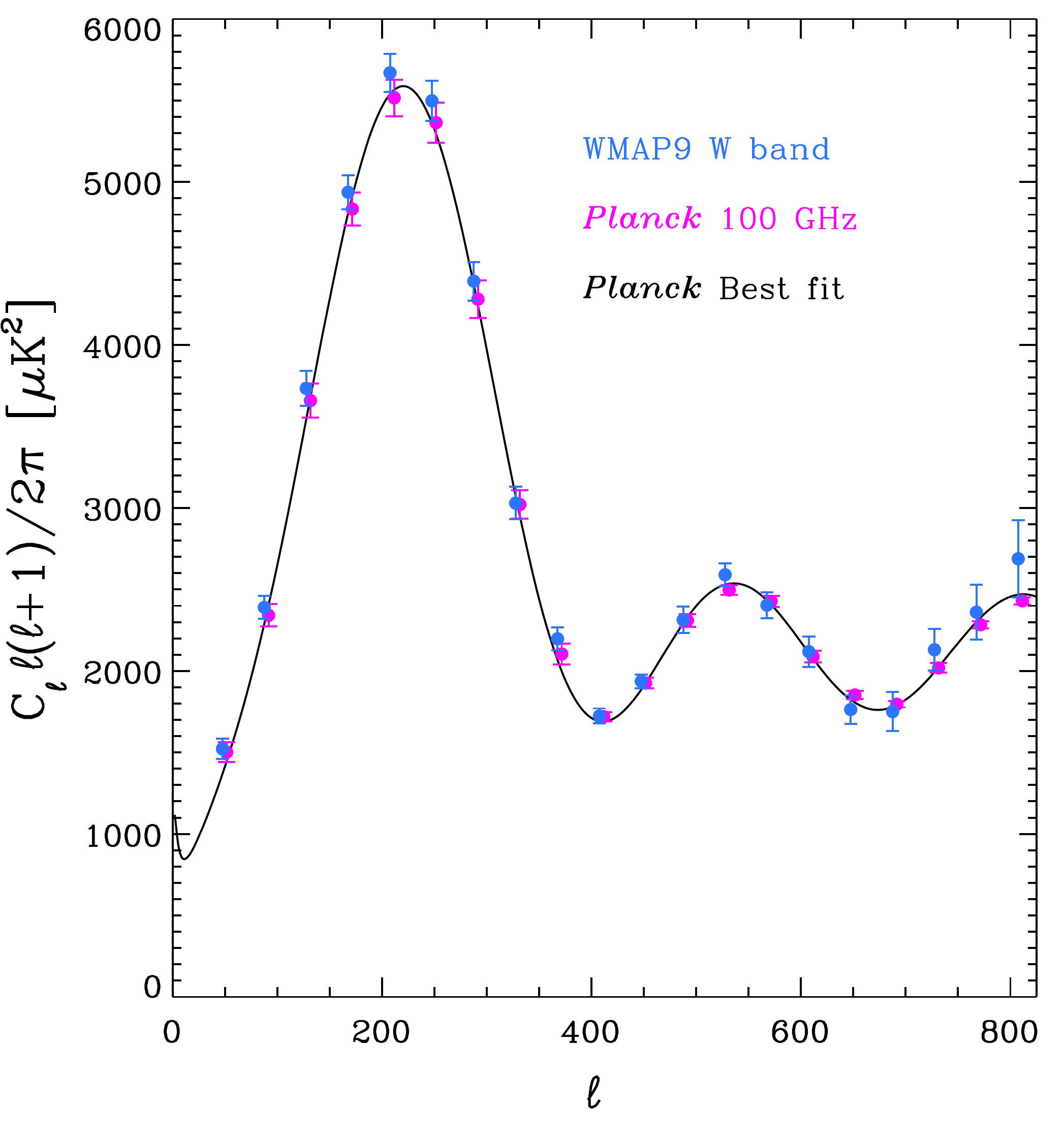}
\caption{{\it Left}:~\WMAP\ V band compared to \Planck\ 70\,GHz and best-fit model.  {\it Right}:~\WMAP\ W band compared to \Planck\ 100\,GHz and best-fit model.  The joint \Planck/KQ75 sky mask + \Planck\ point source mask ($\fsky = 56.7\,\%$; see Fig.~\ref{fig:Planck06Ukq75_mask}) is used.  Because the frequencies are so close, no corrections for foregrounds are made.}
\label{fig:planck_spectra_70_100_vs_V_W}
\end{figure*}

For \WMAP, there are nine yearly sky maps for each differential assembly V1, V2, W1, W2, W3, and W4, at $N_{\rm side} = 1024$.  Because the \WMAP\ V~band and the \Planck\ 70\,GHz band are so close in frequency, as are W~band and 100\,GHz, we use maps not corrected for foregrounds for the comparison.  All possible cross spectra from the yearly maps and differential assemblies are computed (630 at W band, 153 at V), and corrected for the mask, beam (using \WMAP\ beam response functions, different for each differential assembly), and pixel-smoothing.  The corrected spectra are averaged, and the error on the mean is computed for each $C_\ell$.  These average differential-assembly spectra are then co-added with inverse noise weighting to form one V~band and one W~band spectrum.  These are binned ($\ell_{\rm min} = 30$, $\Delta \ell = 40$), and rms errors in the bin values are computed.  The resulting spectra are shown in Fig.~\ref{fig:planck_spectra_70_100_vs_V_W}.

\begin{figure*}
\includegraphics[width=9.0cm]{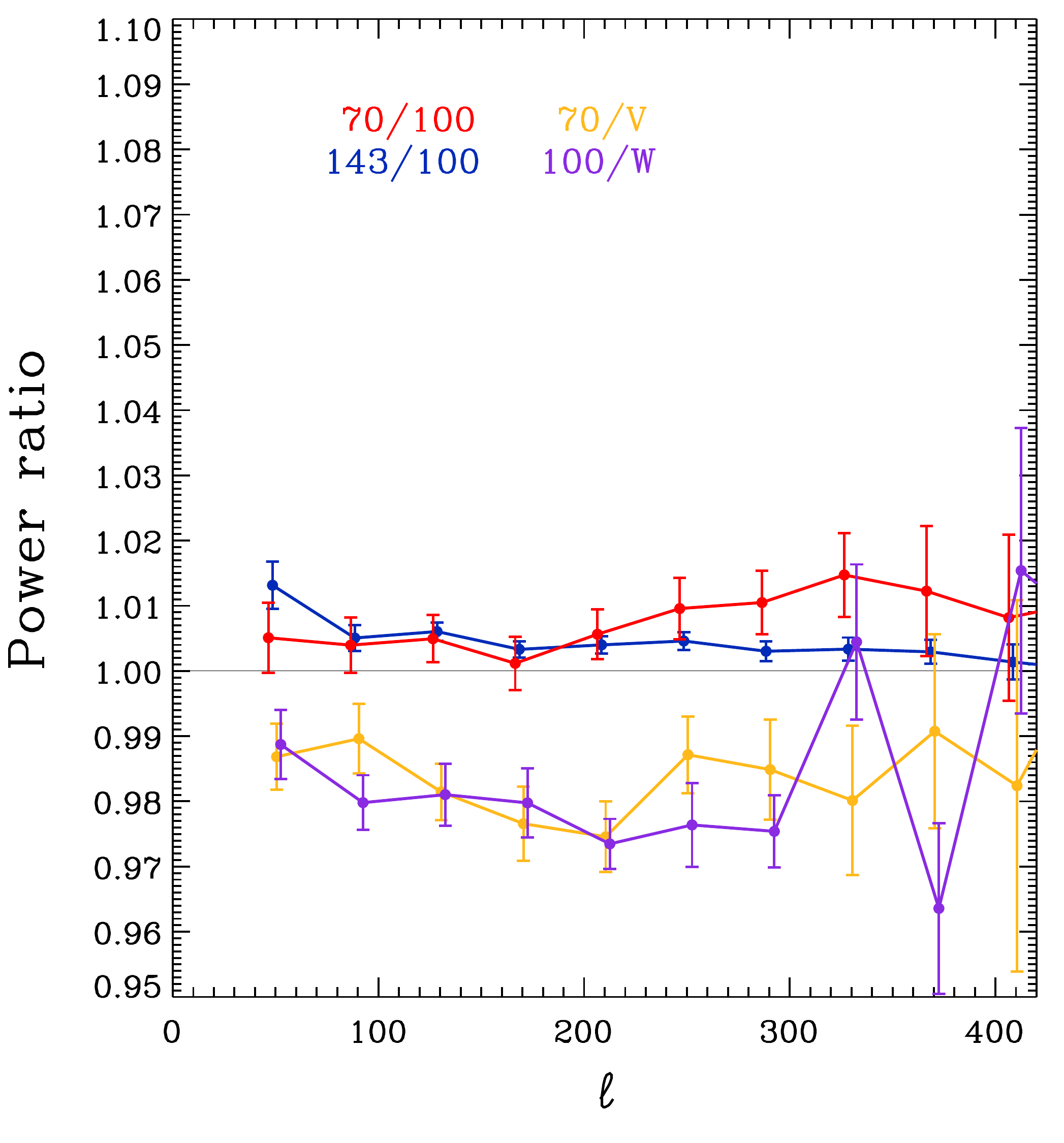}
\includegraphics[width=9.0cm]{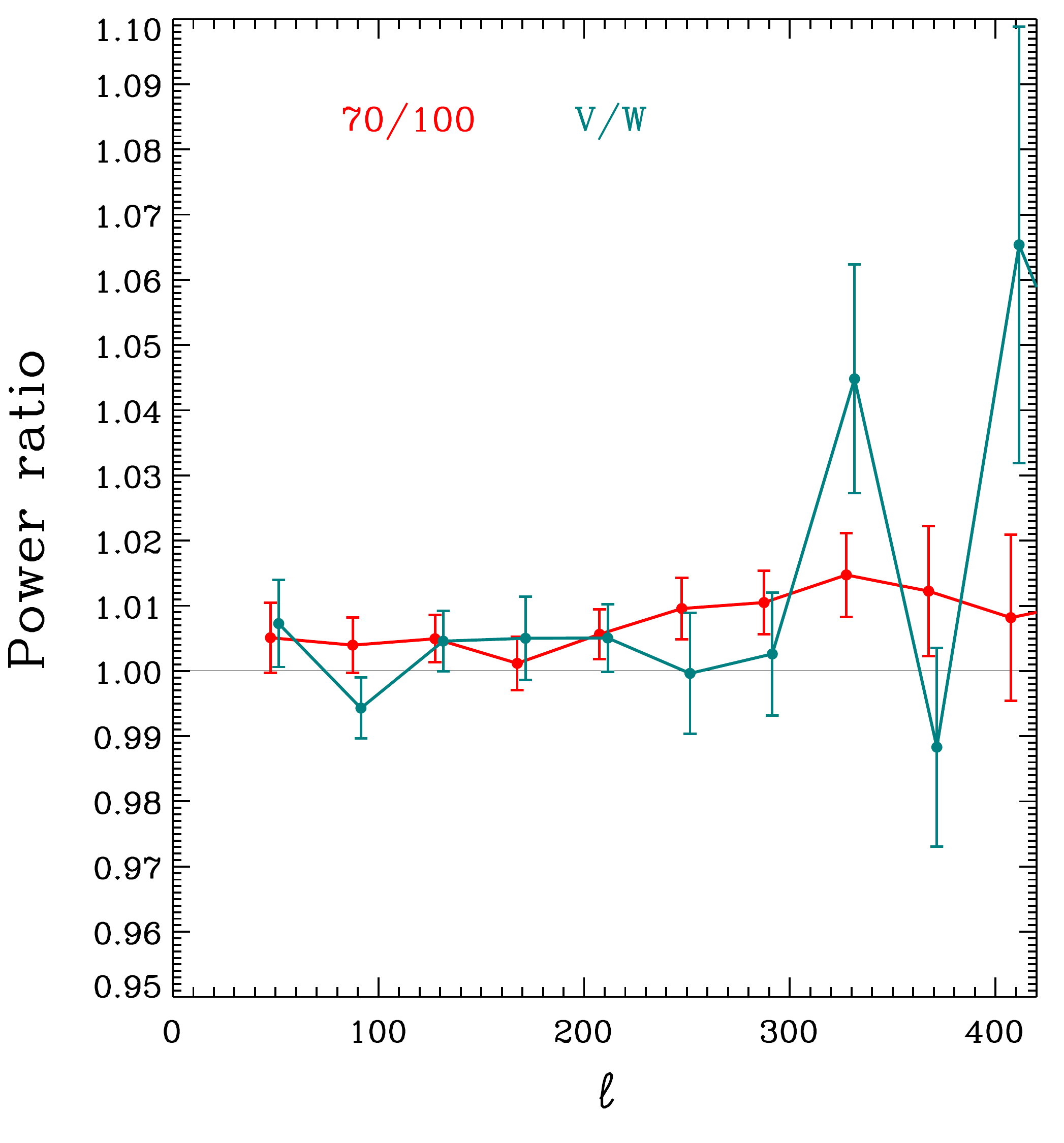}
\caption{Ratios of power spectra for \planck\ and \WMAP\ over the joint \Planck/KQ75+\Planck\ point source mask with $\fsky = 56.7\,\%$.  The \Planck\ 70/100 and 143/100 ratios can be compared to the bottom middle panel in Fig.~\ref{fig:planck_spectra_70_100_143} for $\fsky = 59.6\,\%$.  Here we limit the horizontal scale because the \WMAP\ noise beyond $\ell=400$ makes the ratios uninformative.  The general shape of the \Planck\ ratios is the same; however, it is clear (as it was in Fig.~\ref{fig:planck_spectra_70_100_143}) that changes in the sky fraction change the ratios significantly within the uncertainties.  This underscores the importance of strict equality of all factors in such comparisons.}
\label{fig:planck_vs_WMAP_power_ratios}
\end{figure*}

\begin{figure*}
\includegraphics[width=9.0cm]{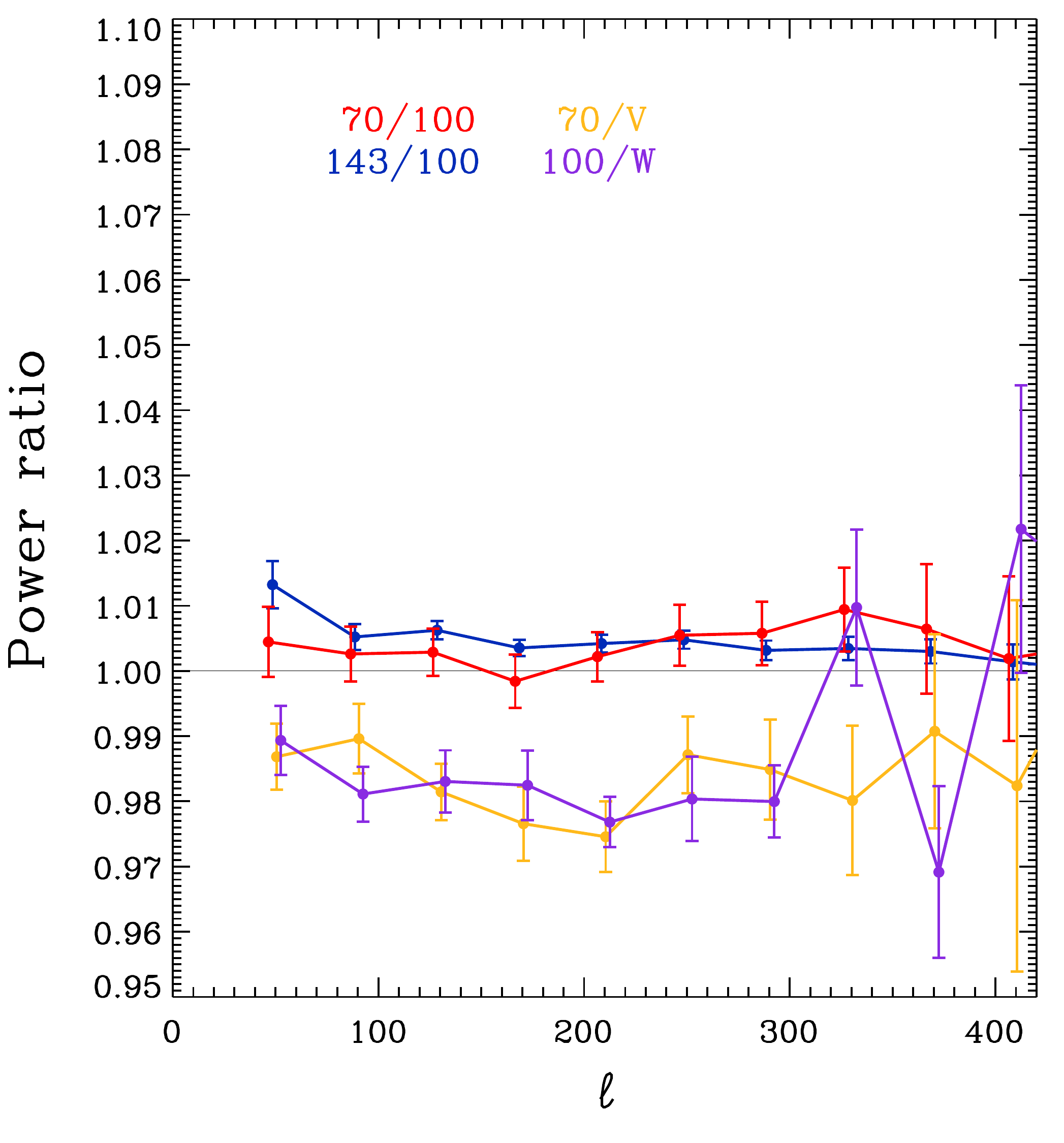}
\includegraphics[width=9.0cm]{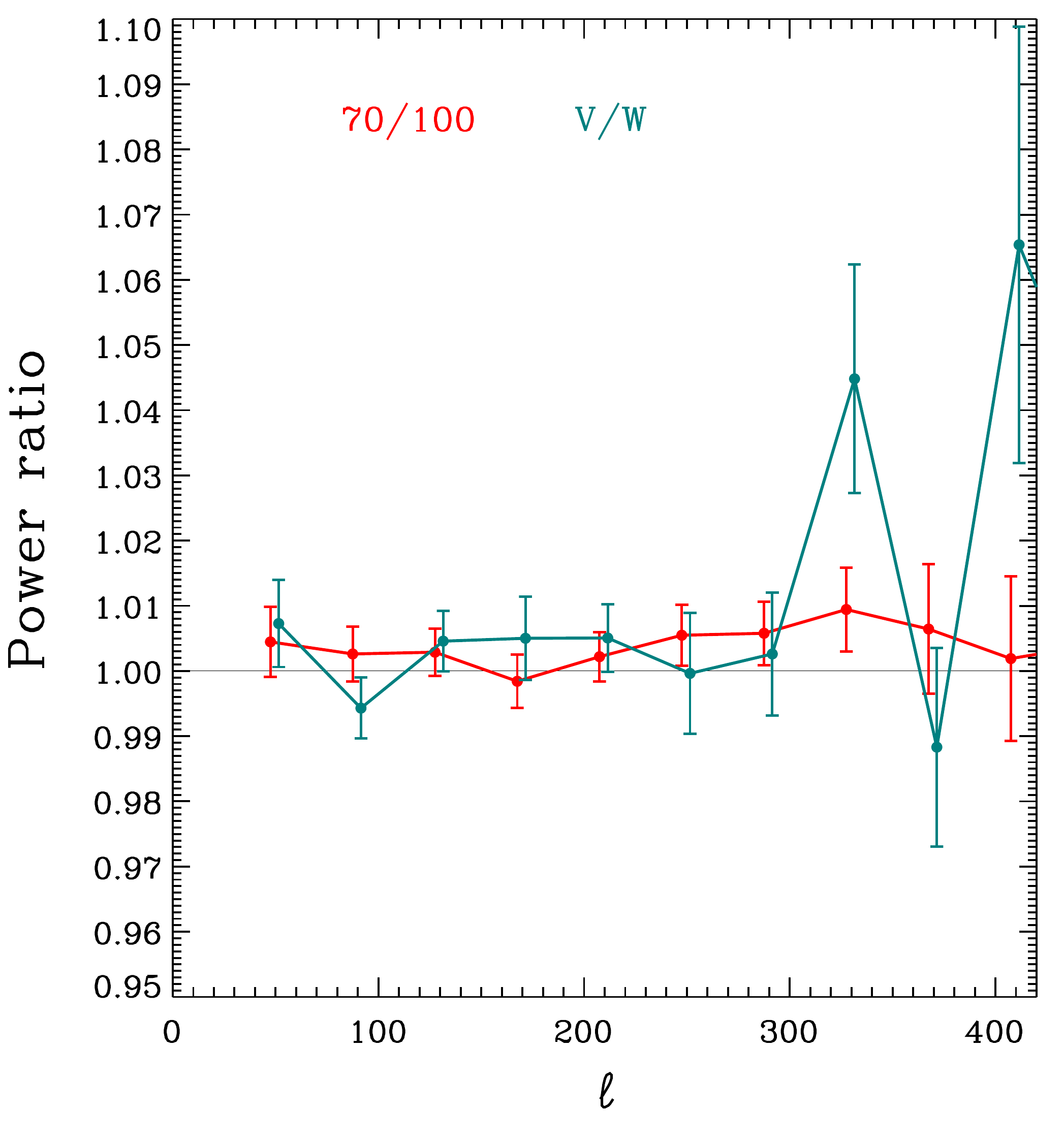}
\caption{Same as Fig.~\ref{fig:planck_vs_WMAP_power_ratios}, but the \Planck\  70/100 and 143/100 ratios are corrected for beam power at 100 and 143\,GHz that was not included in the effective beam window function used in the 2013 results.} 
\label{fig:planck_vs_WMAP_power_ratios_beam_corrected}
\end{figure*}

The 70, 100, and 143\,GHz \Planck\ spectra and spectral ratios in Figs.~\ref{fig:planck_spectra_70_100_vs_V_W}--\ref{fig:planck_vs_WMAP_power_ratios_beam_corrected} are determined as before, but using the new mask, starting from the 70\,GHz $N_{\rm side} = 1024$ half-ring PLA maps and the 100 and 143\,GHz $N_{\rm side} = 2048$ maps degraded to $N_{\rm side} = 1024$.  Thus all spectra are evaluated with the identical mask, at the same resolution.  Spectral binning and the estimation of rms bin errors proceed in exactly the same way as for the \WMAP\ spectra and for previous  \Planck-only comparisons.

\begin{figure*}
\includegraphics[width=9.0cm]{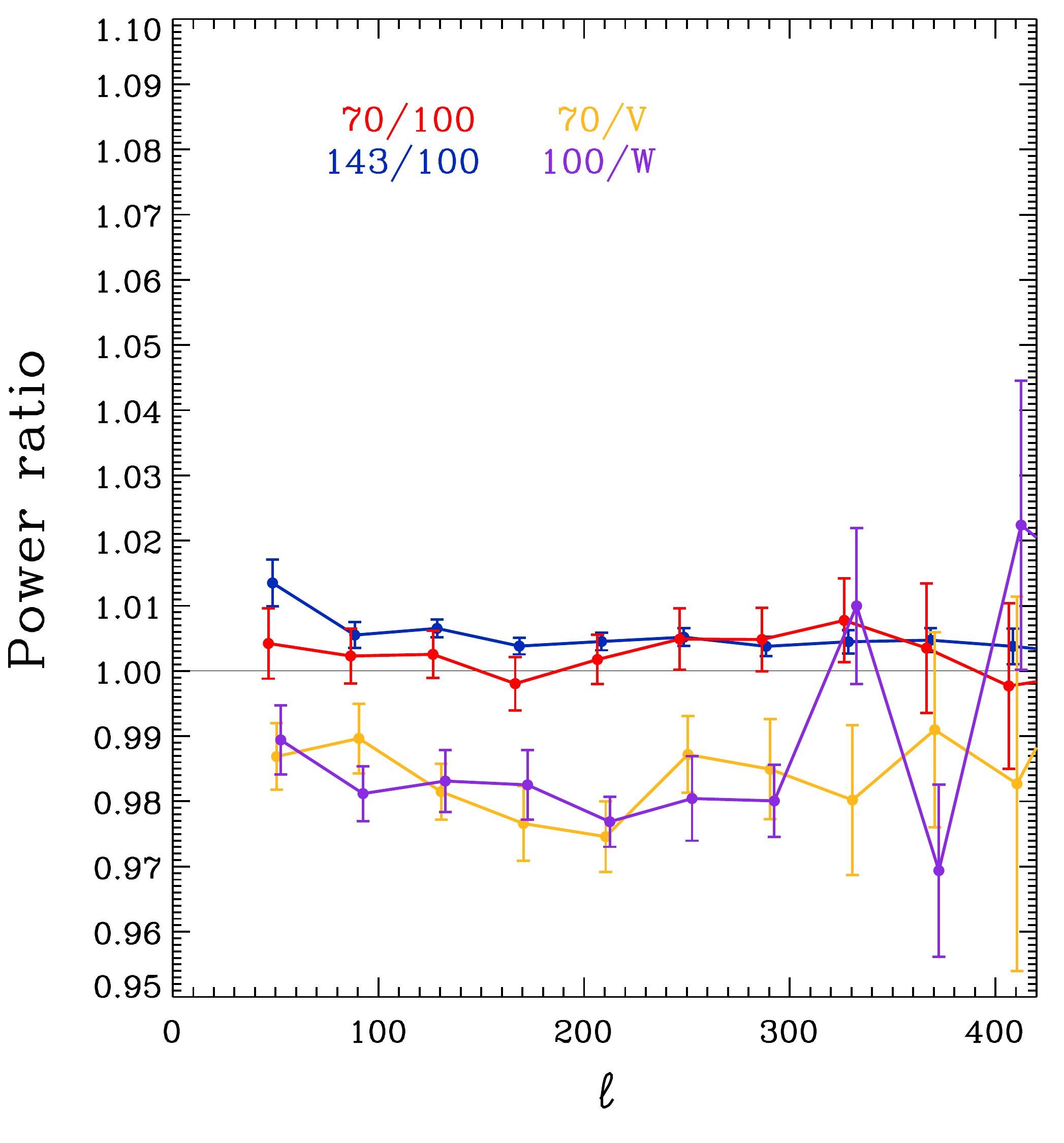}
\includegraphics[width=9.0cm]{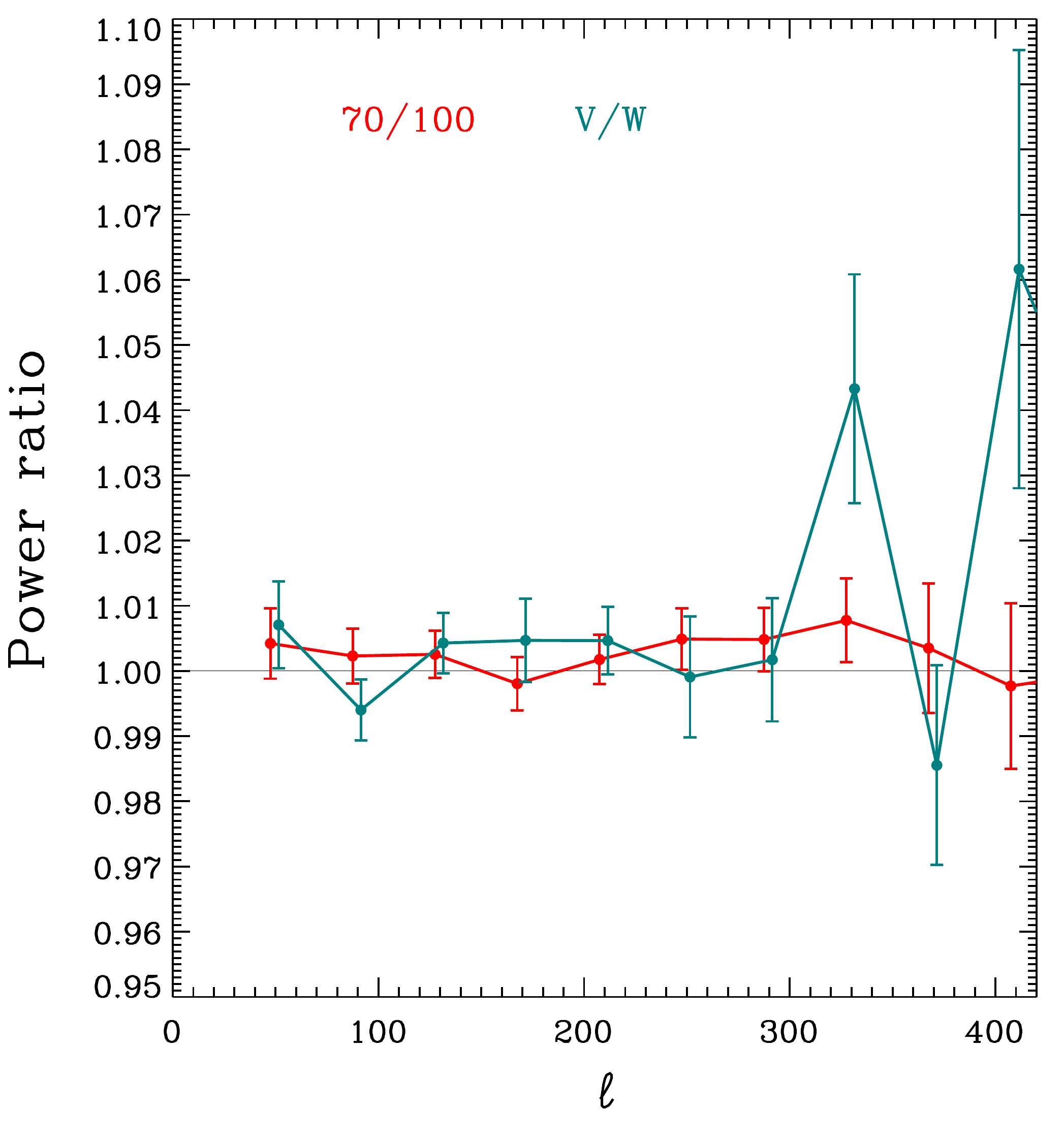}
\caption{Same as Fig.~\ref{fig:planck_vs_WMAP_power_ratios}, but including corrections for both \Planck\ beams and for \Planck\ and \WMAP\  unresolved-source residuals.} 
\label{fig:planck_vs_WMAP_power_ratios_beam+source_corrected}
\end{figure*}

Figure~\ref{fig:planck_spectra_70_100_vs_V_W} compares the \Planck\ 70\,GHz power spectrum with the \WMAP\ V-band spectrum, and the \Planck\ 100\,GHz power spectrum with the \WMAP\ W-band spectrum.  The \Planck\ 2013 best-fit model is shown for comparison.  The \planck\ and \WMAP\/9 spectra disagree noticeably in the $\ell$-range of the first two peaks.  Ratios of spectra in Fig.~\ref{fig:planck_vs_WMAP_power_ratios} show this disagreement directly.  In Figs.~\ref{fig:planck_vs_WMAP_power_ratios}--\ref{fig:planck_vs_WMAP_power_ratios_beam+source_corrected} the 70/100 and 143/100 ratios are the same as in Sects.~\ref{sec:PSestimation} and \ref{sec:calibration-and-transfer-function}, except for the small change in the mask.

Figure~\ref{fig:planck_vs_WMAP_power_ratios_beam_corrected} is the same as Fig.~\ref{fig:planck_vs_WMAP_power_ratios}, but with the \Planck\ 70/100 and 143/100 ratios corrected for the missing near sidelobe power in the 100 and 143\,GHz channels, discussed in Sect.~\ref{sec:calibration-and-transfer-function}.

Figure~\ref{fig:planck_vs_WMAP_power_ratios_beam+source_corrected} is the same as Fig.~\ref{fig:planck_vs_WMAP_power_ratios_beam_corrected}, but with all spectra additionally corrected for residual unresolved sources, as described in Sect.~\ref{sec:residualsources}.  Mean values of the ratios over specified multipole ranges are given in Table~\ref{tab:WMAPratios}.

\begin{table*}[t]
\caption{Summary of ratios of \Planck\ 70 and 100\,GHz and \WMAP\ V- and W-band power spectra appearing in this paper.}
\label{tab:WMAPratios}
\vskip -6.5mm
\tiny
\setbox\tablebox=\vbox{
\newdimen\digitwidth
\setbox0=\hbox{\rm 0}
\digitwidth=\wd0
\catcode`*=\active
\def*{\kern\digitwidth}
\newdimen\pointwidth
\setbox0=\hbox{\rm .}
\signwidth=\wd0
\catcode`!=\active
\def!{\kern\pointwidth}
\halign{\hbox to 1.2in{#\leaderfil}\tabskip=2em&
    \vtop{\hsize=2.4in\hangafter=1\hangindent=1em\noindent\strut#\strut\par}\tabskip=3em& 
    \hfil#\hfil\tabskip=3em& 
    \hfil#\hfil\tabskip=3em& 
    \hfil#\hfil\tabskip=1em& 
    \hfil#\hfil\tabskip=1em& 
    \hfil#\hfil\tabskip=0pt\cr
\noalign{\doubleline}
\omit&&&&\multispan3\hfil Spectrum Ratios\hfil\cr
\noalign{\vskip -3pt}
\omit&&&&\multispan3\hrulefill\cr
\noalign{\vskip 3pt}
\omit\hfil Location\hfil&\omit\hfil Features\hfil&$\fsky$&$\ell$ Range&70/V&100/W&100/(V+W)\cr
\noalign{\vskip 3pt\hrule\vskip 5pt}
Sect.~6.1, Fig.~20&No corrections&               $56.7\,\%$&$*70\le\ell\le390$&0.983&0.979&\dots\cr
\omit&                                                    &&$110\le\ell\le310$&0.981&0.977&\dots\cr
\noalign{\vskip 3pt}
Sect.~6.1, Fig.~21&Near-sidelobe (NS) correction, 100\,GHz&      $56.7\,\%$&$*70\le\ell\le390$&0.983&0.983&\dots\cr
\omit&                                                    &&$110\le\ell\le310$&0.981&0.981&\dots\cr
\noalign{\vskip 3pt}
Sect.~6.1, Fig.~22&Discrete-source residual+NS corrections& $56.7\,\%$&$*70\le\ell\le390$&0.983&0.983&\dots\cr
\omit&                                                    &&$110\le\ell\le310$&0.981&0.981&\dots\cr
\noalign{\vskip 3pt}
Sect.~6.2&Likelihood ``calibration factor''; \WMAP\ full-mission maps at $N_{\rm side}=512$ rather than yearly maps at $N_{\rm side} = 1024$; \Planck\ detector sets (see Sect.~\ref{sec:PSlate}) rather than full-frequency half-ring data.&
                                                 $56.7\,\%$&$*50\le\ell\le400$&0.978&0.976&0.974\cr
\noalign{\vskip 5pt\hrule\vskip 3pt}}}
\endPlancktablewide
\end{table*}

\subsection{Likelihood analysis}
\label{subsec:WMAPlike}

The likelihood analysis here is slightly different from the analysis presented in Sect.~\ref{sec:PSlate}.  First, we created a point source mask by concatenating the \WMAP/70/100/143/217 point source catalogues.  We present here only the results comparing \WMAP, LFI 70\,GHz, and HFI 100\,GHz. Restricting the frequencies to a range close to the diffuse foreground minimum has the advantage that we can increase the sky area used.  We therefore present results for mask G56 (unapodized), which leaves 56\,\% of the sky.  This mask, combined with the point source mask, is degraded in resolution from $N_{\rm side} = 2048$ to the natural $N_{\rm side}$ for \WMAP\ (512) and LFI (1024). 

Beam-corrected spectra for the \WMAP\ V, W, and V+W bands, and for the LFI 70\,GHz bands were computed.  Errors on these spectra were estimated from numerical simulations.  At 70\,GHz we have three maps of independent subsets of detectors, therefore we can estimate three pseudo-spectra by cross-correlating them by pairs.  Pseudo-spectra are then mask- and beam-deconvolved. The final 70\,GHz spectrum is obtained as a noise-weighted average of the cross-spectra.  Noise variances are estimated from anisotropic, coloured noise MC maps for each set of detectors.  We use the {\tt FEBeCoP} effective beam window functions for each subset of detectors (the three pairs 18-23, 19-22, and 20-21 as defined in \citealt{planck2013-p02}).  In the present plots we do not show beam uncertainties, which are bounded to be $\Delta B_\ell/B_\ell \la 0.2\,\%$ in the multipole range considered here \citep{planck2013-p03}.

We also derive power spectra for the \WMAP\ V and W bands, using coadded nine year maps per DA, for which no foreground cleaning has been attempted.  There are two DA maps for V-band and four for W-band. The V-band spectrum is obtained by cross-correlating the two available maps, whereas the W spectrum is the noise-weighted average of the six spectra derived by correlating pairs of the four DA maps.  We have produced Monte Carlo simulations of noise in order to assess the error bars of the \WMAP\ spectra.  We generate noise maps according to the pixel noise values provided by the \WMAP\ team rescaled by the number of observations per pixel.  Beam transfer functions per DA are those provided by the \WMAP\ team. In the present error budget we did not include beam uncertainties, which would be $\Delta B_\ell/B_\ell \approx 0.4\,\%$ for V and 0.5\,\% for W, over $\ell < 400$.

\begin{figure}
\centering
\includegraphics[width=79mm,angle=0]{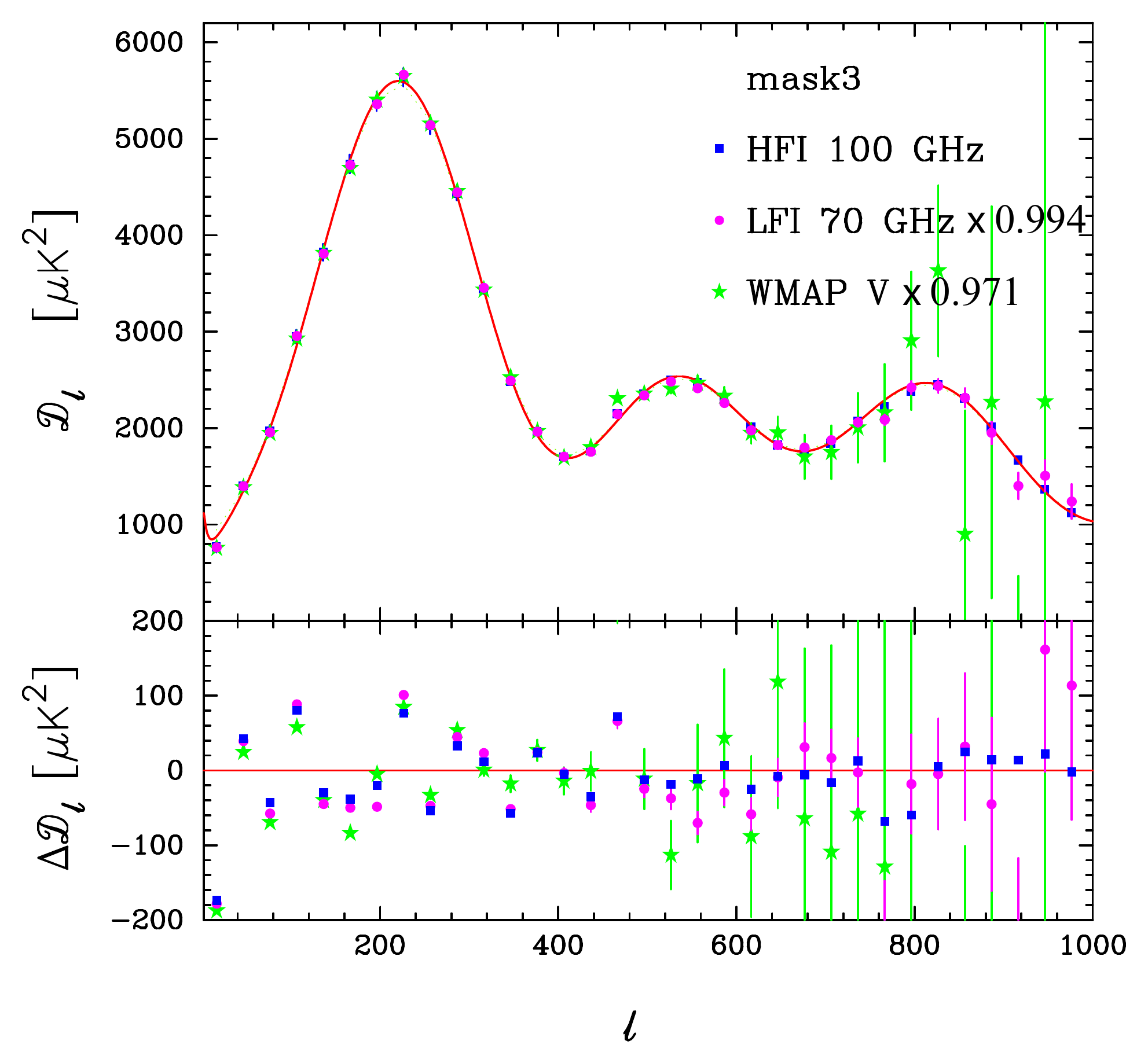}
\includegraphics[width=79mm,angle=0]{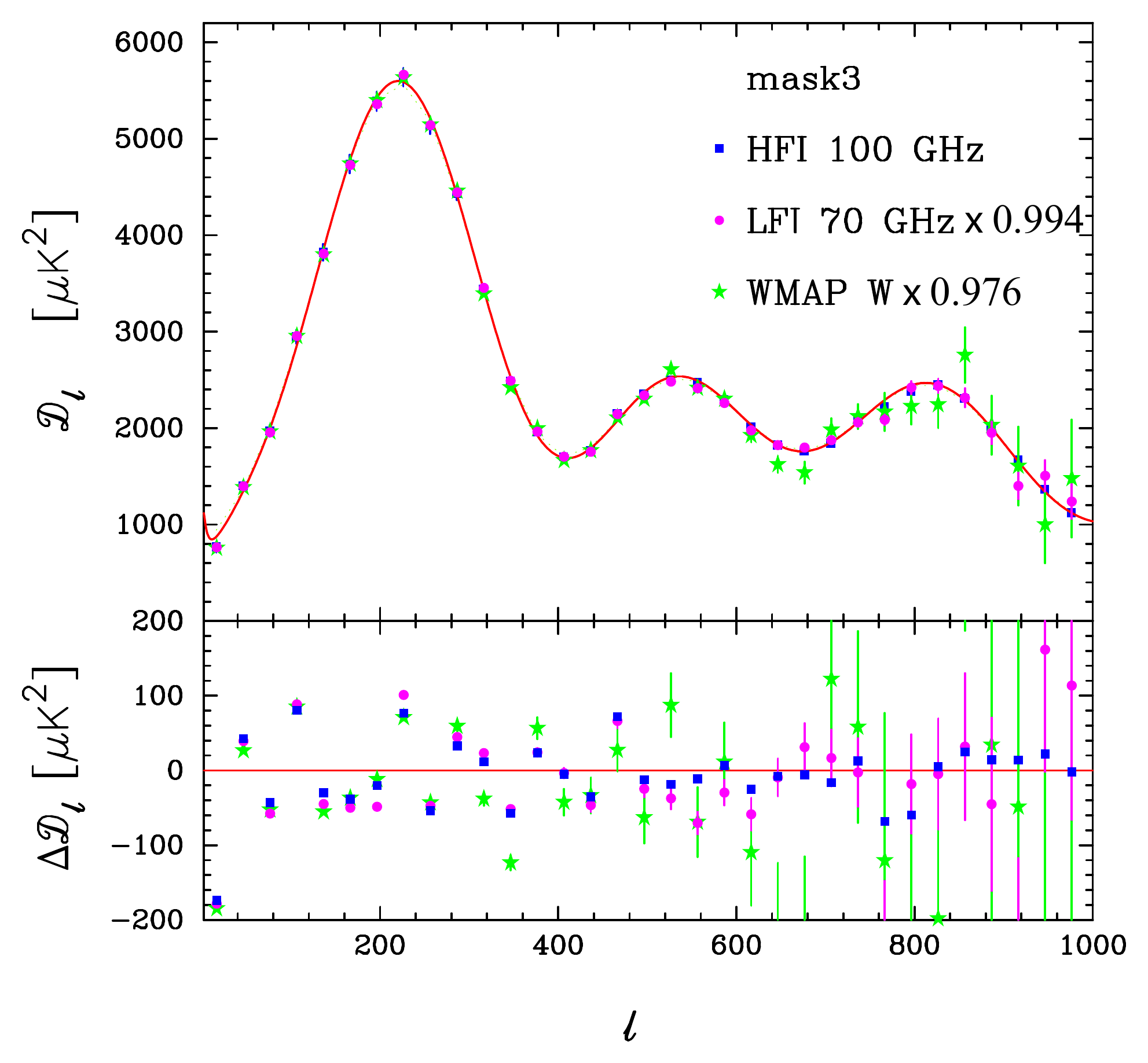}
\includegraphics[width=79mm,angle=0]{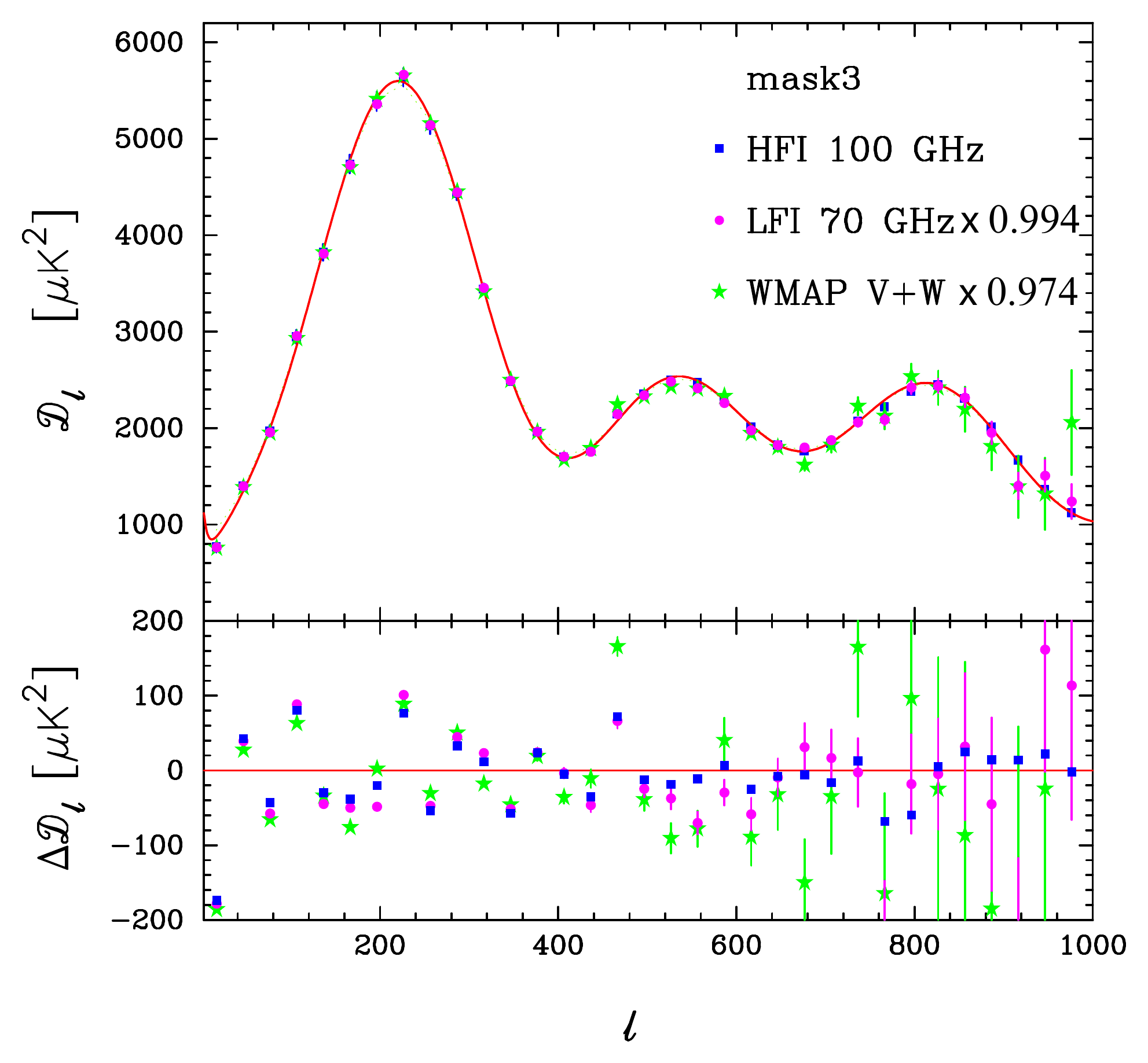}
\vglue -2.1mm
\caption{Foreground-corrected spectra computed using mask G56, with calibration factors $c_X$ from Eq.~\ref{D1} applied to 70\,GHz, V, W, and V+W. The solid red lines in the upper panels show the best-fit model CMB spectrum \citep{planck2013-p11}.  The lower panels show the residuals with respect to this model. The error bars on the LFI and \WMAP\ points show noise errors only.}
\label{HFILFIWMAPspectra}
\end{figure}

For LFI and \WMAP, we subtract an unresolved thermal SZ template with the amplitude derived from the {\tt CamSpec} likelihood analysis, which we extrapolate to the central LFI and \WMAP\ frequencies.  We also subtract a Poisson point source term with an amplitude chosen to minimize the variance between the measured spectra and the best-fit theoretical model at high multipoles.  We then determine calibration factors $c_X$ for the \WMAP\ and 70\,GHz LFI spectrum relative to the 100\,GHz HFI spectrum by minimizing 
\begin{equation}
\chi^2 = \sum_b{( {\cal D}_b^{100} - c_X{\cal D}_b^X)^2 \over 
(\sigma_{Xb}^2 + \sigma_{\rm I}^2) },  \label{D1}
\end{equation}
until convergence,  at each iteration determining $\sigma_{\rm I}$, an ``excess scatter'' term,  by requiring that the reduced $\chi^2$ be unity.  (The excess scatter comes primarily from foreground-CMB correlations, as discussed in \citealt{planck2013-p08}.)  The sum extends over bins with central multipole in the range $50 \le \ell \le 400$.  The index $X$ denotes the spectrum (70, V, W, or V+W), and $\sigma_{Xb}$ is the noise contribution to the error in band $b$.  The spectra ${\cal D}$ in Eq.~(\ref{D1}) are foreground-corrected.  Calibration factors $c_X$ and resulting spectra for 70\,GHz, V, W, and V+W-band relative to 100\,GHz are given in Fig.~\ref{HFILFIWMAPspectra} and Table~\ref{tab:WMAPratios}.   The value of $c_{70} = 0.994$ seen in Fig.~\ref{HFILFIWMAPspectra} is entirely consistent with the 70/100 ratios given in Sect.~\ref{sec:PSlate}, taking into account mask and dataset variations as discussed in Sect.~\ref{sec:PSlate}.  A calibration factor for V relative to 70\,GHz calculated the same way (${\cal D}_b^{100}$ is replaced by ${\cal D}_b^{70}$ in Eq.~\ref{D1}) is also included in Table~\ref{tab:WMAPratios}.   

To compare the likelihood results with the map-based ones, we need to identify the closest cases for which results are given. No correction for near sidelobes at 100\,GHz (Sect.~\ref{subsec:keyfindings}) was included in the likelihood results, but corrections for residual unresolved sources were.  On the other hand, from the results shown in Table~\ref{tab:WMAPratios}, residual-source corrections make negligible difference.  In the likelihood approach, $\ell$ values beyond about 300 will be significantly down-weighted due to increased variance.  Thus the most reasonable comparison is between the ratios from the map-based approach in the $\ell$ range $110\le\ell\le310$ with no corrections, and the corresponding calibration factors from the likelihood approach.  Values in Table~\ref{tab:WMAPratios} show good agreement between the two analyses.  We have furthermore noticed that using the \WMAP\ full-mission maps at resolution $N_{\rm side}=512$ rather than the yearly maps at $N_{\rm side}=1024$, and \Planck\ detector sets rather than full-frequency half-ring maps, can lower the spectral ratio by 0.2\,\% in the multipole range $110\le\ell\le310$.

\subsection{Assessment}
\label{subsec:WMAPassessment}

Both the direct comparison of power spectra from the 2013 results maps and the likelihood analysis show a discrepancy between \Planck\ and \WMAP\ across the region of the first peak, where the S/N for \WMAP\ is good.  This difference is about 2\,\% in power, corresponding to 1\,\% in the maps.  These numbers quantify what can be seen by eye in Fig.~\ref{fig:planck_spectra_70_100_vs_V_W}.  The result is roughly the same for comparisons of 70\,GHz with V-band and 100\,GHz with W-band, where the frequency differences are small enough to rule out foregrounds as the cause.  There is some variation of the ratio with $\ell$ (perhaps seen more easily in Fig.~\ref{HFILFIWMAPspectra}), suggesting that the cause is not simply a result of calibration errors, but neither does the shape correspond obviously to what would be expected from missing power in beams.  

The 70/V and 100/W ratios differ from unity by more than expected from the uncertainties in absolute calibration determined for \Planck\ and \WMAP.  Calibration of \WMAP\/9 is based on the orbital dipole (i.e., the modulation of the solar dipole due to the Earth's orbital motion around the Sun), with overall calibration uncertainty 0.2\,\% \citep{bennett2012}.  The absolute calibration of the \Planck\ 2013 results is based on the solar dipole (Sect.~\ref{sec:introduction}), assuming the \WMAP\/5 value of $(369.0\pm0.9)$\,km\,s\mo, where the 0.24\,\% uncertainty includes the 0.2\,\% absolute calibration uncertainty \citep{hinshaw2009}.   The overall calibration uncertainty is 0.62\,\% in the 70\,GHz maps and 0.54\,\% in the 100 and 143\,GHz maps (\citealt{planck2013-p01}, Table~6).  When comparing \Planck\ and \WMAP\ calibrated maps, one must remove from these uncertainties the 0.2\,\% contribution of the \WMAP\ absolute calibration uncertainty to the \WMAP\ dipole, which thus affects both experiments equally.  The remaining 0.14\,\% uncertainty in the \WMAP\ dipole (mostly due to foregrounds; \citealt{hinshaw2009}) affects \Planck\ but not \WMAP, because its absolute calibration is from the orbital dipole.  In the planned 2014 release, the \Planck\ absolute calibration will be based on the orbital dipole, bypassing uncertainties in the solar dipole.  At the power spectrum level for comparison with \WMAP, then, the \Planck\ uncertainty would be between $2\times (0.54-0.2)\,\% = 0.68\,\%$ and $2\times(0.62-0.2)\,\% = 0.84$\,\%, and the \WMAP\ uncertainty 0.4\,\%.  The power spectrum ratios in Table~\ref{tab:WMAPratios} from Fig.~\ref{fig:planck_vs_WMAP_power_ratios_beam_corrected} and Sect.~\ref{subsec:WMAPlike} then represent a 1.5--2$\sigma$ difference.

As the primary calibration reference used by \Planck\ in the 2013 results is the \WMAP\ solar dipole, the inconsistency between \Planck\ and \WMAP\ is unlikely to be the result of simple calibration factors.  Reinforcing this conclusion is the fact that the intercalibration comparison given in Fig.~35 of \citet{planck2013-p03} for CMB anisotropies shows agreement between channels to better than 0.5\,\%  over the range 70--217\,GHz, and 1\,\% over all channels from 44 to 353\,GHz, using 143\,GHz as a reference.  Problems with transfer functions are more likely to be the cause. The larger deviations at higher multipoles in the \Planck\ intercalibration comparison just referred to also point towards transfer function problems.  

Comparisons between \WMAP- and LFI-derived brightness temperatures of Jupiter presented in figure~21 of \citet{planck2013-p02b} provide a potentially useful clue:  the two instruments seem to agree (for a simple linear spectral model for Jupiter) within 1\,\% , and much better than this at 70\,GHz.  The LFI points are derived from just two Jupiter transits.  We now have eight transits in hand.  If the full analysis of all the Jupiter crossings shows similar consistency with \WMAP\ brightness temperatures, subtle effects from beam asymmetries and near sidelobes in the two instruments, rather than gross errors in main beam solid angles, may be the cause.

At present, we do not have an explanation for the $\approx 2\%$ calibration difference between \WMAP\ and \Planck. The differences between \WMAP\ and \Planck\ are primarily multiplicative in the power spectra, and so have little impact on cosmological parameters other than on the amplitude of the primordial spectrum $A_{\rm s}$.   Figure~\ref{HFILFIWMAPspectra} shows that after the spectra are rescaled by multiplicative factors, there is excellent {\it point-to-point\/} agreement between the LFI, HFI, and \WMAP\ power spectra.  Appendix~A of \citet{planck2013-p11} also shows explicitly that by restricting the Planck 2013 likelihood to $\ell_{\rm max}=1000$, we recover almost identical cosmological parameters to those from \WMAP\, apart from a small shift in $A_s$ reflecting the calibration differences between the two experiments.   Any shape differences between the power spectra of the two experiments therefore have little impact on cosmological parameters.

\section{Conclusions}
\label{sec:summary}

Data consistency in CMB experiments is an important indicator of the validity of the results derived from these data.  We have compared calibrated data from different \Planck\ channels in three ways: (1)~comparison of power spectra calculated on masked, publicly-released maps at 70, 100, and 143\,GHz, including corrections for residual unresolved sources; (2)~comparison of calibration factors determined to minimize differences between the CMB in the maps; and (3)~comparison of power spectra and calibration factors obtained in the likelihood analysis used to determine cosmological parameters.  We have emphasized particularly the comparison of 70\,GHz and 100\,GHz, as these frequencies come from different instruments and are analysed independently, and are near the diffuse foreground minimum.  The internal consistency of the \Planck\ data is remarkable (relative calibration inconsistencies between CMB channels of only a few tenths of a percent), and is at a level consistent with the uncertainties based on the 2013 solar dipole calibration scheme.  In the future, calibration based on the orbital dipole should lead to  more accurate absolute calibration.

The estimate of the contribution of the near sidelobes to the HFI beam solid angles in March 2013 implicitly assumed azimuthal symmetry of the near sidelobes. Subsequent analyses of the beam and calibration procedures suggest that this assumption underestimated near-sidelobe power by as much as 0.1\,\%.  Corrections for this missing power result in improved agreement between 70 and 100\,GHz, measured by ratios of power spectra, at a level of typically 0.3\,\%, well below the levels of uncertainty estimated for LFI and HFI calibration, and therefore having little effect on the cosmological parameters determined from the 2013 data.  

Similar analysis applied to \WMAP\ data shows a roughly 2\,\% difference between \Planck\ and \WMAP\ over the region of the first acoustic peak.  At present, the explanation for this discrepancy with \WMAP\ is not known.  

Precision calibration is an ongoing challenge.
Future analyses will incorporate the lessons of this study of consistency, further reducing the size of uncertainties and errors on the scientific results from \Planck.

\begin{acknowledgements}
The development of \Planck\ has been supported by: ESA; CNES and CNRS/INSU-IN2P3-INP (France); ASI, CNR, and INAF (Italy); NASA and DoE (USA); STFC and UKSA (UK); CSIC, MICINN and JA (Spain); Tekes, AoF and CSC (Finland); DLR and MPG (Germany); CSA (Canada); DTU Space (Denmark); SER/SSO (Switzerland); RCN (Norway); SFI (Ireland); FCT/MCTES (Portugal); and DEISA (EU). A description of the Planck Collaboration and a list of its members, including the technical or scientific activities in which they have been involved, can be found at \url{http://www.rssd.esa.int/Planck}.  Some of the results in this paper have been derived using the {\tt HEALPix} package.  We acknowledge the use of the Legacy Archive for Microwave Background Data Analysis (LAMBDA), part of the High Energy Astrophysics Science Archive Center (HEASARC); HEASARC/LAMBDA is a service of the Astrophysics Science Division at the NASA Goddard Space Flight Center.
\end{acknowledgements}

\bibliographystyle{aa}
\bibliography{Planck_bib,planck_map_consistency}

\appendix

\section{Difference maps}
\label{sec:mapappendix}

\begin{landscape}
\begin{figure*}
\hglue -2.4in\includegraphics[width=225mm]{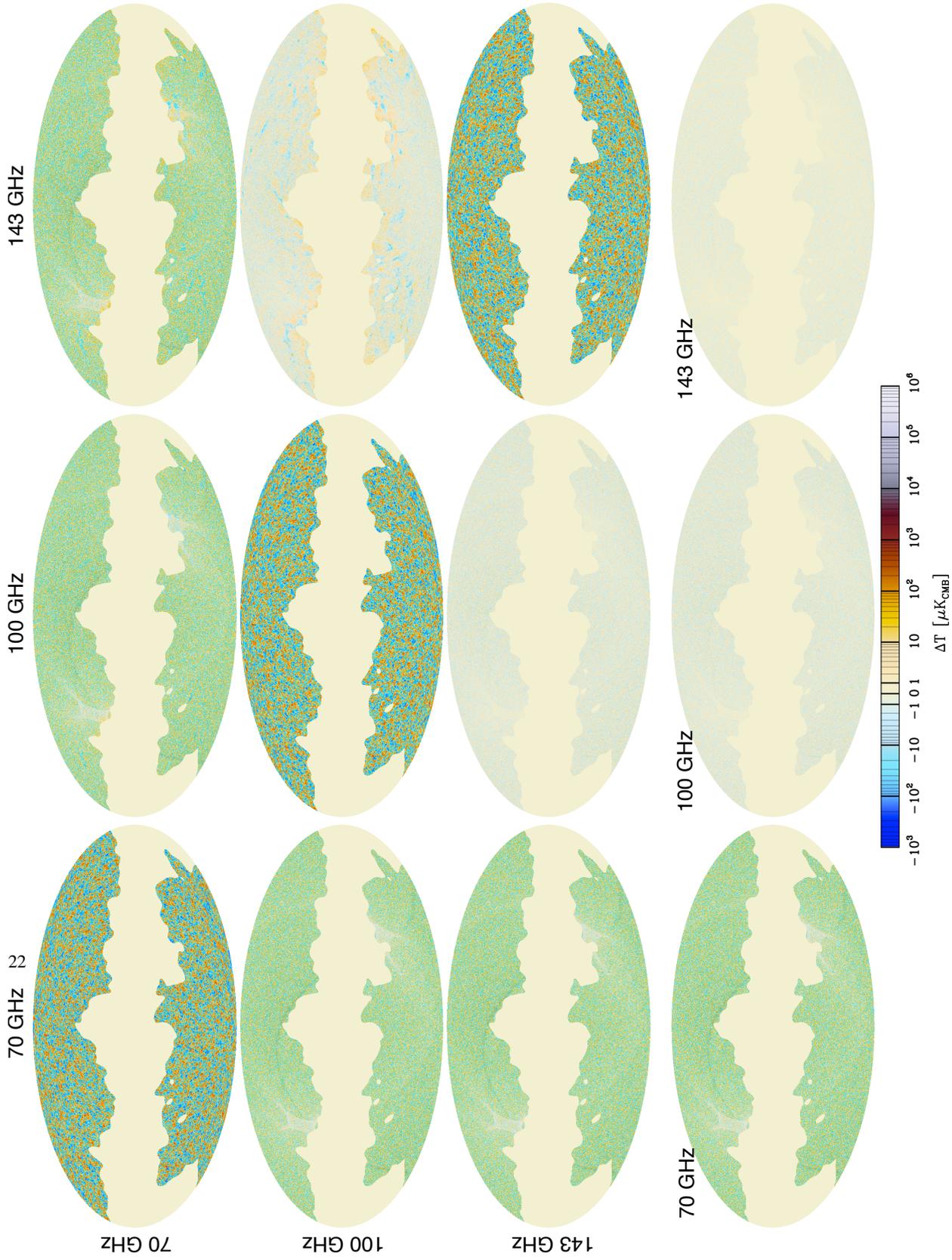}
\vskip -0.6in
\hglue -2.4in\caption{Frequency maps and difference maps using the GAL060 mask.  Rms values are given in Table~1; histograms are shown in Fig.~\ref{fig:stats_histograms}.  In the top $3\times3$ group, the frequency maps are on the diagonal.  The three frequency difference maps are in the upper right block.  The noise difference maps are in the lower left block.  The bottom row gives the single-frequency noise maps.  The caption of Fig.~\ref{fig:stats_histograms} describes characteristics of the histograms that can also be seen quite clearly in these maps.  The three upper right maps are the same as the three in Fig.~\ref{fig:Maps2}, but with structure on scales larger than 8\deg\ removed, as described in the text.}
\label{fig:stats_maps}
\end{figure*}
\end{landscape}

Figure~\ref{fig:stats_maps} shows the frequency maps and difference maps used to calculate the rms values given in Table~\ref{tab:rms}.  Frequency maps are half-ring sums.  Difference maps are half-ring differences.  The maps have been processed as described in Sect.~\ref{sec:planckskymaps} to retain structure between angular scales of 15\arcm\ and 8\deg.  Figure~\ref{fig:stats_histograms} gives the corresponding histograms.

\section{Nominal beam calibration: definitions, approximations, and rescaling factors}
\label{sec:calibration-convolution-details} 

In the \Planck\ scanning strategy, for each pointing period, the spin axis remains fixed and the output voltage sees a spin-synchronous modulation as the telescope points at direction $\xhat$ along a scan circle. The output signal (in arbitrary units) measured by a given \Planck\ detector, either LFI or HFI, can be written as

\begin{equation}
\Vout(\xhat) = G\big[P\ast(T_{\rm sky} + D)\big](\xhat) + M(\phat), 
\label{eq:Vout}
\end{equation}
where $P$ is the angular response (``beam''), normalized to unit solid angle ($\int_{4\pi} P\,d\Omega = 1$), $D$ is the dipole signal (including the solar and orbital terms), \Tsky\ is the sky brightness temperature (CMB and foregrounds, i.e., everything {\em other} than the dipole), and $M$ is a small monopole term assumed to be constant for each pointing period \phat.  We use the convolution notation $A\ast P = \int_{4\pi} AP\,d\Omega$. 

For each pointing period, the output voltage measures a spin-synchronous modulation dominated by the dipole, which is used to recover a gain calibration factor. The dipole term, $P\ast D$, is dominant (by two orders of magnitude) once strong foreground regions are masked, so for calibration purposes we have $\Vout = G(P\ast D)$. For each pointing period, calibration is performed by fitting $\Vout$ to a model of the beam-convolved dipole:

\begin{equation}
\Gest = {\Vout\over(P\ast D)_{\rm model}},
\end{equation}
where $\Gest$ is the estimated gain, while $G =\Vout / (P\ast D)$ is
the true gain. For the 2013 release, we rely on the a priori knowledge of the solar dipole from \WMAP\ \citep{hinshaw2009}.  Relative changes of the gain $G(\phat)$ between sky circles contribute to the monopole term $M(\phat)$, and are removed by destriping. Assuming an error-free knowledge of the true dipole and of the full beam pattern, then $\Vout = (P\ast D)_{\rm model}$ and $\Gest = G$, so one would recover the true value of the gain, preserving the entire power entering the beam. However, full beam convolution was not implemented in the LFI and HFI pipelines for the 2013 release, and we adopted a ``nominal beam approximation" approach.

\subsection{Nominal beam approximation}

We approximate the beam pattern with a ``pencil beam," i.e., a normalized delta-function:
\begin{equation}
(P\ast D)_{\rm model} = \int_{4\pi} \Ppencil D_{\rm model} d\Omega =
\Ppencil\ast D_{\rm model}.
\end{equation}
Neglecting intrinsic errors in the model dipole ($D_{\rm model} \approx
D$), the estimated gain is
\begin{equation}
\Gest = G{(P\ast D)\over(\Ppencil\ast D)}.
\end{equation}
We can write $P$ as the sum of two terms, $P = \Pnominal + \Pside$, where \Pnominal\ represents the contribution of the nominal beam (defined as the portion of the beam defining the window function, see Sect.~\ref{sec:calibration-and-transfer-function}) and \Pside\ is the contribution from the (near and far) sidelobes. We have
\begin{equation}
\Gest = G{(\Pnominal\ast D) + (\Pside\ast D)\over(\Ppencil \ast D)}.
\end{equation}
The convolution of the nominal beam to the dipole is nearly identical to an ideal pencil beam convolution, except that only a fraction $(1-\fsl)$ of the antenna gain is contained in the nominal beam, so we have
\begin{equation}
 \Pnominal \ast D \approx (1 - \fsl) \Ppencil \ast D = (1 - \fsl) D,
 \label{eq:mainVersusPencil}
\end{equation}
where \fsl\ is the solid angle fraction of the sidelobes.  For the 70, 100, and 143\,GHz channels, $\fsl \approx 0.7$, 0.5, and 0.5\% (see Sect.~\ref{subsec:beamdefinitions}).  In this formalism, we assume that the time response of the bolometers has been accurately deconvolved; thus no additional time constant corrections are needed.  This approximation is true to high accuracy ($<0.2\,\%$), since the dipole changes very little within the nominal beam solid angle.  Rearranging the terms, we find
\begin{equation}
 \Gest = G(1-\fsl)(1+\phiD),
 \label{eq:CorrectionFactor}
\end{equation}
where $\phiD \equiv (\Pside\ast D)/(\Pnominal\ast D)$ represents the coupling of the dipole with sidelobes and depends on the detailed structure of the sidelobes and the scanning strategy.  Equation~\ref{eq:CorrectionFactor} gives the correction factor that one should apply to recover the proper gain factor if one models the beam as a pencil beam.

\subsection{Rescaling factor}

To obtain the reconstructed sky signal after calibration, \Tskyest(\xhat), the dipole term needs to be subtracted from the timelines (Eq.~\ref{eq:Vout}).  Since we are interested in the signal from the sky entering the nominal beam, the coupling of \Tsky\ with sidelobes represents a spurious additive signal.  So we write the timeline as
\begin{equation}
\Vout(\xhat) = G(\Pnominal\ast\Tsky + \Pside\ast\Tsky)(\xhat) + M(\phat).
\end{equation}
In the ideal case of an error-free, full-convolution model, $\Gest = G$, one gets
\begin{equation}
 \Tskyest = {\Vout\over\Gest} = \Pnominal\ast\Tsky + \Pside\ast\Tsky.
 \label{eq:Tskyest}
\end{equation}
For angular scales larger than the nominal beam, a relation similar to Eq.~(\ref{eq:mainVersusPencil}) holds for \Tsky, i.e., $\Pnominal\ast\Tsky \approx (1-\fsl) \Tsky$. With this and Eq.~(\ref{eq:Tskyest}), we obtain
\begin{equation}
\Tsky =\Tskyest{1-\phisky\over 1 - \fsl},
\label{eq:Tskycoupling}
\end{equation}
where $\phisky \equiv (\Pside\ast\Tsky)/ \Tskyest$ represents the coupling of the sidelobes with the sky signal (other than the dipole), i.e., the straylight pickup.

As mentioned, however, in the 2013 release both LFI and HFI used the nominal beam approximation. In this case, the estimated gain is given by Eq.~(\ref{eq:CorrectionFactor}), and we have
\begin{equation}
\Tskyest = {\Vout\over \Gest} = {(\Pnominal\ast\Tsky) + (\Pside\ast\Tsky) \over (1-\fsl)(1+\phiD)}.
\end{equation}
Solving for the true sky temperature we find
\begin{equation}
\Tsky = \Tskyest \left[ 1 - {\phisky \over (1-\fsl)} + \phiD\right] \approx \Tskyest \left(1 - \phisky + \phiD\right).
\label{eq:Tsky_corr--appendix}
\end{equation}
This expression does not contain the term $\fsl$ to first order; however, the (smaller) correction terms $\phiD$ and $\phisky$ need to be evaluated to derive the proper normalization of the temperature maps and of the power spectra.

In summary, to evaluate the true sky temperature from the estimated one, some correction factors need to be evaluated to properly account for the beam. These correction terms depend on the model assumptions.  For a full convolution model, the rescaling factor is $\eta = (1-\phisky)(1-\fsl)$.  For a pencil beam approximation, currently applied by both \Planck\ instruments, the rescaling factor is $\eta = (1 - \phisky + \phiD)$. 

If one were to apply a correction, a rescaling factor $\eta$ could be directly applied to the temperature map to give a properly calibrated map. Alternatively, the window function should be rescaled by $\eta^{-2}$.

\subsection{Semi-analytical expression for $\phi_{\rm D}$}
\label{sec:Expression-for-phiD}

A useful expression for $\phiD$ can be obtained as follows. 
The dipole seen through any particular line of sight (LOS) will be proportional to the dot product of the dipole vector with that of the line of sight:
\begin{equation}
D = D_0\left[
 \sin \Theta_\mathrm{LOS}\cos\left(\Phi_\mathrm{LOS}-\Phi_\mathrm{D}\right) + \cos \Theta_\mathrm{D}\sin \Theta_\mathrm{LOS}
       \right],
\label{eq:fullDipole}
\end{equation}
where $D_0 = 3.35$\,mK, the amplitude of the dipole, $\Theta_\mathrm{D}$ and $\Phi_\mathrm{D}$ are the polar and azimuthal angles of the direction of the dipole maximum on the sky\footnote{Note that these angles should include the change to the dipole induced by the satellite motion with respect to the Sun, in addition to that induced by the Sun's motion with respect to the CMB. We have neglected such boosting effects so far -- see~\cite{planck2013-pipaberration}.},
and $\Theta_\mathrm{LOS}$ and $\Phi_\mathrm{LOS}$ are the same for the direction of the nominal beam observation. For each ring, choosing a coordinate system for which the $z$-axis is aligned with the satellite spin axis, and around which both the LOS and the dipole axis will rotate, ensures that the last term in Eq.~(\ref{eq:fullDipole}) is constant.  Since a constant is removed for each stable pointing period~\citep{planck2013-p03f}, we ignore this and write
\begin{equation}
 D 
 = 
 D_0\sin \Theta_\mathrm{LOS}\cos\left(\Phi_\mathrm{LOS}-\Phi_\mathrm{D}\right).
 \label{eq:constantlessDipole}
\end{equation}
Specializing, when the line of sight (LOS) is the nominal beam, the signal seen will be
\begin{equation}
 D_\mathrm{main} 
 = 
 D_0\sin \Theta_\mathrm{main}\cos\left(\Phi_\mathrm{main}-\Phi_\mathrm{D}\right).
 \label{eq:mainDipole}
\end{equation}
Similarly, when we observe a point on the sky through a sidelobe, we have
\begin{equation}
 D_\mathrm{sl} 
 = 
 D_0\sin \Theta_\mathrm{sl}\cos\left(\Phi_\mathrm{main}+\Delta \Phi_\mathrm{sl}-\Phi_\mathrm{D}\right),
 \label{eq:fsl2Dipole}
\end{equation}
where $\Delta \Phi_\mathrm{sl} = \Phi_\mathrm{sl} - \Phi_\mathrm{main}$ is the difference in azimuthal angles between the nominal beam and the sidelobe of interest. This yields:
\begin{eqnarray}
 D_\mathrm{sl} 
  && =
  D_0\sin \Theta_\mathrm{sl}
  \cos\left(\Delta \Phi_\mathrm{sl}\right)\cos\left(\Phi_\mathrm{main}-\Phi_\mathrm{D}\right)
  \nonumber
  \\
  && \quad - 
  D_0\sin \Theta_\mathrm{sl}
  \sin\left(\Delta \Phi_\mathrm{sl}\right)\sin\left(\Phi_\mathrm{main}-\Phi_\mathrm{D}\right).
 \label{eq:fsl3Dipole}
\end{eqnarray}

Because \Planck\ calibrates by ``locking in'' on, or fitting to, a sinusoidal signal of the form $\cos\left(\Phi_\mathrm{main}-\Phi_\mathrm{D}\right)$ (Eq.~\ref{eq:mainDipole}), the second term in Eq.~(\ref{eq:fsl3Dipole}) is irrelevant, and the far sidelobes will make a contribution to the gain according to
\begin{equation}
 \phiD
 =
 {\Pside\ast D\over \Pnominal\ast D}
 =
 \frac
 {\int_{4\pi}d\Omega P\left(\Theta_\mathrm{sl},\Phi_\mathrm{sl}\right)\sin \Theta_\mathrm{sl}\cos\left(\Delta \Phi_\mathrm{sl}\right)}
  {4\pi\sin \Theta_\mathrm{main}}.
\end{equation}

Integrations over {\tt GRASP} models of the HFI far sidelobes indicate that $\phi_{\rm D} \approx f_\mathrm{FSL}/3$.

\section{Assessment of LFI and HFI with respect to beams and calibration}
\label{sec:LFIHFIassessment}

\subsection{LFI}
\label{subsec:statusLFI}

\subsubsection{Evaluation of $\phi_{\rm sky}$}

The term $\phisky$ turns out to be small, but it is difficult to quantify because it requires a full convolution of a sky model with the full beam, and it is frequency dependent.  We have performed a simplified simulation for a 70\,GHz channel (LFI18M), and find values for \phisky\ ranging between 0.05\,\% and 0.2\,\% throughout a full survey.  More realistic simulations have been produced recently on an LFI 30\,GHz channel (LFI27M), for both Survey~1 and Survey~2.  We find:

Survey~1, $\langle\phisky\rangle  = 0.05\,\%$,	$0.01\,\% <\phisky< 0.23\,\%$;

Survey~2, $\langle\phisky\rangle  = 0.06\,\%$,	$0.01\,\% <\phisky< 1.11\,\%$.
 
\noindent (Differences between odd and even Surveys are to be expected, given the differences in scanning described in \citet{planck2013-p01} section 4.1.) These results are consistent with our preliminary findings, and indicate a typical value of $\phisky \approx 0.05\,\%$ at all \Planck\ frequencies.

\subsubsection{Evaluation of $\phi_{D}$}
\label{sec:phiD}

We can estimate $\phiD$ by taking average values of the convolutions involved in its definition in Eq.~(\ref{eq:Tsky_corr--appendix}). The denominator $\langle\Pnominal\ast D\rangle = 3.35$\,mK is the dipole amplitude. To estimate $\langle\Pside\ast D\rangle$, we have computed the convolution of the LFI sidelobes, modelled with {\tt GRASP}, with the dipole.  

Note that \phiD\ contains the effects of both near and far sidelobes; however, it is the dipole modulation through the near sidelobes that contributes most to its value.  Destriping removes the contribution from sidelobes nearly aligned with the spacecraft spin axis (such as the ``primary spillover,'' see~\citealt{planck2013-p01}), which produces a nearly constant offset for each pointing period. After destriping, we find that $\langle\Pside\ast D\rangle$ is approximately $3\microK$ at 70\,GHz and $5\microK$ at 30\,GHz, leading to $\langle\phiD\rangle \approx 0.1\,\%$ and $\langle\phiD\rangle \approx 0.15\,\%$, respectively. 

\begin{figure}
\includegraphics[width=88mm]{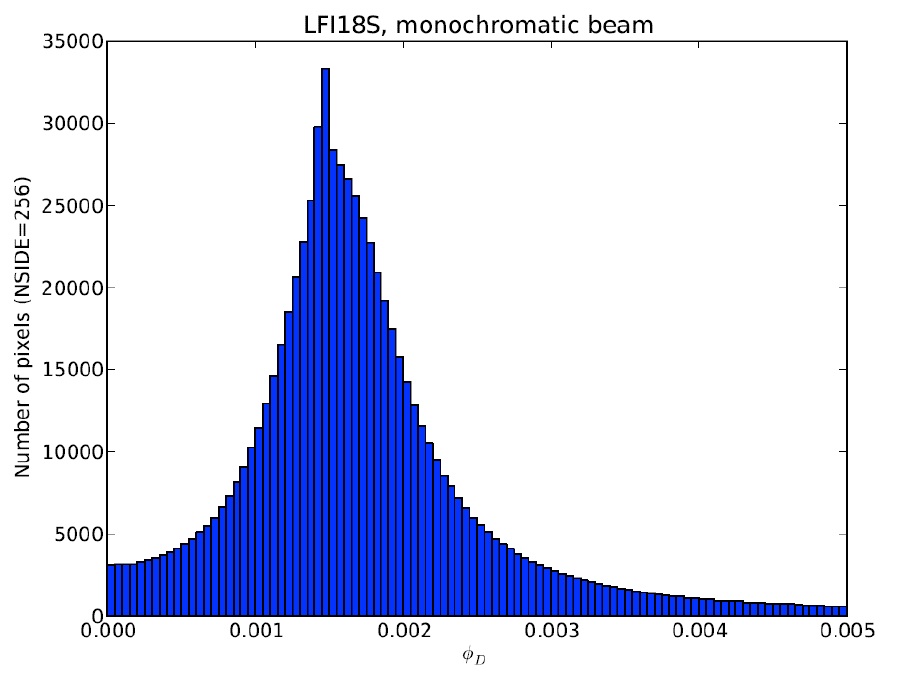}
\caption{\phiD\ for a simulation using a 70\,GHz beam.}
\label{fig:Marco2}
\end{figure}

Figure \ref{fig:Marco2} shows the pixel histogram for a simulation where the value of \phiD\ was computed in every sky direction for an LFI 70\,GHz channel.  The median value, $\phiD \approx 0.16\,\%$, is in good agreement with our first-order estimate.

In conclusion, the correction terms \phiD\ and \phisky\ are of the same order (0.15\,\% and 0.05\,\%, respectively) and tend to cancel each other in Eq.~(\ref{eq:Tsky_corr--appendix}), resulting in a net effect of $\approx 0.1\,\%$ on the LFI calibrated map. For the 2013 release, we have not carried out a systematic simulation for all channels; rather we have assumed a conservative residual uncertainty of 0.2\,\% in the LFI calibrated maps due to sidelobe convolution effects \citep[Table~8]{planck2013-p02b}.  More detailed analysis is planned for the 2014 release.

\subsection{HFI}
\label{subsec:statusHFI}

The effective beam window function, including the effects of both calibration and map reconstruction, is based on measurements of the scanning beam \citep{planck2013-p03c}. These are obtained from Mars observations after deconvolution of the time response of the detection chain.  The effective beam window function is used to correct the cross-power spectra used in the cosmological analysis.  Figure~10 of \citet{planck2013-p03c} shows this scanning beam, and provides estimates of the near sidelobe contribution to the solid angle between 0.1 and 0.2\,\% at 100 and 143\,GHz, respectively. In the subsections below, we estimate the impact of the approximations used in estimating the nominal beam and near sidelobe region (up to 5\deg\ from the beam centre) through the use of Saturn and Jupiter observations.  We estimate the bias introduced by these approximations, as well as the bias in the calibration factor due to the far sidelobes. As previously stated, we make no attempt to recalibrate the 2013 data release, but only show that the LFI--HFI differences at intermediate multipoles are understood within the present uncertainties. 

\subsubsection{Nominal beam and near sidelobes}
\label{subsubsec:mainandnear}

The \emph{nominal beam} for HFI is measured to roughly 15\arcm\ with Mars observations. This provides most of the information needed for the high-$\ell$ analysis used to obtain the 2013 \Planck\ cosmological results. Jupiter observations show shoulders extending to 20\arcm\  for all channels~\citep[see fogire~10 of][]{planck2013-p03c}. Features in this region are small and difficult to measure. Their integrated contribution, however, is not negligible at the level of a few tenths of a percent. Table~2 and figure~10 of \citet{planck2013-p03c} show estimates of the near sidelobe contribution to the solid angle of between 0.1 and 0.22\,\%. 

In addition, pre-launch optical calculations show that at a few FWHM from the beam centre, the HFI beam will have a diffraction pattern from the secondary mirror edges (the pupil of the \Planck\ optical system).  Figure~14 of \citet{planck2013-p03c} shows that at 353\,GHz there are near sidelobes with a diffraction signature, which have not been accounted for. These features are expected from optical calculations, and they were seen in the pre-launch test campaign. A reanalysis of Saturn and Jupiter data has been carried out for all frequencies, and the near sidelobe wing due to diffraction by the edge of the secondary is seen at least at 143\,GHz and higher frequencies.  Additionally, the sidelobe profiles of \cite{planck2013-p03c} led to a removal of the residual time response effects with a weighting that also removed optical near sidelobe signal around 20\arcmin.  

\begin{figure} 
 \includegraphics[width=88mm]{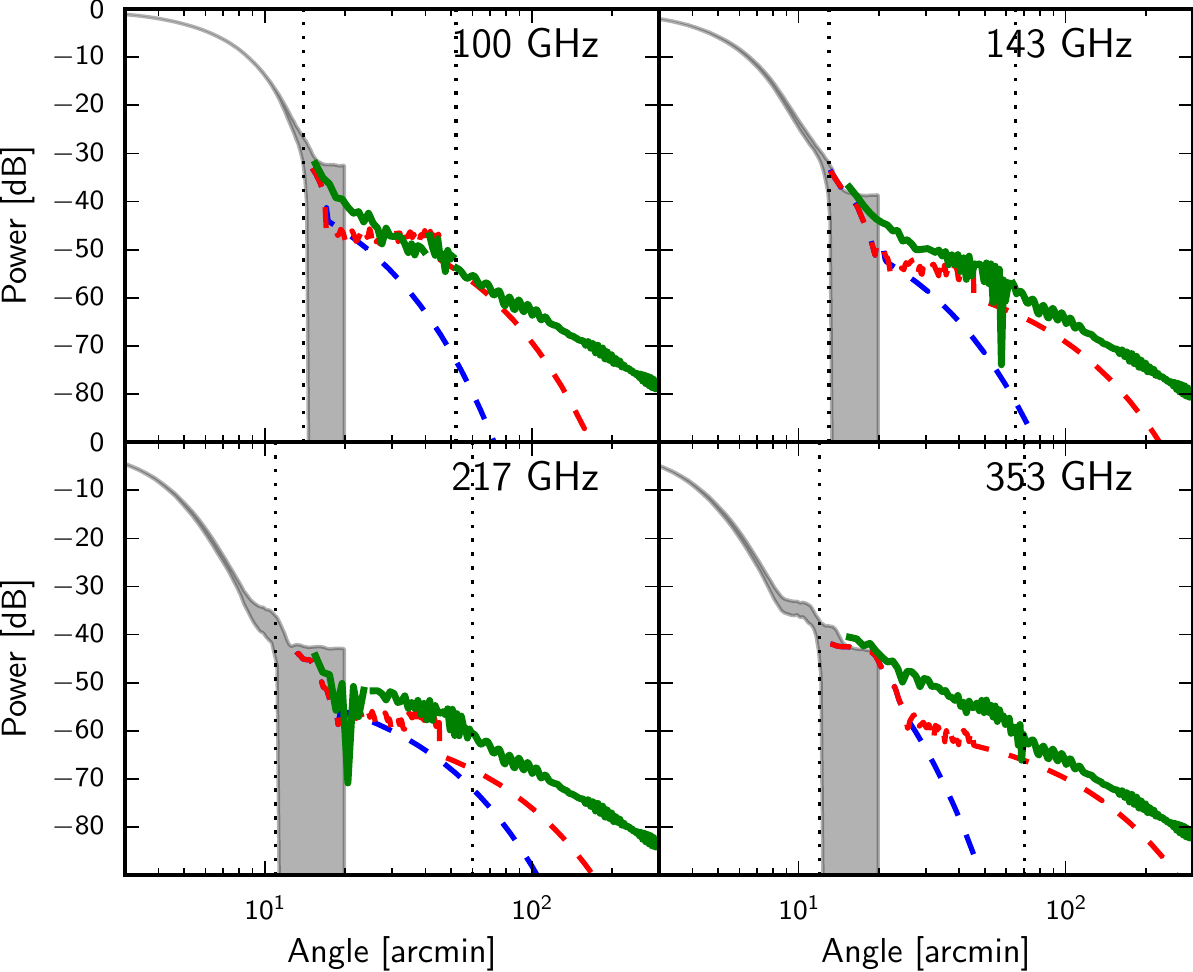}
 \caption{HFI beam and planet measurement profiles.  The grey shaded region shows the $1\,\sigma$ errors in the nominal beam profile used to make the effective beam window function.  The dashed blue and red lines are the best-case and worst-case near sidelobe estimates from Jupiter, presented in figure~10 of~\cite{planck2013-p03c}.  The solid green curve is the hybrid beam profile used to derive correction factors here.  The vertical dotted lines indicate the radial extent of each portion of the hybrid beam:  the nominal beam is at small angles; at intermediate angles Jupiter data are used; and at large angles the scaled {\tt GRASP} simulation is used.}
 \label{fig:HFI_nsl}
\end{figure}

In order to assess the impact of a more refined accounting of the near sidelobes, we have constructed a new hybrid beam profile that consists of the nominal beam, azimuthally averaged Jupiter observations, and {\tt GRASP} simulations.  The azimuthally averaged Jupiter observations extend to 2\deg, and include all residual time response and mirror dimpling effects.  The astrophysical background is subtracted from each observation, and we fit a baseline and a scaling factor to match the Jupiter profile to the Mars profile at an overlap angle.  The {\tt GRASP} simulations consist of all eight 100\,GHz detectors simulated at five frequencies across the band.  A 100\,GHz full-frequency beam is constructed by weighting the simulations with the map-making weights, detector bandpass, and planet spectral energy distribution.  At large angle, the {\tt GRASP} calculations are dominated by diffraction around the secondary mirror.  The diffraction pattern has a similar angular dependence at all HFI frequencies, and its normalization is set by the edge taper of the secondary.  We therefore scale the 100\,GHz simulations to higher frequencies by fitting to the Jupiter profile.  Figure~\ref{fig:HFI_nsl} shows the radial profiles of the hybrid beams, compared with those given in~\citet{planck2013-p03c}.

The Saturn-Jupiter combination (Fig.~\ref{fig:HFI_nsl}) gives a diffraction pattern similar to that predicted and measured preflight \citep{tauber2010b}.  This increases the estimate of the contribution of the near sidelobes to the total solid angle of beam, as given in figure~10 of \citet{planck2013-p03c}.  However, the dominant contribution to the increase in solid angle at 100 and 143\,GHz occurs around 20\arcm, and is due to removing the assumption of azimuthal symmetry and applying no time response correction when computing the radial profile.

\begin{figure}
\includegraphics[width=88mm]{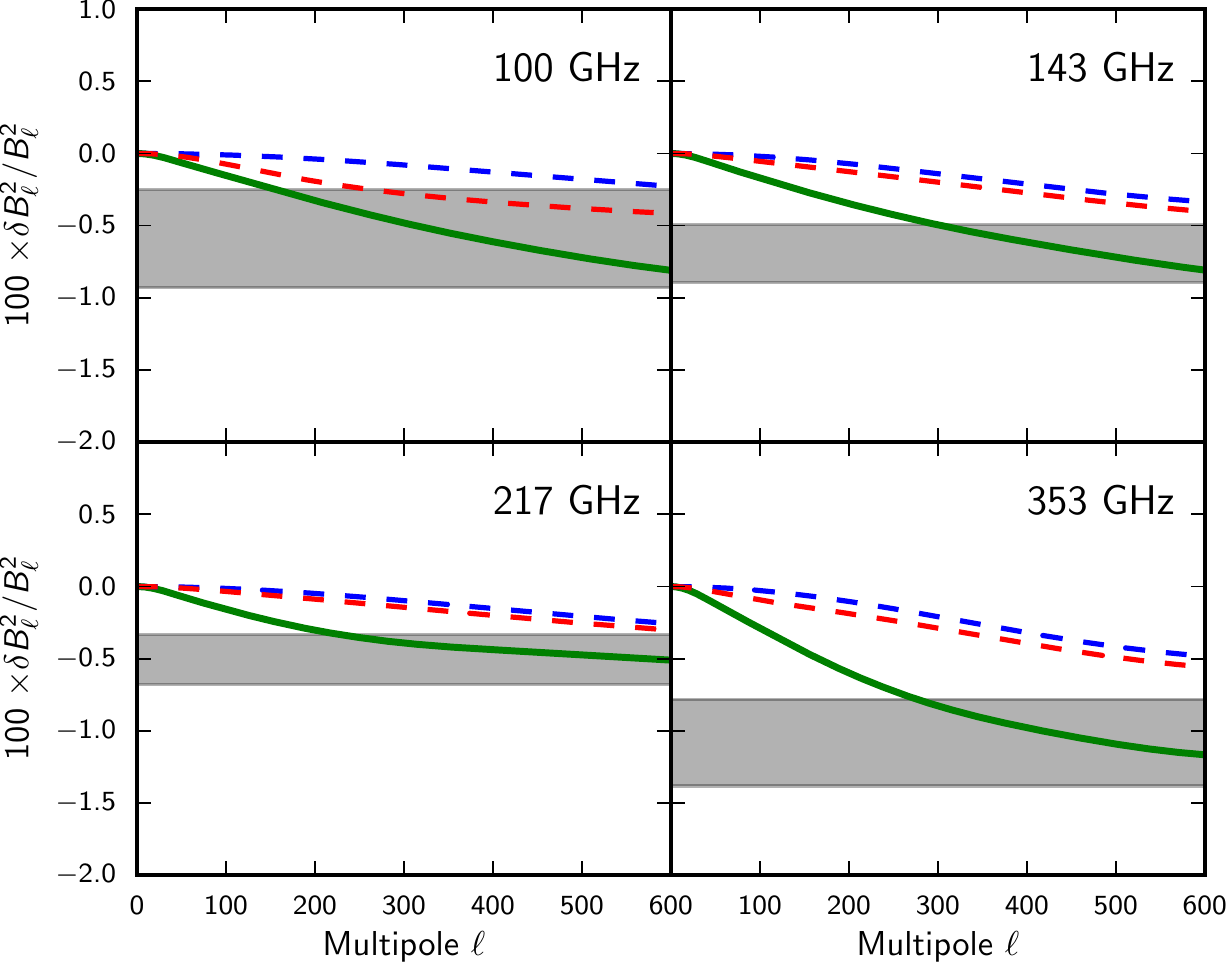}
\caption{Effective beam window function corrections based on the hybrid beam profile.  The green curve shows the correction based on the hybrid model presented here.  The dashed blue and red lines are the best-case and worst-case near sidelobe estimates from Jupiter, as shown in Fig.~10 of \cite{planck2013-p03c}.  The grey shaded region shows the $\pm1\,\sigma$ errors in the correction to the solid angle.}
 \label{fig:HFI_nsl_wfn}
\end{figure}

This also affects the shape of the effective beam window function, with a decrease similar to that illustrated in the green curve in Fig.~19 of \cite{planck2013-p03c}, although with a larger amplitude and large error bars (Fig.~\ref{fig:HFI_nsl_wfn}).  The effect of these corrections for near-sidelobe power missing in the beams used in the 2013 results is illustrated in Fig.~\ref{fig:beamcorrected}. The error in the integral of the beam profile is dominated by the background removal and inter-calibration of the Mars and Jupiter data.

\subsubsection{Far sidelobes}

Estimates from \citet{tauber2010b} put the integrated solid angle of the far sidelobes at roughly 0.3\,\% of the total beam solid angle. While this is small, a significant fraction of this solid angle appears near the satellite spin axis. This means that the signal seen through this part of the sidelobes is not modulated by the satellite rotation, and therefore does not contribute to the calibration.  Using calculations such as those in Sect.~\ref{sec:phiD}, we estimate the far sidelobe effect on the calibration to be of order one-third the total solid angle for the far sidelobes. For HFI, this constitutes about 0.1\,\%  at 100 and 143\,GHz.

\end{document}